\documentclass[review,authoryear,3p,times,10pt]{elsarticle}

\usepackage{multirow,setspace,times,amssymb,amsmath,graphicx,color,rotating,subfigure,url}
\usepackage{lineno,color}
\usepackage{natbib}
\usepackage{booktabs}%
\usepackage{longtable}%
\usepackage{rotating}
\usepackage{lscape}
\usepackage{ulem}
\usepackage{threeparttable}
\graphicspath{{Figures/}}
\usepackage[table]{xcolor}
\usepackage{tabularx}
\usepackage{graphicx} 
\usepackage{epstopdf}
\usepackage{mathrsfs}
\usepackage{makecell}
\usepackage[bookmarks=true,colorlinks,linkcolor=blue,anchorcolor=blue,citecolor=blue,unicode]{hyperref}
\usepackage{bookmark}

\usepackage{todonotes}
\usepackage{ragged2e}
\usepackage[font=small]{caption}

\hypersetup{CJKbookmarks=true}%
\makeatletter
\def\ps@pprintTitle{%
\let\@oddhead\@empty 
\let\@evenhead\@empty
}

\journal{Energy Economics}

\begin{document}

\begin{frontmatter}

\title{Moment connectedness and driving factors in the energy-food nexus: A time-frequency perspective}

\author[SB,UVSQ,EMLV]{Yun-Shi Dai}
\author[WHUT1,WHUT2]{Peng-Fei Dai}
\author[UVSQ]{St{\'e}phane Goutte}
\author[EMLV,ISVNU]{Duc Khuong Nguyen}
\author[SB,RCE,DM]{Wei-Xing Zhou\corref{CorAuth}}
\ead{wxzhou@ecust.edu.cn}
\cortext[CorAuth]{Corresponding author.} 

\address[SB]{School of Business, East China University of Science and Technology, Shanghai 200237, China}
\address[UVSQ]{UMI SOURCE, University Paris-Saclay, UVSQ, IRD, France}
\address[EMLV]{De Vinci Research Center, De Vinci Higher Education, Paris, France}
\address[WHUT1]{School of Management, Wuhan University of Technology, Wuhan 430070, China}
\address[WHUT2]{Research Institute of Digital Governance and Management Decision Innovation, Wuhan University of Technology, Wuhan 430070, China}
\address[ISVNU]{International School, Vietnam National University, Hanoi, Vietnam}
\address[RCE]{Research Center for Econophysics, East China University of Science and Technology, Shanghai 200237, China}
\address[DM]{School of Mathematics, East China University of Science and Technology, Shanghai 200237, China}

\begin{abstract}
With escalating macroeconomic uncertainty, the risk interlinkages between energy and food markets have become increasingly complex, posing serious challenges to global energy and food security. This paper proposes an integrated framework combining the GJRSK model, the time-frequency connectedness analysis, and the random forest method to systematically investigate the multimoment connectedness within the energy-food nexus and explore the key drivers of various spillover effects. The results reveal significant multidimensional risk spillovers with pronounced time variation, heterogeneity, and crisis sensitivity. Return and skewness connectedness are primarily driven by short-term spillovers, kurtosis connectedness is more prominent over the medium term, while volatility connectedness is dominated by long-term dynamics. Notably, crude oil consistently serves as a central transmitter in diverse connectedness networks. Furthermore, the spillover effects are influenced by multiple factors, including macro-financial conditions, oil supply-demand fundamentals, policy uncertainties, and climate-related shocks, with the core drivers of connectedness varying considerably across different moments and timescales. These findings provide valuable insights for the coordinated governance of energy and food markets, the improvement of multilayered risk early-warning systems, and the optimization of investment strategies.
\end{abstract}


\end{frontmatter}


\section{Introduction}

Energy and food are fundamental resources for human survival, social stability, and economic development. However, the current situation of global energy and food security is severe and complex, facing multiple risks and challenges. According to the {\textit{Global Energy Review 2025}}\footnote{\url{https://www.iea.org/reports/global-energy-review-2025}} published by the International Energy Agency, global energy demand grew by 2.2\% in 2024, significantly surpassing the average annual growth rate of 1.3\% over the past decade, with substantial increases across all energy sources, including oil and natural gas. The {\textit{WFP 2025 Global Outlook}}\footnote{\url{https://www.wfp.org/publications/wfp-2025-global-outlook}}, released by the World Food Programme, points out that approximately 343 million people faced acute food insecurity in 2024, with the number of people suffering from catastrophic hunger reaching a record high. While demand for energy and food continues to rise, factors such as climate change, frequent extreme weather events, escalating geopolitical conflicts, and the global economic slowdown have jointly exacerbated supply-side uncertainties, exposing the potential vulnerability of global energy and food systems. In particular, during the COVID-19 pandemic and the Russia-Ukraine conflict, prices of energy and food experienced frequent spikes and crashes, further intensifying the global energy and food crisis \citep{Behnassi-ElHaiba-2022-NatHumBehav,Shepard-Pratson-2022-NatEnergy,Shumilova-Tockner-Sukhodolov-Khilchevskyi-DeMeester-Stepanenko-Trokhymenko-HernandezAgueero-Gleick-2023-NatSustain}.

Stable and efficient markets are essential for safeguarding energy and food security, yet the dramatic and frequent price fluctuations in recent years have significantly increased risks within both energy and food markets. With the accelerating process of commodity financialization and market integration, the interdependence between energy and food has intensified, displaying a more intricate relationship of both competition and cooperation \citep{Ji-Bouri-Roubaud-Shahzad-2018-EnergyEcon,Han-Jin-Wu-Zeng-2020-IntRevFinancAnal,Iqbal-Naeem-Karim-Haseeb-2024-EurJFinanc}. On the one hand, energy serves as a critical input in agricultural production. Price fluctuations of oil and its derivatives directly affect the cost structures of farming, transportation, and food processing, thereby generating linkages with food prices. On the other hand, the rapid development of biofuels has incorporated certain food commodities, such as corn and soybeans, into the energy system, leading to direct competition for resources and further strengthening the coupling between the two markets. Furthermore, energy and food markets exhibit similar sensitivities to external shocks, including macroeconomic policy changes, shifts in financial market sentiment, extreme weather events, and geopolitical conflicts. As a result, risk spillovers between the two markets tend to become more pronounced during major global crises \citep{FernandezAviles-Montero-SanchisMarco-2020-EurJFinanc,Cui-Maghyereh-2023-IntRevFinancAnal}. Against this backdrop, a comprehensive investigation into risk transmission within and across energy and food markets, as well as the identification of their key driving factors, is crucial for enhancing policy effectiveness, improving early-warning systems, and maintaining global resource security.

Most traditional risk analyses focus primarily on returns and volatility, often ignoring higher-order information such as skewness and kurtosis in return distributions. However, financial asset returns in practice typically exhibit non-normality, asymmetry, and volatility clustering—stylized facts that are closely associated with higher-order moments \citep{Ang-Timmermann-2012-AnnuRevFinancEcon,Paolella-Polak-2015-JEconom,Ding-2023-JEconom}. Skewness (the third-order moment) captures the asymmetry of return distributions and reflects potential directional risks in the market, while kurtosis (the fourth-order moment) measures the probability and magnitude of extreme events, representing tail risks \citep{Leon-Rubio-Serna-2005-QRevEconFinanc,Nakagawa-Uchiyama-2020-Mathematics}. As important statistical characteristics of return distributions, higher-order moments play indispensable roles in asset pricing, risk management, and portfolio optimization \citep{Zhou-Wu-Liu-Rognone-2023-NatCommun,Cui-Maghyereh-2023-IntRevFinancAnal,Hao-Pham-2024-EnergyEcon}. In addition, conventional rolling-window-based dynamic connectedness analyses are highly sensitive to the subjectively chosen window length and suffer from the loss of initial observations. In contrast, the connectedness approaches based on the time-varying parameter vector autoregression (TVP-VAR) model introduce a parameter evolution mechanism to flexibly capture the dynamic trends in connectedness structures over time and effectively mitigate the impact of outliers and data loss \citep{Koop-Korobilis-2014-EurEconRev,Antonakakis-Chatziantoniou-Gabauer-2020-JRiskFinancManag,Chatziantoniou-Gabauer-Gupta-2023-ResourPolicy}.

With the time-varying higher-order moments computed by the GJRSK model, this study employs both the TVP-VAR-DY and TVP-VAR-BK approaches to systematically examine the time-frequency connectedness of return, volatility, skewness, and kurtosis between energy and food markets. Furthermore, the random forest model is applied to identify the key drivers behind various spillover effects. Our findings reveal significant multidimensional risk spillovers within the energy-food system, with pronounced time variation, heterogeneity, and crisis sensitivity. Return connectedness is dominated by short-term spillovers, while volatility connectedness is dominated by long-term spillovers. Skewness connectedness exhibits strong short-term spillovers during market turbulence, suggesting that asymmetries in return distributions are primarily transmitted over short horizons during extreme events. Kurtosis connectedness shows more prominent spillovers in the short and medium term, with weaker long-term effects. Additionally, the net pairwise time-frequency connectedness networks effectively capture the complex spillover relationships and critical transmission pathways, highlighting the central role of crude oil in most network structures. The factor importance analysis based on the random forest model indicates multifactorial influences on the energy-food nexus, with macroeconomic and financial conditions, crude oil supply-demand fundamentals, external uncertainties, and climate shocks showing high importance. Moreover, the key driving factors of connectedness vary significantly across moments and timescales, further uncovering the multidimensionality and heterogeneity of spillover effects.

This study aims to comprehensively investigate risk transmission between international energy and food markets and to clarify the core drivers of various risk spillovers. Compared with the existing literature, our contributions lie in three key aspects. First, building on prior research that primarily focuses on the first-order (return) and second-order (volatility) moments, we incorporate time-varying skewness and kurtosis into the analytical framework to capture asymmetric and extreme tail risk spillovers in energy and food markets. The findings deepen the understanding of risk interactions within the energy-food nexus, offering new evidence on multidimensional risk transmission and enriching cross-market spillover analysis. Second, this study utilizes a novel TVP-VAR-based connectedness methodology from both time- and frequency-domain perspectives, effectively overcoming the subjectivity and information loss inherent in traditional rolling-window approaches. In addition, the connectedness network analysis reveals heterogeneous characteristics of higher-order spillovers across short-, medium-, and long-term horizons, and explicitly delineates market roles and transmission pathways within different connectedness networks. Third, this research innovatively integrates the random forest model to explore the key drivers of risk spillovers. By combining econometric and machine learning techniques, we enhance the robustness of factor identification and provide valuable insights into the underlying mechanisms and sources of multidimensional moment connectedness. Overall, this study offers important theoretical and practical implications for improving risk early-warning systems, formulating multilayered policy interventions, and optimizing asset allocation and risk management strategies.

The remainder of this paper is organized as follows. Section~\ref{S1:LitRev} reviews the related literature, focusing on higher-order moment connectedness. Section~\ref{S1:Methodology} introduces the GJRSK approach, the TVP-VAR-based connectedness framework, and the random forest model. Section~\ref{S1:Data} presents the data sources and descriptive statistics. Section~\ref{S1:EmpAnal} discusses the empirical results, including higher-order moment measures, time-frequency connectedness analysis, and factor importance assessment. Section~\ref{S1:Conclude} concludes and provides policy implications based on the main findings.

\section{Literature review}
\label{S1:LitRev}

The phenomenon whereby the risk of one market or asset is transmitted to others is commonly referred to as risk spillover or risk contagion. Risk spillover has become an important theoretical construct for understanding financial market interdependence and systemic risk. Early related studies mainly focused on the covariance structure or correlation-based co-movement measures between markets or assets, analyzing risk relationships from a static perspective \citep{King-Wadhwani-1990-RevFinancStud,Solnik-Boucrelle-LeFur-1996-FinancAnalJ,Longin-Solnik-2001-JFinanc,Hartmann-Straetmans-deVries-2004-RevEconStat}. However, with the deepening of economic globalization and financial integration, such static methods have proven inadequate in capturing the dynamic evolution of risk spillovers, especially after the onset of the global financial crisis in 2008. In response, \cite{Diebold-Yilmaz-2009-EJ,Diebold-Yilmaz-2012-IntJForecast,Diebold-Yilmaz-2014-JEconom} proposed a dynamic connectedness framework based on the vector autoregressive (VAR) model. This methodology utilizes variance decomposition to measure the contribution of each variable to system-wide spillovers and develops directional and net spillover indices, offering powerful tools for modeling and quantifying cross-market and cross-asset risk transmission \citep{Brunetti-Harris-Mankad-Michailidis-2019-JFinancEcon,Chen-Zheng-Hao-2022-JBusRes,Diebold-Yilmaz-2023-JEconom}.

Considering that shocks to economics may affect variables with varying intensities across different frequencies, \cite{Barunik-Krehlik-2018-JFinancEconom} introduced frequency dynamics into connectedness analysis and proposed an innovative connectedness framework based on the spectral representation of variance decomposition, which enables the differentiation of short-, medium-, and long-term interlinkages between markets or assets. This method not only reveals the frequency-domain dynamics of connectedness but also helps uncover its underlying sources, significantly enhancing the precision and depth of risk transmission analysis \citep{Barunik-Bevilacqua-Tunaru-2022-RevEconStat,Cotter-Hallam-Yilmaz-2023-JIntMoneyFinan,Naeem-Qureshi-Farid-Tiwari-Elheddad-2024-AnnOperRes}. However, although the Diebold-Yilmaz and Barun{\'{i}}k-K{\v{r}}ehl{\'{i}}k models perform well in describing connectedness in the time and frequency domains, their dynamic analyses rely on rolling window estimation, which involves subjective choices of window length and may cause data loss. To address these limitations, \cite{Antonakakis-Chatziantoniou-Gabauer-2020-JRiskFinancManag} developed a time-domain connectedness approach based on the TVP-VAR model \citep{Koop-Korobilis-2014-EurEconRev}, which allows for more flexible and robust modeling of potential changes in data structures. Moreover, \cite{Chatziantoniou-Gabauer-Gupta-2023-ResourPolicy} further extended this method to the frequency domain, proposing the TVP-VAR-BK model, which enables dynamic depiction of risk transmission across different timescales within a time-varying parameter framework.

Most of the existing studies on connectedness primarily focus on the first-order moment (return) and second-order moment (volatility), neglecting the potential role of higher-order moments in return distributions \citep{Barunik-Kocenda-Vacha-2016-JFinancMark,Ferrer-Shahzad-Lopez-Jareno-2018-EnergyEcon,Ando-GreenwoodNimmo-Shin-2022-ManageSci}. However, asset returns in financial markets usually deviate from normality \citep{Harvey-Siddique-1999-JFinancQuantAnal,Christoffersen-Diebold-2006-ManageSci}, making risk analyses based solely on mean and variance insufficient to reflect the true risk structure. Therefore, considering higher-order moments in return distributions is critical for asset pricing and risk management \citep{Christoffersen-Heston-Jacobs-2006-JEconom,Amaya-Christoffersen-Jacobs-Vasquez-2015-JFinancEcon,Nakagawa-Uchiyama-2020-Mathematics}. Skewness, the third-order moment, quantifies the asymmetry of return distributions and reflects asymmetric risks in market upswings and downswings. Kurtosis, the fourth-order moment, measures the degree of peakedness and fat tails, offering insights into the probability of extreme events and tail risks under extreme shocks \citep{Jondeau-Rockinger-2003-JEconDynControl,Leon-Rubio-Serna-2005-QRevEconFinanc}. Hence, skewness and kurtosis provide important information on asymmetry and tail risks, contributing to a more nuanced understanding and more accurate assessment of market risks under unexpected events.

In recent years, some scholars have begun to incorporate higher-order moments into connectedness frameworks to explore higher-order risk spillovers across markets, such as in the stock market \citep{Finta-Aboura-2020-JFinancMark,He-Hamori-2021-JIntMoneyFinan} and cryptocurrency market \citep{Apergis-2023-JIntFinancMarkInstMoney,He-Hamori-2024-IntRevFinancAnal}. With the growing financialization of commodity markets, attention has increasingly turned to higher-order moment connectedness both within commodity markets and between them and other markets. \cite{Bouri-Lei-Jalkh-Xu-Zhang-2021-ResourPolicy}  compute realized volatility, skewness, kurtosis, and jumps using 5-minute data from U.S. stock, oil, and gold markets, finding significant higher-order risk spillovers across these markets. Similarly, studies by \cite{Cui-Maghyereh-2023-JCommodMark}, \cite{Bouri-Lei-Xu-Zhang-2023-Energy}, \cite{Hao-Pham-2024-EnergyEcon}, and \cite{Chu-Fan-Zhou-2024-EnergyEcon} examine the higher-order moment connectedness among various sectors, including global oil markets, precious metals and energy, crude oil and clean energy, as well as carbon and energy markets. Moreover, \cite{Zhang-Yang-Li-Hao-2023-JFuturesMark} employ a novel network topology approach to analyze contemporaneous and non-contemporaneous spillovers of low- and high-order risks across energy, precious metals, and agricultural futures, highlighting that the total spillovers of higher-order risks are stronger than those of lower-order risks.

A few studies have further extended the analysis of higher-order connectedness by exploring its potential driving factors. \cite{Cui-Maghyereh-2023-IntRevFinancAnal} examine the influence of the COVID-19 pandemic and the Russia-Ukraine conflict on higher-order connectedness between oil, metal, and agricultural futures markets. \cite{Zhou-Wu-Liu-Rognone-2023-NatCommun} combine time-frequency analysis and quantile methods to reveal the asymmetric impact of climate risk on higher-order connectedness among carbon, metal, and energy markets. From the perspective of systemic risk, \cite{Li-Pei-Zhang-2024-RiskAnal} explore the risk contagion through higher-order moments between climate policy uncertainty, economic policy uncertainty, geopolitical risk, and epidemic risk. \cite{Xie-Bi-Xi-Xu-2025-EnergyEcon} further demonstrate that geopolitical risks significantly intensify higher-order connectedness in global energy markets. When identifying the sources of risk transmission, traditional approaches—such as regression analysis and Granger causality tests—are often constrained by strong model assumptions, particularly when dealing with multi-factor and high-dimensional datasets. In contrast, machine learning techniques have emerged as powerful tools in financial risk research due to their flexible non-parametric modeling and excellent feature identification, which help to uncover the intrinsic mechanisms and multidimensional pathways of risk spillovers \citep{Gu-Kelly-Xiu-2020-RevFinancStud,Karim-Shafiullah-Naeem-2024-IntRevFinancAnal}. For instance, \cite{Zhang-Tang-Tang-2025-JEnvironManage} apply machine learning to explore the macroeconomic drivers of asymmetric spillovers in China's green finance market.

Recent empirical work has substantially advanced our understanding of the energy-food nexus by documenting rich patterns of return and volatility transmission across fossil fuels, renewables, biofuels, and agricultural commodities. Early time-frequency network studies show pronounced heterogeneity in spillovers across horizons. For example, \cite{Kang-Tiwari-Albulescu-Yoon-2019-EnergyEcon} highlight distinct short- and long-run connectedness between crude oil and agricultural commodities, while \cite{Adeleke-Awodumi-2022-JApplEcon} extend these insights by jointly modeling energy, agricultural raw materials, and food markets in a time-frequency framework. Several contributions emphasize the role of extreme events and geopolitical shocks. \cite{Wu-Ren-Wan-Liu-2023-FinancResLett} and \cite{Farid-Naeem-Paltrinieri-Nepal-2022-EnergyEcon} analyze the dynamic links between energy and agricultural commodities during the COVID-19 pandemic and the Russia-Ukraine conflict, uncovering temporary intensifications of spillovers. Methodologically, scholars have expanded beyond standard spectral decompositions. \cite{Polat-Ertugrul-Sakarya-Akgul-2024-ApplEnergy} implement TVP-VAR approaches to better capture evolving dependence, and \cite{Deng-Fang-Zhang-Ma-2023-JKnowlEcon} combine quantile dependence analysis with deep learning techniques to reveal non-linear and tail-specific transmissions. These studies establish a strong empirical foundation but also point to the need for further investigation of higher-order moment connectedness and potential risk drivers.

After reviewing the relevant literature, we identify several critical research gaps despite the substantial progress. First, although some studies have explored higher-order risk spillovers in commodity markets, the systemic higher-order moment connectedness between energy and food markets remains largely underexplored—an issue of increasing importance given the dual global challenges of energy and food security. Second, most existing analyses of dynamic connectedness still rely on rolling-window estimation, which limits the ability to fully capture the structural shifts and evolving patterns of risk spillovers. There remains ample room for further development of dynamic time-frequency analysis based on the TVP-VAR framework. Third, research on the drivers of moment connectedness has so far involved only a limited number of explanatory factors, with few studies adopting machine learning techniques to account for high-dimensional data and heterogeneous effects. To fill these gaps, this paper comprehensively examines the time-frequency connectedness of return, volatility, skewness, and kurtosis between energy and food markets, and incorporates the random forest method to identify the key drivers of different spillovers, thereby offering novel insights into the multidimensional risk transmission and underlying risk sources of the energy-food nexus.

\section{Methodology}
\label{S1:Methodology}

\subsection{GJRSK model} 

To quantify the time-varying volatility, skewness, and kurtosis of energy and staple food markets, we employ the GJRSK model proposed by \cite{Nakagawa-Uchiyama-2020-Mathematics} to construct higher-order moment risk measures. By incorporating the GJR framework into the GARCHSK model \citep{Leon-Rubio-Serna-2005-QRevEconFinanc}, the GJRSK model takes into account not only stylized facts commonly observed in financial time series, such as non-normality and serial correlation, but also possible leverage effects that allow for asymmetric responses to positive and negative shocks.

Given a sequence of asset prices $\{P_{0}, P_{1}, \dots, P_{T}\}$, we define the return of the financial asset at time $t$ as $r_{t} = \ln \left(P_{t} / P_{t-1}\right), t=1,\dots,T$. Let $h_{t}$, $s_{t}$, and $k_{t}$ denote the conditional volatility, conditional skewness, and conditional kurtosis, respectively. Then, the GJRSK model is specified as follows:
\begin{subequations}
  \begin{equation}
    r_{t} = \alpha_{1} r_{t-1} + \varepsilon_{t},
  \label{Eq:function_mean}
  \end{equation}
  \begin{equation}
  h_{t} = \beta_{0} + \beta_{1} \varepsilon_{t-1}^{2} + \beta_{2} h_{t-1} + \beta_{3} \varepsilon_{t-1}^{2} 1_{\{\eta_{t-1}<0\}},
  \label{Eq:function_volatility}
  \end{equation}
  \begin{equation}
  s_{t} = \gamma_{0} + \gamma_{1} \eta_{t-1}^{3} + \gamma_{2} s_{t-1} + \gamma_{3} \eta_{t-1}^{3} 1_{\{\eta_{t-1}<0\}},
  \label{Eq:function_skewness}
  \end{equation}
  \begin{equation}
  k_{t} = \delta_{0} + \delta_{1} \eta_{t-1}^{4} + \delta_{2} k_{t-1} + \delta_{3} \eta_{t-1}^{4} 1_{\{\eta_{t-1}<0\}},
  \label{Eq:function_kurtosis}
  \end{equation}
  \begin{equation}
  \eta_{t} = h_{t}^{-1/2} \varepsilon_{t}, \quad \eta_{t} \big| I_{t-1} \sim g\left(0, 1, s_{t}, k_{t}\right),
  \label{Eq:function_error}
  \end{equation}
\label{Eq:function_GJRSK}%
\end{subequations}
where $\varepsilon_{t}$ denotes the error term, $I_{t-1}$ is the information set available at time $t-1$, and $g\left(0, 1, s_{t}, k_{t}\right)$ represents the probability density function with mean zero, variance one, skewness $s_{t}$, and kurtosis $k_{t}$. The indicator function $1_{\{\eta_{t-1}<0\}}$ equals 1 if $\eta_{t-1}<0$, and 0 otherwise.

The probability density function $g\left(0, 1, s_{t}, k_{t}\right)$ in the GJRSK model is derived from a Gram-Charlier expansion of the Chebyshev-Hermite polynomial, which is expressed as
\begin{subequations}
  \begin{equation}
    g\left(\eta_{t} \big| I_{t-1}\right) = \frac{\varPhi \left(\eta_{t} \right) \varPsi^{2} \left(\eta_{t}\right)}{\Gamma_{t}},
  \label{Eq:function_GJRSK_pdf_g}
  \end{equation}
  \begin{equation}
  \varPhi \left(\eta_{t} \right) = \frac{1}{\sqrt{2\pi h_{t}}} \exp \left( \eta_{t}^{2} - h_{t} \right),
  \label{Eq:function_GJRSK_pdf_varPhi}
  \end{equation}
  \begin{equation}
  \varPsi \left(\eta_{t}\right) = 1 + \frac{s_{t}}{3!} \left( \eta_{t}^{3} - 3\eta_{t} \right) + \frac{k_{t}-3}{4!} \left( \eta_{t}^{4} - 6\eta_{t}^{2} + 3 \right),
  \label{Eq:function_GJRSK_pdf_varPsi}
  \end{equation}
  \begin{equation}
  \Gamma_{t} = 1 + \frac{s_{t}^{2}}{3!} + \frac{\left( k_{t}-3 \right)^{2}}{4!}.
  \label{Eq:function_GJRSK_pdf_Gamma}
  \end{equation}
\label{Eq:function_GJRSK_pdf}%
\end{subequations}

The parameters of the GJRSK model can be estimated by maximizing the log-likelihood function
\begin{equation}
    L_{t} = -\frac{1}{2} \ln h_{t} - \frac{1}{2}\eta_{t}^{2} + \ln \left( \varPsi^{2} \left(\eta_{t}\right) \right) - \Gamma_{t}.
    \label{Eq:function_GJRSK_Loglikelihood}
\end{equation}

\subsection{TVP-VAR-based connectedness approach}

Following \cite{Antonakakis-Chatziantoniou-Gabauer-2020-JRiskFinancManag} and \cite{Chatziantoniou-Gabauer-Gupta-2023-ResourPolicy}, we incorporate the TVP-VAR model developed by \cite{Koop-Korobilis-2014-EurEconRev} into the DY approach \citep{Diebold-Yilmaz-2014-JEconom} and the BK approach \citep{Barunik-Krehlik-2018-JFinancEconom}, thereby constructing the TVP-VAR-DY model in the time domain and the TVP-VAR-BK model in the frequency domain. The TVP-VAR framework overcomes the inherent limitations of the rolling-window VAR approach, as it does not require the arbitrary specification of window sizes. It enables more accurate detection of potential parameter changes, reduces the impact of outliers, and avoids loss of observations. Therefore, we adopt these novel TVP-VAR-based connectedness methods to comprehensively examine the dynamic risk transmission between energy and food markets at the time- and frequency-domain levels.

The TVP-VAR model with $p$-order lags is given by
\begin{subequations}
  \begin{equation}
    \mathbf{y}_{t} = \boldsymbol{\Phi}_{1t} \mathbf{y}_{t-1} + \boldsymbol{\Phi}_{2t} \mathbf{y}_{t-2} + \dots + \boldsymbol{\Phi}_{pt} \mathbf{y}_{t-p} + \boldsymbol{\epsilon}_{t}, \quad \boldsymbol{\epsilon}_{t} \big| \Omega_{t-1} \sim N\left( \mathbf{0}, \boldsymbol{\Sigma}_{t} \right),
  \label{Eq:function_VAR}
  \end{equation}
  \begin{equation}
    vec \left(\boldsymbol{\Phi}_{t} \right) = vec \left(\boldsymbol{\Phi}_{t-1} \right) + \boldsymbol{\xi}_{t}, \quad \boldsymbol{\xi}_{t} \big| \Omega_{t-1} \sim N\left( \mathbf{0}, \boldsymbol{\Xi}_{t} \right),
  \label{Eq:function_TVP}
  \end{equation}
  \label{Eq:function_TVP_VAR}%
\end{subequations}
where $\boldsymbol{\Phi}_{t} = \left( \boldsymbol{\Phi}_{1t}, \boldsymbol{\Phi}_{2t}, \cdots, \boldsymbol{\Phi}_{pt} \right)$, $\Omega_{t-1}$ denotes the information set available at time $t-1$, $\mathbf{y}_{t}$ and $\boldsymbol{\epsilon}_{t}$ are $N \times 1$ vectors,  $\boldsymbol{\xi}_{t}$ and $vec \left(\boldsymbol{\Phi}_{t} \right)$ are $N^{2}p \times 1$ vectors, $\boldsymbol{\Phi}_{it}$ represents the $N \times N$ time-varying coefficient matrix, and $\boldsymbol{\Sigma}_{t}$ and $\boldsymbol{\Xi}_{t}$ are respectively the $N \times N$ and $N^{2}p \times N^{2}p$ variance-covariance matrices.

The generalized forecast error variance decomposition (GFEVD) is utilized to measure directional risk spillovers among variables. To compute GFEVD, we transform the TVP-VAR model into a time-varying parameter infinite-order vector moving average (TVP-VMA) representation based on the Wold representation theorem, that is, $\mathbf{y}_{t} = \sum_{i=1}^{p} \boldsymbol{\Phi}_{it} \mathbf{y}_{t-i} + \boldsymbol{\epsilon}_{t} = \sum_{j=0}^{\infty} \boldsymbol{\Psi}_{jt}\boldsymbol{\epsilon}_{t-j}$. Accordingly, the $H$-step-ahead GFEVD can be expressed as
\begin{equation}
    \theta_{jkt} \left(H\right) = \frac{\left( \boldsymbol{\Sigma}_{t} \right)_{kk}^{-1} \sum_{h=0}^{H} \left( \left( \boldsymbol{\Psi}_{h} \boldsymbol{\Sigma}_{t} \right)_{jkt} \right)^{2} }{ \sum_{h=0}^{H} \left( \boldsymbol{\Psi}_{h} \boldsymbol{\Sigma}_{t} \boldsymbol{\Psi}_{h}^{\prime} \right)_{jj} },
\label{Eq:function_GFEVD}
\end{equation}
\begin{equation}    
  \tilde{\theta}_{jkt}\left(H\right) = \frac{\theta_{jkt}\left(H\right)}{\sum_{k=1}^{N} \theta_{jkt}\left(H\right)},
\label{Eq:function_GFEVD_Normalization}
\end{equation}
where $\sum_{j=1}^{N} \tilde{\theta}_{jkt}\left(H\right) = 1$ and $\sum_{k=1}^{N} \sum_{j=1}^{N} \tilde{\theta}_{jkt}\left(H\right) = N$. $\tilde{\theta}_{jkt}$ can be interpreted as the contribution of variable $k$ to the forecast error variance of variable $j$.

With the GFEVD, we calculate the connectedness spillover indices introduced by \cite{Diebold-Yilmaz-2012-IntJForecast,Diebold-Yilmaz-2014-JEconom}, including the total directional spillovers to others from variable $j$ ($TO_{jt} \left(H\right)$), the total directional spillovers from others to variable $j$ ($FROM_{jt} \left(H\right)$), the net total directional spillovers of variable $j$ ($NET_{jt} \left(H\right)$), the net pairwise directional connectedness from variable $k$ to variable $j$ ($NPDC_{jkt} \left(H\right)$), and the total connectedness index ($TCI_{t} \left(H\right)$). The corresponding formulas are as follows:
\begin{subequations}
  \begin{equation}
    TO_{jt} \left(H\right) = \sum_{k=1,j \neq k}^{N} \tilde{\theta}_{kjt}\left(H\right),
    \label{Eq:function_DY_TO}
  \end{equation}
  \begin{equation}
    FROM_{jt} \left(H\right) = \sum_{k=1,j \neq k}^{N} \tilde{\theta}_{jkt}\left(H\right),
    \label{Eq:function_DY_FROM}
  \end{equation}
  \begin{equation}
    NET_{jt} \left(H\right) = TO_{jt} \left(H\right) - FROM_{jt} \left(H\right),
    \label{Eq:function_DY_NET}
  \end{equation}
  \begin{equation}
    NPDC_{jkt} \left(H\right) = \tilde{\theta}_{jkt}\left(H\right) - \tilde{\theta}_{kjt}\left(H\right),
    \label{Eq:function_DY_NPDC}
  \end{equation}
  \begin{equation}
    TCI_{t} \left(H\right) = N^{-1} \sum_{j=1}^{N} TO_{jt} \left(H\right)  = N^{-1} \sum_{j=1}^{N} FROM_{jt} \left(H\right),
    \label{Eq:function_DY_TCI}
  \end{equation}
\label{Eq:function_DY_Connectedness_Measures}%
\end{subequations}
where $NET_{jt} \left(H\right)>0$ ($<0$) indicates that variable $j$ serves as a net risk transmitter (receiver). Similarly, $NPDC_{jkt} \left(H\right)>0$ ($<0$) implies that the spillover from $k$ to $j$ is greater (less) than the spillover from $j$ to $k$, suggesting the dominance of variable $k$ (variable $j$). The total connectedness index $TCI_{t} \left(H\right)$ is regarded as a proxy for market risk, with larger values reflecting higher risk spillovers in the system.

After the time-domain connectedness measures, we further introduce the frequency-domain connectedness measures. The frequency response function is defined by the Fourier transform of $\boldsymbol{\Psi}_{h}$:
\begin{equation}
    \boldsymbol{\Psi} \left( e^{-\mathrm{i}\omega} \right) = \sum\limits_{h=0}^{\infty} e^{-\mathrm{i}\omega h} \boldsymbol{\Psi}_{h},
\label{Eq:function_frequency_response}
\end{equation}
where $\mathrm{i}=\sqrt{-1}$ and $\omega$ denotes the frequency. The spectral density function of $\mathbf{y}_{t}$ at frequency $\omega$ can be expressed as the Fourier transform of the TVP-VMA:
\begin{equation}
    \boldsymbol{S}_{\mathbf{y}} \left( \omega \right) = \sum\limits_{h=-\infty}^{\infty} E \left( \mathbf{y}_{t}\mathbf{y}_{t-h}^{\prime} \right) e^{-\mathrm{i}\omega h} = \boldsymbol{\Psi}_{t} \left( e^{-\mathrm{i}\omega} \right) \boldsymbol{\Sigma}_{t} \boldsymbol{\Psi}_{t}^{\prime} \left( e^{+\mathrm{i}\omega} \right).
\label{Eq:function_spectral_density}
\end{equation}

By combining the spectral density with the GFEVD, we obtain the frequency-domain GFEVD, the expression of which is
\begin{equation}
    \theta_{jkt} \left( \omega \right) = \frac{\left( \boldsymbol{\Sigma}_{t} \right)_{kk}^{-1} \Big| \left( \boldsymbol{\Psi}_{t} \left( e^{-\mathrm{i}\omega} \right) \boldsymbol{\Sigma}_{t} \right)_{jkt} \Big| ^{2} }{ \left( \boldsymbol{\Psi}_{t} \left( e^{-\mathrm{i}\omega} \right) \boldsymbol{\Sigma}_{t} \boldsymbol{\Psi}_{t}^{\prime} \left( e^{+\mathrm{i}\omega} \right)\right)_{jj} },
\label{Eq:function_frequency_GFEVD}
\end{equation}
\begin{equation}    
  \tilde{\theta}_{jkt}\left( \omega \right) = \frac{\theta_{jkt}\left( \omega \right)}{\sum_{k=1}^{N} \theta_{jkt}\left( \omega \right)},
\label{Eq:function_frequency_GFEVD_Normalization}
\end{equation}
where $\tilde{\theta}_{jkt}\left( \omega \right)$ represents the spectrum of variable $j$ at frequency $\omega$ that can be attributed to shocks from variable $k$.

In risk transmission analysis, we are more interested in evaluating connectedness at different timescales—such as short-, medium-, and long-term—rather than at a specific frequency. To that end, we define a frequency band $d = \left( a, b \right)$ and aggregate all frequencies within $d$ to obtain:
\begin{equation}
    \tilde{\theta}_{jkt}\left( d \right) = \int\nolimits_{a}^{b} \tilde{\theta}_{jkt}\left( \omega \right) \mathrm{d}\omega,
\label{Eq:function_frequency_GFEVD_aggregation}
\end{equation}
where $a,b \in \left(-\pi, \pi\right)$ and $a<b$.

Within a given frequency band $d$, the total directional connectedness to others from variable $j$ ($TO_{jt} \left(d\right)$), the total directional connectedness from others to variable $j$ ($FROM_{jt} \left(d\right)$), the net total connectedness of variable $j$ ($NET_{jt} \left(d\right)$), the net pairwise directional connectedness from $k$ to $j$ ($NPDC_{jkt} \left(d\right)$), and the total connectedness index ($TCI_{t} \left(d\right)$) are as follows:
\begin{subequations}
  \begin{equation}
    TO_{jt} \left(d\right) = \sum_{k=1,j \neq k}^{N} \tilde{\theta}_{kjt}\left(d\right),
    \label{Eq:function_BK_TO}
  \end{equation}
  \begin{equation}
    FROM_{jt} \left(d\right) = \sum_{k=1,j \neq k}^{N} \tilde{\theta}_{jkt}\left(d\right),
    \label{Eq:function_BK_FROM}
  \end{equation}
  \begin{equation}
    NET_{jt} \left(d\right) = TO_{jt} \left(d\right) - FROM_{jt} \left(d\right),
    \label{Eq:function_BK_NET}
  \end{equation}
  \begin{equation}
    NPDC_{jkt} \left(d\right) = \tilde{\theta}_{jkt}\left(d\right) - \tilde{\theta}_{kjt}\left(d\right),
    \label{Eq:function_BK_NPDC}
  \end{equation}
  \begin{equation}
    TCI_{t} \left(d\right) = N^{-1} \sum_{j=1}^{N} TO_{jt} \left(d\right)  = N^{-1} \sum_{j=1}^{N} FROM_{jt} \left(d\right).
    \label{Eq:function_BK_TCI}
  \end{equation}
\label{Eq:function_BK_Connectedness_Measures}%
\end{subequations}

It is worth noting that the time-domain connectedness measures from the TVP-VAR-DY model and the frequency-domain measures from the TVP-VAR-BK model are theoretically related as
\begin{subequations}
  \begin{equation}
    TO_{jt} \left(H\right) = \sum_{d} TO_{jt} \left(d\right),
    \label{Eq:function_DYBK_TO}
  \end{equation}
  \begin{equation}
    FROM_{jt} \left(H\right) = \sum_{d} FROM_{jt} \left(d\right),
    \label{Eq:function_DYBK_FROM}
  \end{equation}
  \begin{equation}
    NET_{jt} \left(H\right) = \sum_{d} NET_{jt} \left(d\right),
    \label{Eq:function_DYBK_NET}
  \end{equation}
  \begin{equation}
    NPDC_{jkt} \left(H\right) = \sum_{d} NPDC_{jkt} \left(d\right),
    \label{Eq:function_DYBK_NPDC}
  \end{equation}
  \begin{equation}
    TCI_{t} \left(H\right) = \sum_{d} TCI_{t} \left(d\right).
    \label{Eq:function_DYBK_TCI}
  \end{equation}
\label{Eq:function_DYBK_Connectedness_Measures}%
\end{subequations}

In this study, we determine the optimal lag order $p$ of the TVP-VAR model based on the Akaike information criterion (AIC) and the Bayesian information criterion (BIC). Referring to \cite{Chatziantoniou-Gabauer-Gupta-2023-ResourPolicy} and \cite{Zhou-Wu-Liu-Rognone-2023-NatCommun}, we set the forecast horizon $H$ to 100. Inspired by \cite{Barunik-Krehlik-2018-JFinancEconom} and \cite{Naeem-Qureshi-Farid-Tiwari-Elheddad-2024-AnnOperRes}, three different timescales are considered in our frequency-domain analysis: high frequency (short term) spans from 1 to 5 trading days (i.e., one day to one week), medium frequency (medium term) ranges from 5 to 120 trading days (i.e., one week to six months), and low frequency (long term) covers periods longer than 120 trading days (i.e., six months to infinity). This frequency-band classification enables a more nuanced and comprehensive assessment of the dynamic risk spillovers between energy and staple food markets at different timescales.

\subsection{Random forest model}

Random forest (RF), originally proposed by \cite{Breiman-2001-MachLearn}, is a nonparametric statistical method for handling both classification and regression problems. Compared with traditional models, the random forest model imposes fewer restrictions on the data and exhibits strong robustness against issues such as overfitting, noise, and missing values. Moreover, this method is well-suited for complex datasets, capable of capturing nonlinear relationships among variables and evaluating the importance of multiple factors. Hence, the random forest model has become one of the mainstream machine learning methods and is widely applied across various fields \citep{Athey-Tibshirani-Wager-2019-AnnStat,Lundberg-Erion-Chen-DeGrave-Prutkin-Nair-Katz-Himmelfarb-Bansal-Lee-2020-NatMachIntell,Podgorski-Berg-2020-Science,Wei-Gephart-Iizumi-Ramankutty-Davis-2023-NatSustain}.

The random forest regression is an ensemble learning algorithm based on decision trees, where multiple decision trees are aggregated to perform regression tasks. The basic steps are as follows. First, a subset of samples and a subset of features are randomly selected to form the training and feature sets. Second, a decision tree is constructed based on the selected sets until a predefined number of leaf nodes is reached or no further splitting is possible. Third, the process is repeated to build multiple decision trees. Fourth, a new sample is passed through each decision tree to generate multiple predictions. Fifth, the final prediction is obtained by averaging the results from all trees. Accordingly, for regression problems, the output of the random forest model $Y$ can be expressed as
\begin{equation}
    Y = \frac{1}{K}\sum\limits_{k=1}^{K}f_{k}(x),
    \label{Eq:function_RF}
\end{equation}
where $K$ denotes the number of decision trees and $f_{k}(x)$ represents the output of the $k$-th tree.

To assess the performance of the random forest model, we compute several common evaluation metrics, including the coefficient of goodness-of-fit ($R^{2}$), mean absolute error ($MAE$), mean squared error ($MSE$), relative absolute error ($RAE$), and relative squared error ($RSE$). The corresponding expressions are given by
\begin{subequations}
    \begin{equation}
        R^{2} = 1 - \frac{\sum_{q=1}^{Q}\left( y_{q}-\hat{y}_{q}\right)^{2}}{\sum_{q=1}^{Q}\left( y_{q}-\bar{y}\right)^{2}},
        \label{Eq:function_RF_R2}
    \end{equation}
    \begin{equation}
        MAE = \frac{1}{Q}\sum\limits_{q=1}^{Q} \big| y_{q}-\hat{y}_{q} \big|, \quad MSE = \frac{1}{Q}\sum\limits_{q=1}^{Q} \left( y_{q}-\hat{y}_{q}\right)^{2},
        \label{Eq:function_RF_MAE_MSE}
    \end{equation}
    \begin{equation}
        RAE = \frac{\sum_{q=1}^{Q} \big| y_{q}-\hat{y}_{q} \big|}{\sum_{q=1}^{Q} \big| y_{q}-\bar{y} \big|}, \quad RSE = \frac{\sum_{q=1}^{Q} \left( y_{q}-\hat{y}_{q}\right)^{2}}{\sum_{q=1}^{Q} \left( y_{q}-\bar{y}\right)^{2}},
        \label{Eq:function_RF_RAE_RSE}
    \end{equation}
    \label{Eq:RF_Evaluation_Metrics}%
\end{subequations}
where $Q$ is the number of samples, $y_{q}$ and $\hat{y}_{q}$ denote the true and predicted values of the $q$-th sample, respectively, and $\bar{y}$ is the mean of the true values. A higher $R^{2}$ and lower $MAE$ and $MSE$ indicate better model performance. If the $RAE$ and $RSE$ values are less than 1, the model outperforms the mean benchmark model.

The random forest regression model is adopted to explore the key drivers of higher-order connectedness and measure the importance of each factor in explaining risk spillovers. Following \cite{Bergstra-Bengio-2012-JMachLearnRes} and \cite{Wei-Gephart-Iizumi-Ramankutty-Davis-2023-NatSustain}, we employ the grid search approach to determine the optimal model hyperparameters, including the number of decision trees, maximum tree depth, minimum number of leaf samples, and minimum number of samples and maximum number of features for splitting. These hyperparameters effectively control model complexity and mitigate the risk of overfitting. Furthermore, the time-series cross-validation is utilized to further enhance the reliability and robustness of the model.

In the factor importance analysis, we calculate both the Gini and permutation importance to simultaneously quantify the relative and absolute importance of the influencing factors. Gini importance reflects the relative importance of features based on the average decrease in impurity of the decision trees, while permutation importance measures the absolute importance by computing the drop in model performance when feature values are randomly permuted. By jointly validating variable importance through both methods, we can more reliably identify the essential drivers behind multidimensional risk spillover effects.

\section{Data description}
\label{S1:Data}

To investigate the higher-moment risk spillovers within the energy-food system, and considering data availability, we select wheat, corn, soybean, and rice futures from the Chicago Board of Trade (CBOT) as representatives of the food market, and WTI crude oil, heating oil, and Henry Hub natural gas futures from the New York Mercantile Exchange (NYMEX), along with Brent crude oil futures from the Intercontinental Exchange (ICE), as representatives of the energy market. As one of the world's leading futures exchanges, CBOT possesses significant pricing power in agricultural futures markets, while NYMEX and ICE are central hubs for global energy futures trading. Therefore, trading activities on these platforms play a crucial role in global price formation in energy and grain markets. 

The selected grain commodities—wheat, corn, soybean, and rice—are the most important staple crops and are closely linked to the energy sector. WTI and Brent crude oil are the two oil benchmarks worldwide, while Henry Hub natural gas serves as the pricing reference for international gas markets. Heating oil also represents an important segment of the global energy market. Thus, these futures are highly representative, and their price movements reflect global trends in the energy and food markets. The daily closing prices of their continuous contracts are collected from the Wind database, covering the period from January 4, 2000, to February 14, 2025. Following \cite{Zhou-Wu-Liu-Rognone-2023-NatCommun}, we remove several singular data points and compute the logarithmic returns for further analysis.

\begin{figure}[!h]
  \centering
  \includegraphics[width=0.245\linewidth]{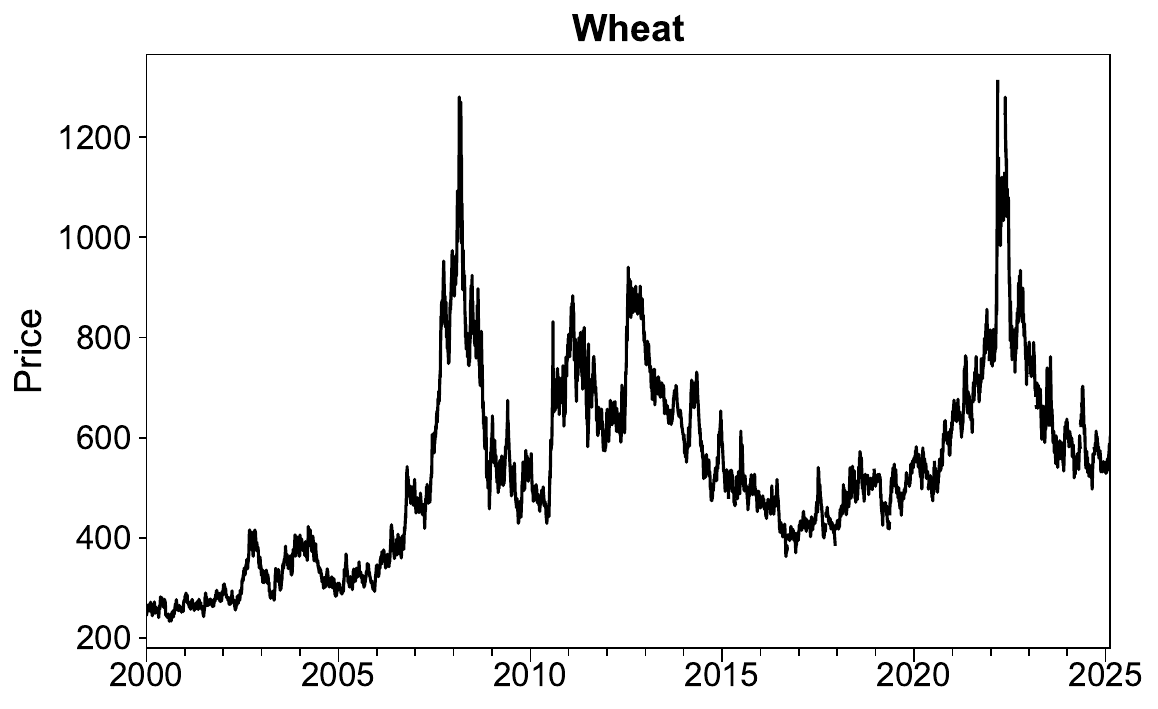}
  \includegraphics[width=0.245\linewidth]{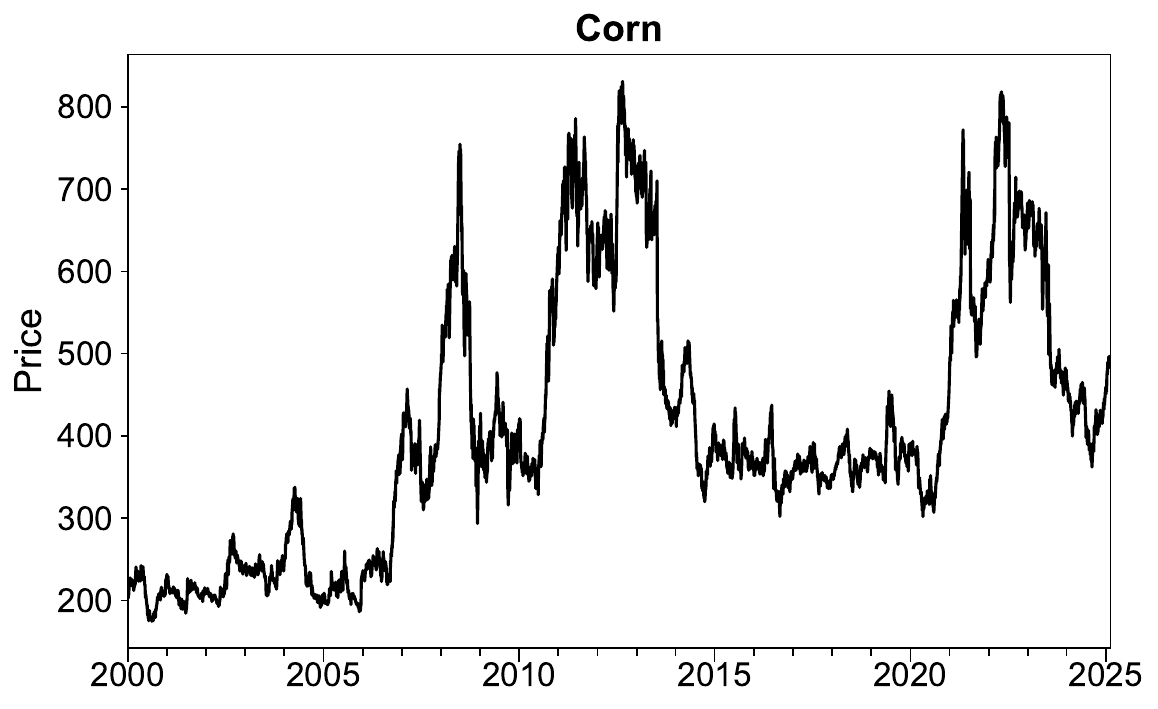}
  \includegraphics[width=0.245\linewidth]{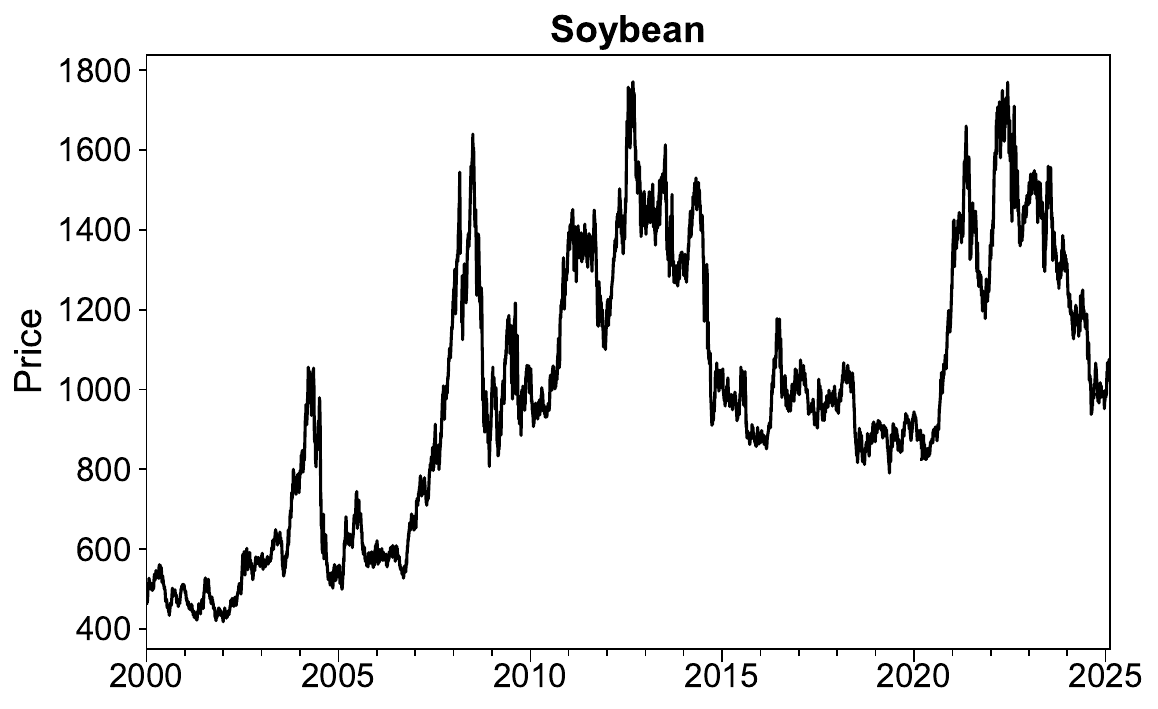}
  \includegraphics[width=0.245\linewidth]{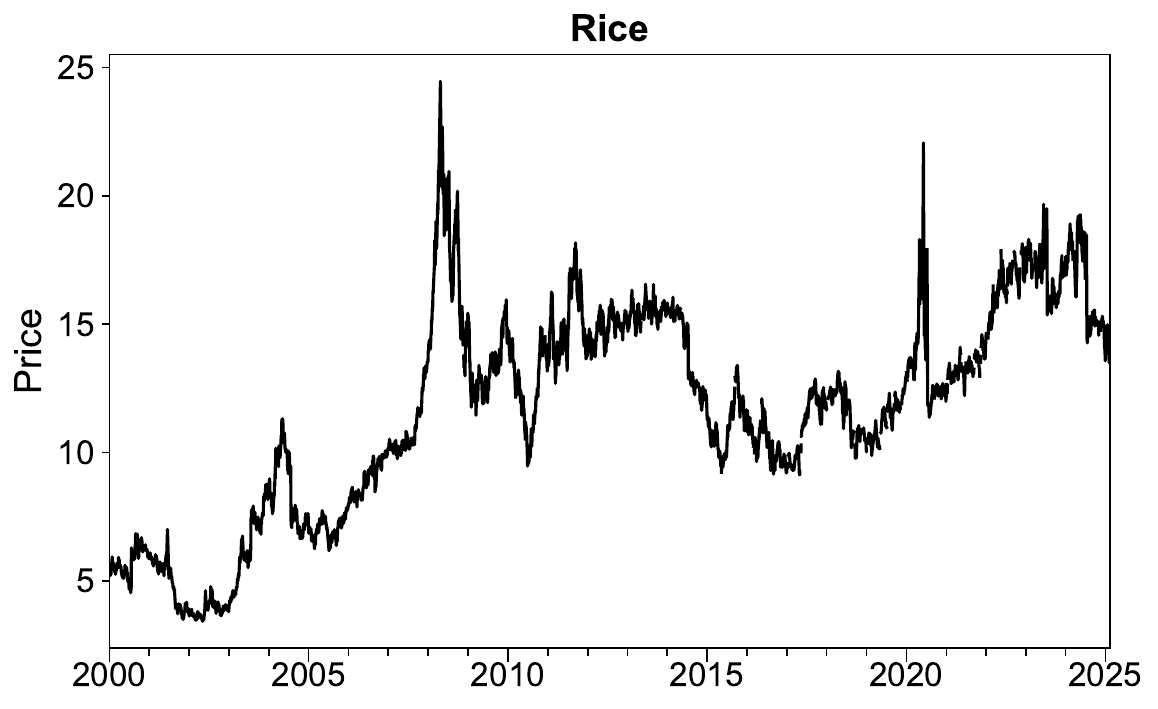}\\
  \includegraphics[width=0.245\linewidth]{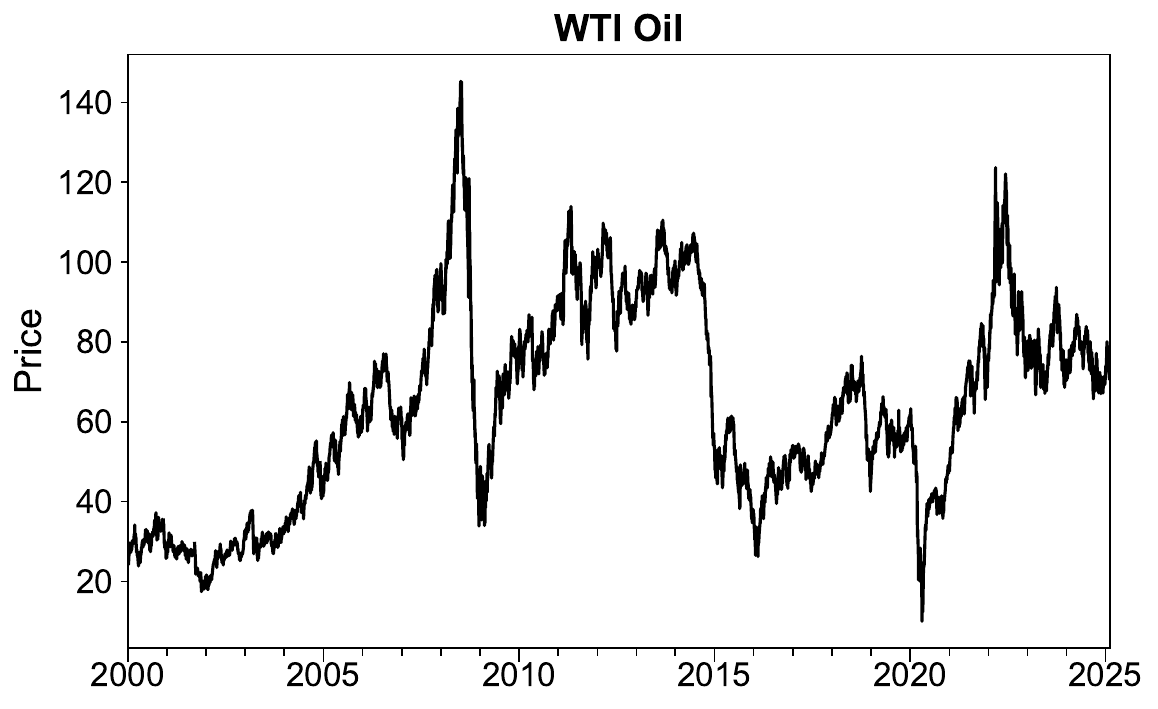}
  \includegraphics[width=0.245\linewidth]{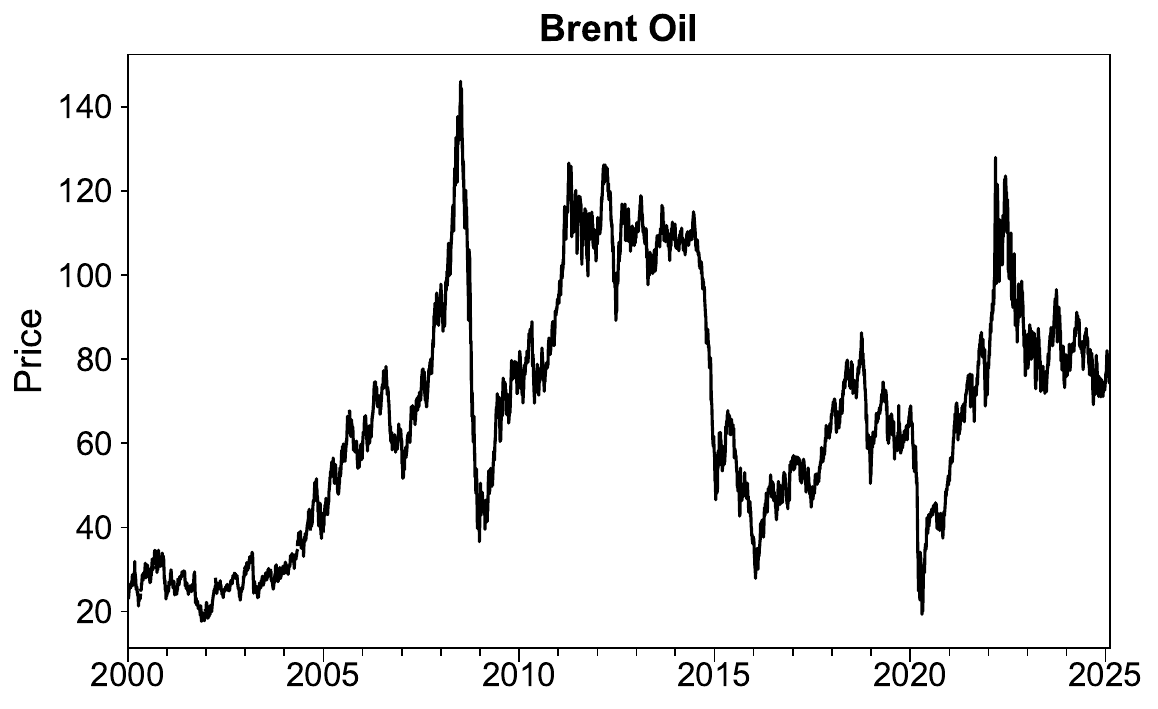}
  \includegraphics[width=0.245\linewidth]{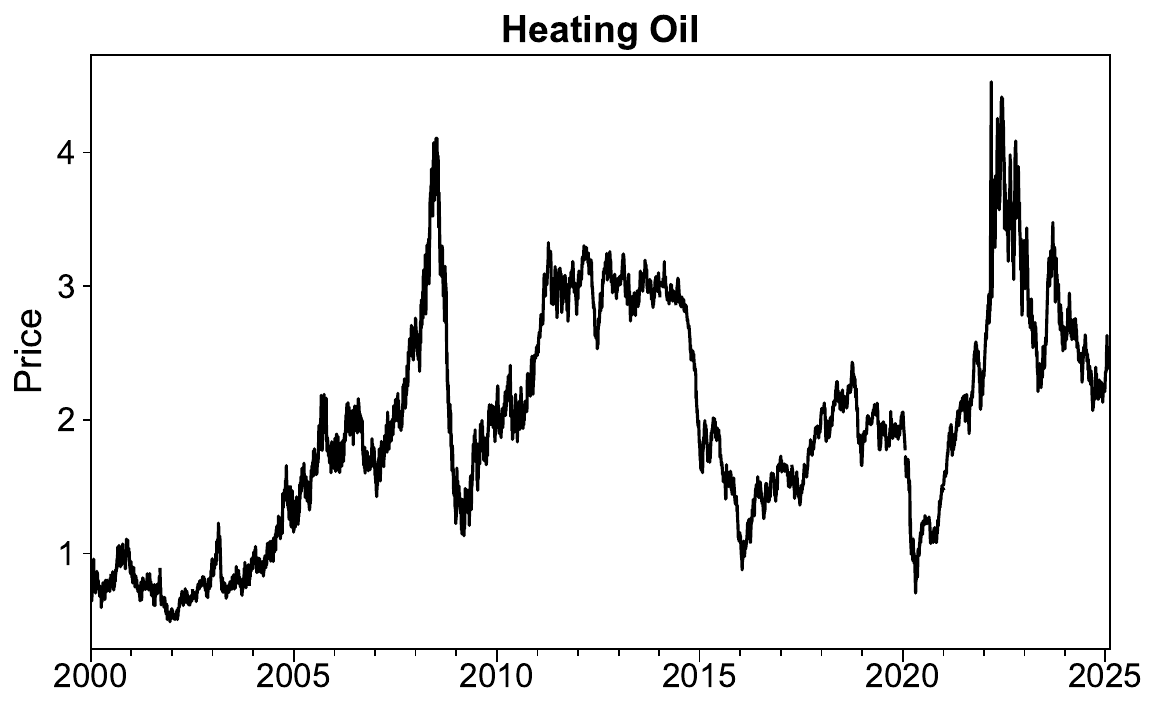}
  \includegraphics[width=0.245\linewidth]{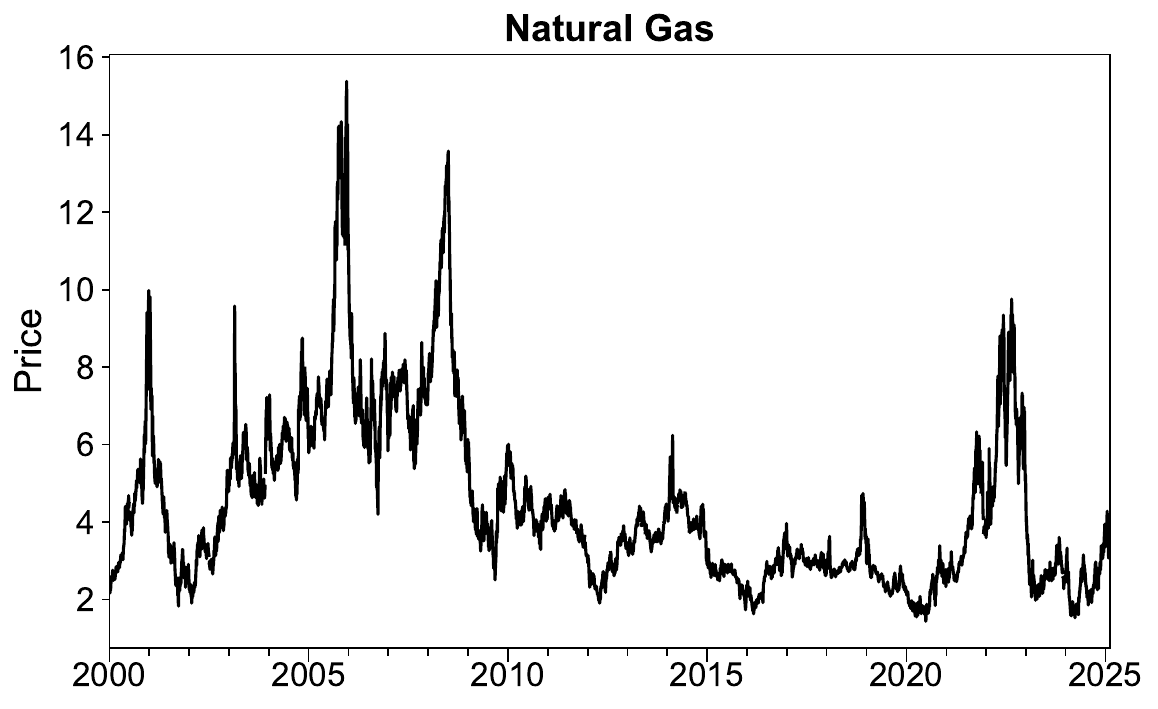}
  \caption{Price evolution of staple food and energy futures.}
\label{Fig:Agro_Price_evolution}
\end{figure}

Figure~\ref{Fig:Agro_Price_evolution} illustrates the price evolution of wheat, corn, soybean, and rice, as well as WTI oil, Brent oil, heating oil, and natural gas. All of these commodities have experienced several rounds of significant price fluctuations. Notably, both food and energy futures prices surged during the 2007–2008 global financial crisis and then fell back, which can be attributed to excessive global liquidity, rising agricultural and energy costs, and substantial speculative capital inflows into commodity markets. In addition, prices also spiked during 2010–2012 and 2020–2022. The former is linked to global climate anomalies and political unrest in the Middle East and North Africa, while the latter is associated with the COVID-19 pandemic and the Russia-Ukraine conflict. These synchronized price fluctuations indicate strong interconnectedness between energy and staple food markets, with energy prices often affecting grain prices through channels such as production costs, logistics transportation, and biofuel substitution demand.

Table~\ref{Tab:Agro_Stat_Test} reports the descriptive statistics and diagnostic tests for futures returns of staple foods and energy. The comparison reveals that energy futures exhibit higher volatility, as evidenced by their larger absolute values of maximum, minimum, and standard deviation. Moreover, the skewness of each series is not 0, and the kurtosis exceeds 3, implying a skewed distribution with leptokurtosis and fat tail. These characteristics are further supported by the results of the Jarque-Bera test, which strongly rejects the null hypothesis of normality for all series. The significant ADF test statistics and the insignificant KPSS test statistics consistently confirm that the return series for both staple food and energy futures are stationary. In addition, the Ljung-Box test suggests that neither the returns nor the squared returns follow white noise processes, reflecting both serial autocorrelation and conditional heteroscedasticity. The statistics of the ARCH-LM test are all significant at the 1\% level, further indicating ARCH effects. These findings justify our construction of the GJRSK model for empirical analysis.

\begin{table}[!ht]
  \centering
  \setlength{\abovecaptionskip}{0pt}
  \setlength{\belowcaptionskip}{10pt}
  \caption{Descriptive statistics and diagnostic tests for futures returns of staple foods and energy}
  \setlength\tabcolsep{5pt}   \resizebox{\textwidth}{!}{ 
    \begin{tabular}{l r@{.}l r@{.}l r@{.}l r@{.}l r@{.}l r@{.}l r@{.}l r@{.}l}
    \toprule
         & \multicolumn{8}{c}{Staple foods} & \multicolumn{8}{c}{Energy}  \\
         \cmidrule(lr){2-9} \cmidrule(lr){10-17}
         & \multicolumn{2}{c}{Wheat} & \multicolumn{2}{c}{Corn} & \multicolumn{2}{c}{Soybean} & \multicolumn{2}{c}{Rice} & \multicolumn{2}{c}{WTI Oil} & \multicolumn{2}{c}{Brent Oil} & \multicolumn{2}{c}{Heating Oil} & \multicolumn{2}{c}{Natural Gas}  \\
        \midrule

    \multicolumn{17}{l}{\textit{Panel A: Descriptive statistics}} \\
    Max & 0&258 & 0&136 & 0&076 & 0&162 & 0&320 & 0&191 & 0&144 & 0&332 \\
    Min & $-$0&226 & $-$0&198 & $-$0&158 & $-$0&157 & $-$0&282 & $-$0&276 & $-$0&245 & $-$0&230 \\
    Mean ($\times 10^{3}$) & 0&180 & 0&212 & 0&154 & 0&192 & 0&259 & 0&216 & 0&209 & 0&085 \\
    Std. Dev. & 0&021 & 0&018 & 0&015 & 0&017 & 0&026 & 0&023 & 0&032 & 0&036 \\
    Skew. & 0&330 &  $-$0&460 &  $-$0&993 &  $-$0&204 &  $-$0&031 &  $-$0&385 &  $-$0&110 & 0&397 \\
    Kurt. & 8&820 & 9&268 & 7&995 & 8&340 & 16&143 & 7&467 & 3&369 & 5&634 \vspace{2mm} \\
    \multicolumn{17}{l}{\textit{Panel B: Diagnostic tests}} \\
    Jarque-Bera & \multicolumn{2}{c}{20584$^{***}$} & \multicolumn{2}{c}{22822$^{***}$} & \multicolumn{2}{c}{17855$^{***}$} & \multicolumn{2}{c}{18346$^{***}$} & \multicolumn{2}{c}{68571$^{***}$} & \multicolumn{2}{c}{14827$^{***}$} & \multicolumn{2}{c}{2999$^{***}$} & \multicolumn{2}{c}{8518$^{***}$} \\ 
    ADF &  $-$81&402$^{***}$ &  $-$44&886$^{***}$ &  $-$25&807$^{***}$ &  $-$31&707$^{***}$ &  $-$19&230$^{***}$ &  $-$81&868$^{***}$ &  $-$25&690$^{***}$ &  $-$30&427$^{***}$ \\
    KPSS & 0&043 & 0&045 & 0&102 & 0&058 & 0&051 & 0&075 & 0&078 & 0&054 \\
    Q(20) & 33&100$^{**}$ & 32&030$^{**}$ & 40&105$^{***}$ & 56&299$^{***}$ & 51&085$^{***}$ & 35&416$^{**}$ & 460&437$^{***}$ & 64&246$^{***}$ \\
    Q$^{2}$(20) & 1143&393$^{***}$ & 292&108 $^{***}$ & 698&708$^{***}$ & 1012&373$^{***}$ & 4205&116$^{***}$ & 2841&847$^{***}$ & 2476&688$^{***}$ & 702&475$^{***}$ \\
    ARCH-LM & 895&711$^{***}$ & 173&501$^{***}$ & 356&268$^{***}$ & 431&149$^{***}$ & 1401&026$^{***}$ & 991&686$^{***}$ & 868&974$^{***}$ & 347&390$^{***}$ \\
  \bottomrule
    \end{tabular}
    }%
  \begin{flushleft}
    \footnotesize
\justifying Note: This table presents the descriptive statistics and diagnostic tests for the return series of staple food and energy futures. The Jarque-Bera test is a normality test, while both the ADF test and the KPSS test are unit root tests. Q(20) and Q$^{2}$(20) refer to the Ljung-Box test for returns and squared returns, respectively, and the ARCH-LM test examines the presence of the ARCH effect. $^{***}$ and $^{**}$ denote statistical significance at the 1\% and 5\% levels.
\end{flushleft} 
  \label{Tab:Agro_Stat_Test}%
\end{table}%

Furthermore, inspired by \cite{Kilian-Murphy-2014-JApplEconom}, \cite{Le-Pham-Do-2023-EnergyEcon}, \cite{Rao-Lucey-Kumar-2023-EnergyEcon}, and \cite{Zhou-Wu-Liu-Rognone-2023-NatCommun}, we select ten potential influencing factors, including the U.S. dollar index (DXY), 10-year treasury yield (TYR), CBOE volatility index (VIX), global economic policy uncertainty (EPU), OPEC+ crude oil production (COP), U.S. crude oil stocks (COS), European emission allowance (EUA), climate policy uncertainty (CPU), global natural disasters (GND), and geopolitical risk index (GPR). The data are sourced from the Wind database, the Federal Reserve Economic Data website, the Policy Uncertainty website, the U.S. Energy Information Administration, and the EM-DAT database. Among these variables, DXY, TYR, VIX, EUA, and GPR are available on a daily basis; COS is weekly; EPU, COP, and CPU are monthly; and GND is annual. Both backward and forward filling are utilized for data alignment, followed by standardization to address variations in the units and scales of the factors. These variables are categorized as economic and financial factors, market supply-demand factors, and climate and risk-related factors.

Bulk commodities, especially energy and food, are typically denominated in U.S. dollars, and DXY measures the dollar's exchange rate against a basket of major currencies. As a benchmark for risk-free interest rates, TYR influences capital costs and trading behavior in commodity markets. VIX captures changes in market sentiment and risk appetite, exhibiting strong co-movements with commodity prices. EPU and CPU reflect uncertainties in economic and climate-related policies, respectively, which are closely tied to both energy and food markets. EUA is expected to affect the cost and demand for energy and food, particularly conventional fossil fuels. Given that crude oil is a core product in commodity markets, both COP and COS are important factors. GND quantifies the frequency of natural disasters related to climatology, hydrology, meteorology, and geophysics, which directly impact agricultural production, transportation, and energy infrastructure. GPR provides a time-varying indicator of adverse geopolitical events and associated risks. The ten factors comprehensively reflect macroeconomic and financial market conditions, policy uncertainties, supply-demand fundamentals, and climate and geopolitical shocks. Therefore, incorporating them into the analysis is conducive to a better understanding of risk sources of the energy-food nexus.

\section{Empirical analysis}
\label{S1:EmpAnal}

\subsection{Higher-order moment measures}

To introduce time-varying skewness and kurtosis, we employ the GJRSK model to calculate the conditional volatility, skewness, and kurtosis for each staple food and energy market. The parameter estimates are presented in Table~\ref{Tab:Agro_GJRSK_estimation}, with standard errors reported in parentheses. In the mean equations, the $\alpha_{1}$ coefficients are significantly negative for all energy markets and the soybean market, suggesting the presence of mean-reversion. In other words, positive (negative) returns in these markets tend to be followed by a decrease (increase) in the next period. Interestingly, the $\alpha_{1}$ coefficient for rice is significantly positive and relatively large, implying strong positive autocorrelation in rice returns. This may be attributed to the smaller market size and lower liquidity of rice futures, which lead to more persistent price movements. From the variance equations, we observe that $\beta_{1}$ is significantly positive across all markets, and $\beta_{2}$ is also significantly positive for WTI and Brent oil, which is in line with the high volatility of crude oil markets. $\beta_{3}$ captures the leverage effect in volatility, the estimates of which are significantly positive at the 1\% level and are close to 1. This indicates that both food and energy markets are more sensitive to bad news, meaning adverse shocks substantially increase the volatility of these markets.

In the skewness equations, the $\gamma_{3}$ coefficients exhibit statistical significance with different signs. Specifically, wheat, soybean, rice, WTI oil, and heating oil correspond to positive coefficients, implying that negative returns in these markets tend to occur more frequently than positive ones, while the opposite is true for corn, Brent oil, and natural gas. Regarding the kurtosis equations, all $\delta_{3}$ estimates are positive, with most being statistically significant except for those of Brent oil and natural gas. This suggests the existence of asymmetric kurtosis effects, where negative return shocks tend to result in higher kurtosis. In other words, staple food and energy markets are prone to extreme negative returns. It is noteworthy that the leverage coefficients in the conditional variance, skewness, and kurtosis equations are different from zero, highlighting the crucial role of leverage in shaping higher-order moments in food and energy markets. These findings reaffirm that the GJRSK model we adopt is more suitable than the GARCHSK model for capturing higher-moment risks within the energy-food nexus.

\begin{table}[!ht]
  \centering
  \setlength{\abovecaptionskip}{0pt}
  \setlength{\belowcaptionskip}{10pt}
  \caption{Parameter estimation of the GJRSK models for staple food and energy futures}
  \setlength\tabcolsep{9pt}   \resizebox{\textwidth}{!}{ 
    \begin{tabular}{l r@{.}l r@{.}l r@{.}l r@{.}l r@{.}l r@{.}l r@{.}l r@{.}l}
    \toprule
         & \multicolumn{8}{c}{Staple foods} & \multicolumn{8}{c}{Energy}  \\
         \cmidrule(lr){2-9} \cmidrule(lr){10-17}
    & \multicolumn{2}{c}{Wheat} & \multicolumn{2}{c}{Corn} & \multicolumn{2}{c}{Soybean} & \multicolumn{2}{c}{Rice} & \multicolumn{2}{c}{WTI Oil} & \multicolumn{2}{c}{Brent Oil} & \multicolumn{2}{c}{Heating Oil} & \multicolumn{2}{c}{Natural Gas}  \\
    \midrule
    \multicolumn{17}{l}{\textit{Panel A: Mean equation}} \\
    $\alpha_1$ & 0&008 & 0&008 & $-$0&022$^{**}$ & 0&187$^{***}$ & $-$0&083$^{***}$ & $-$0&083$^{***}$ & $-$0&166$^{***}$ & $-$0&026$^{***}$ \\
    & (0&011) & (0&009) & (0&009) & (0&009) & (0&009) & (0&009) & (0&009) & (0&009) \vspace{2mm} \\
    \multicolumn{17}{l}{\textit{Panel B: Variance equation}} \\
    $\beta_0$ & 0&000 & 0&000 & 0&000 & 0&000 & 0&000 & 0&000 & 0&000 & 0&000 \\
    & (0&009) & (0&013) & (0&013) & (0&013) & (0&009) & (0&013) & (0&012) & (0&013) \\
    $\beta_1$ & 0&045$^{***}$ & 0&063$^{***}$ & 0&064$^{***}$ & 0&030$^{**}$ & 0&023$^{**}$ & 0&064$^{***}$ & 0&061$^{***}$ & 0&086$^{***}$ \\
    & (0&010) & (0&010) & (0&010) & (0&012) & (0&010) & (0&010) & (0&010) & (0&010) \\
    $\beta_2$ & 0&001 & 0&001 & 0&012 & 0&062$^{***}$ & 0&073$^{***}$ & 0&039$^{***}$ & 0&004 & 0&000 \\
    & (0&010) & (0&010) & (0&010) & (0&010) & (0&010) & (0&010) & (0&010) & (0&010) \\
    $\beta_3$ & 0&940$^{***}$ & 0&921$^{***}$ & 0&917$^{***}$ & 0&896$^{***}$ & 0&903$^{***}$ & 0&898$^{***}$ & 0&935$^{***}$ & 0&907$^{***}$ \\
    & (0&010) & (0&010) & (0&010) & (0&010) & (0&010) & (0&010) & (0&011) & (0&010) \vspace{2mm} \\           
    \multicolumn{17}{l}{\textit{Panel C: Skewness equation}} \\
    $\gamma_0$ & 0&011 & 0&043$^{***}$ & $-$0&018$^{*}$ & 0&031$^{***}$ & $-$0&007 & $-$0&094$^{***}$ & $-$0&020$^{**}$ & 0&124$^{***}$ \\
    & (0&010) & (0&010) & (0&010) & (0&010) & (0&010) & (0&010) & (0&010) & (0&010) \\
    $\gamma_1$ & 0&000 & 0&000 & $-$0&007 & $-$0&005 & $-$0&001 & $-$0&054$^{***}$ & 0&005 & 0&000 \\
    & (0&010) & (0&010) & (0&010) & (0&010) & (0&010) & (0&010) & (0&010) & (0&010) \\
    $\gamma_2$ & 0&003 & 0&000 & 0&034$^{***}$ & 0&019$^{*}$ & $-$0&002 & 0&055$^{***}$ & $-$0&004 & $-$0&001 \\
    & (0&010) & (0&010) & (0&010) & (0&010) & (0&010) & (0&010) & (0&010) & (0&010) \\
    $\gamma_3$ & 0&914$^{***}$ & $-$0&370$^{***}$ & 0&104$^{***}$ & 0&757$^{***}$ & 0&956$^{***}$ & $-$0&153$^{***}$ & 0&040$^{***}$ & $-$0&972$^{***}$ \\
    & (0&009) & (0&009) & (0&010) & (0&009) & (0&009) & (0&009) & (0&009) & (0&009) \vspace{2mm} \\
    \multicolumn{17}{l}{\textit{Panel D: Kurtosis equation}} \\
    $\delta_0$ & 0&446$^{***}$ & 2&403$^{***}$ & 0&763$^{***}$ & 0&855$^{***}$ & 2&231$^{***}$ & 3&300$^{***}$ & 3&045$^{***}$ & 3&367$^{***}$ \\
    & (0&010 )& (0&012) & (0&010) & (0&012) & (0&010) & (0&013) & (0&011) & (0&012) \\
    $\delta_1$ & 0&001 & 0&000 & 0&001 & 0&005 & 0&010 & 0&004 & 0&001 & 0&004 \\
    & (0&010) & (0&011) & (0&010) & (0&010) & (0&010) & (0&010) & (0&010) & (0&010) \\
    $\delta_2$ & 0&001 & 0&000 & 0&000 & 0&006 & 0&019$^{*}$ & 0&000 & 0&000 & 0&000 \\
    & (0&010) & (0&010) & (0&010) & (0&010) & (0&010) & (0&010) & (0&010) & (0&010) \\
    $\delta_3$ & 0&865$^{***}$ & 0&318$^{***}$ & 0&775$^{***}$ & 0&753$^{***}$ & 0&306$^{***}$ & 0&007 & 0&068$^{***}$ & 0&000 \\
    & (0&008) & (0&008) & (0&008) & (0&010) & (0&008) & (0&008) & (0&012) & (0&008) \\
   \bottomrule
    \end{tabular}
    }%
  \begin{flushleft}
    \footnotesize
    \justifying Note: This table reports the parameter estimates of the GJRSK models for staple food and energy futures, where standard errors are listed in parentheses. $^{***}$, $^{**}$, and $^{*}$ refer to significance at the 1\%, 5\%, and 10\% levels, respectively.
  \end{flushleft}
  \label{Tab:Agro_GJRSK_estimation}%
\end{table}%

\begin{figure}[!htp]
  \centering
  \includegraphics[width=0.245\linewidth]{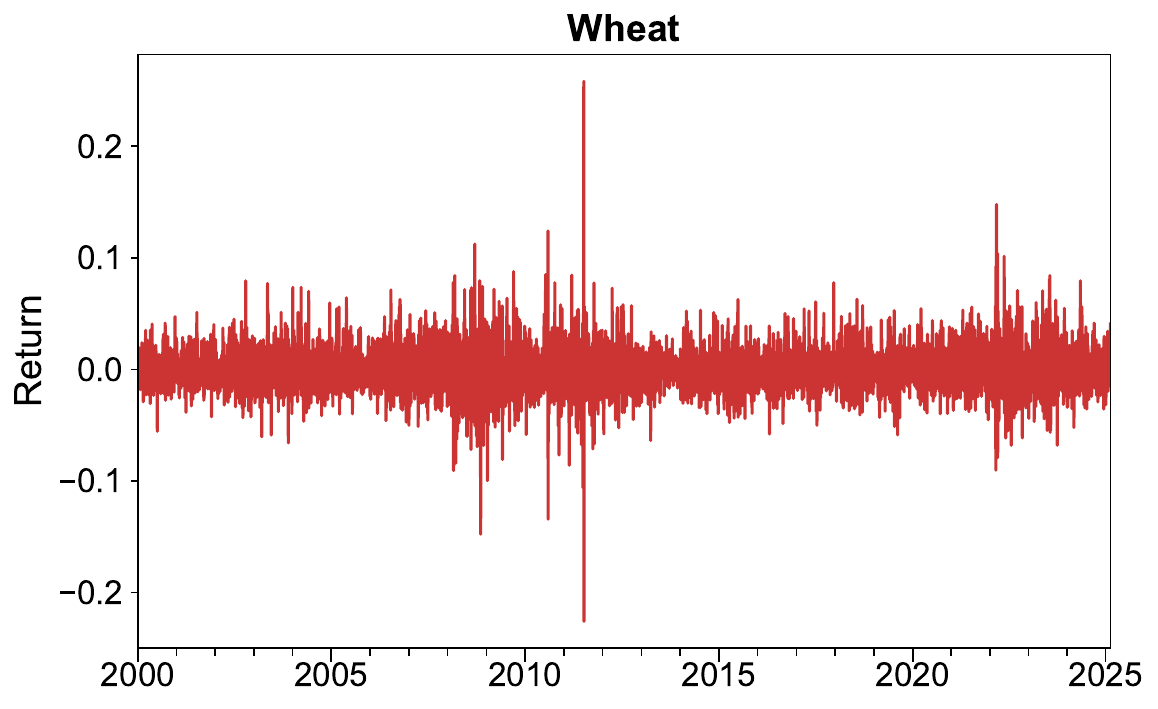}
  \includegraphics[width=0.245\linewidth]{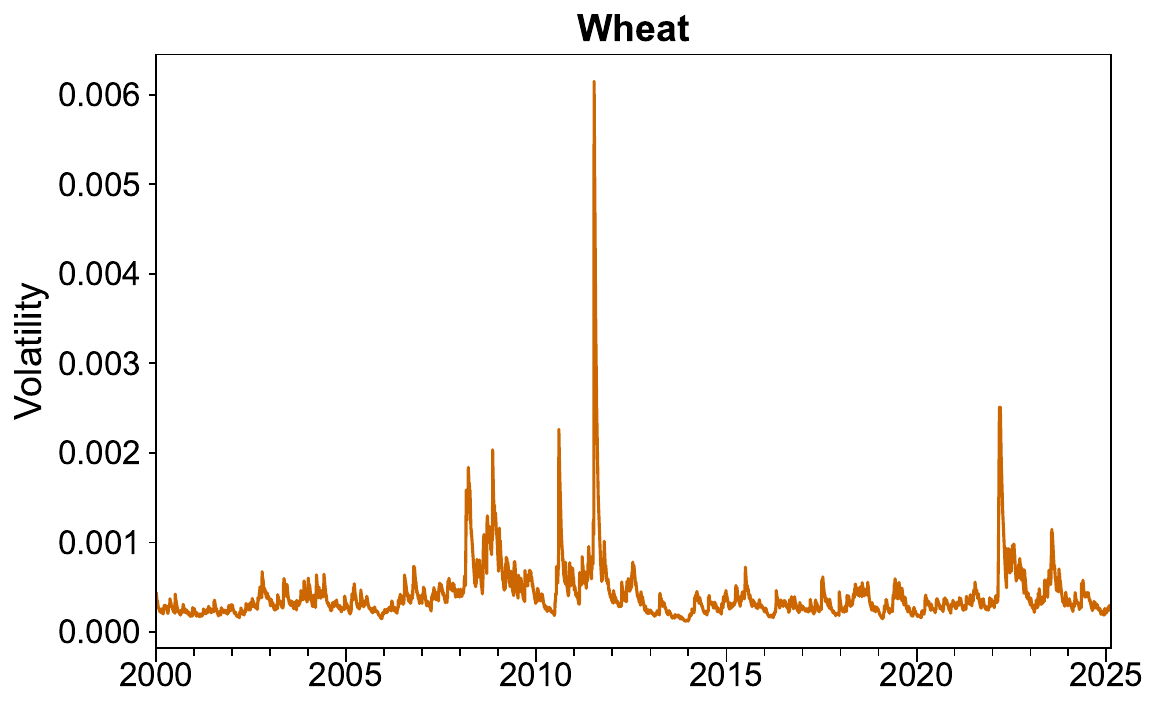}
  \includegraphics[width=0.245\linewidth]{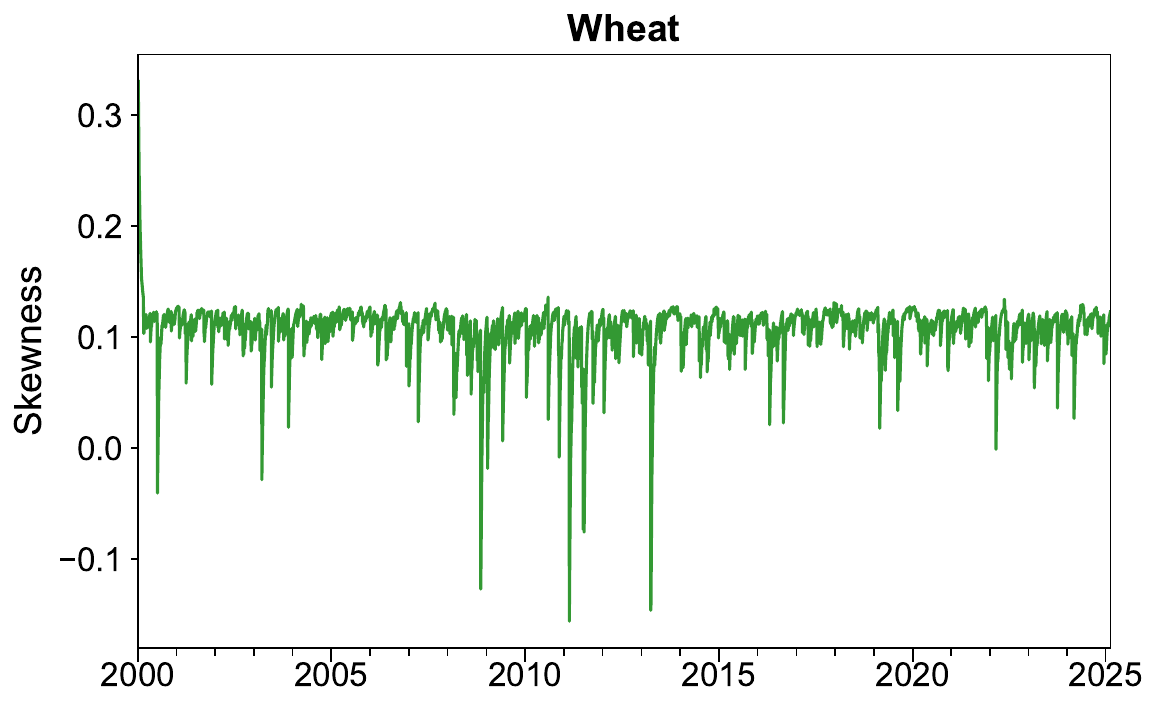}
  \includegraphics[width=0.245\linewidth]{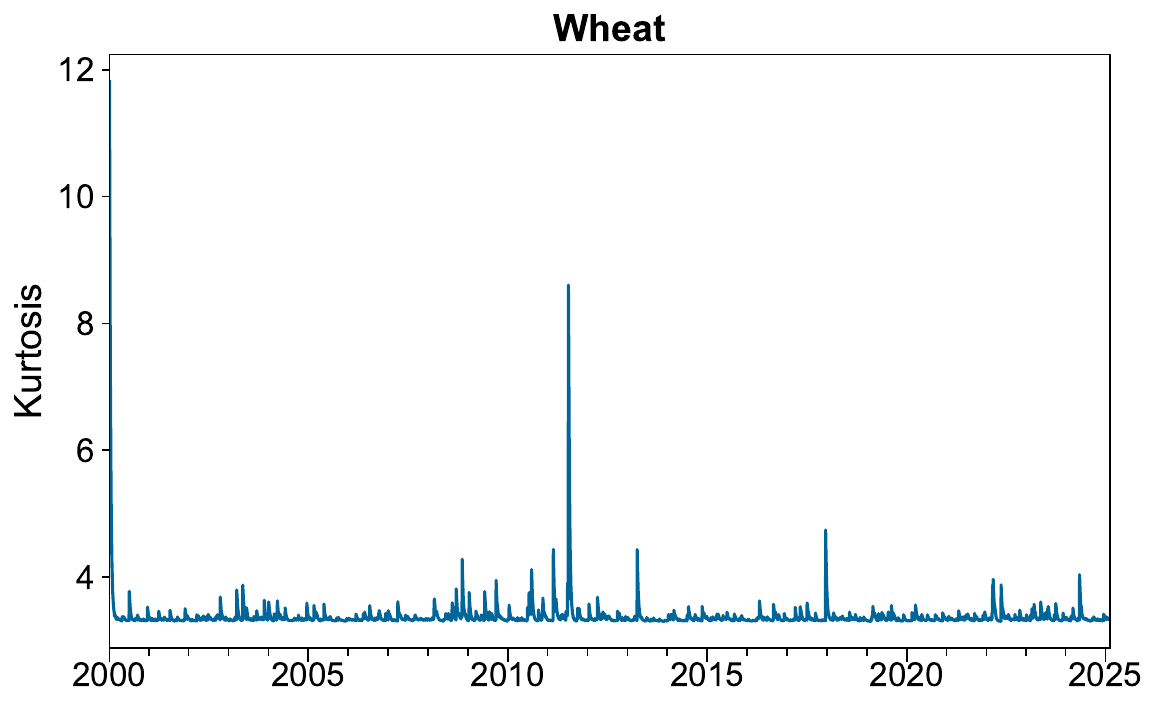}\\
  \includegraphics[width=0.245\linewidth]{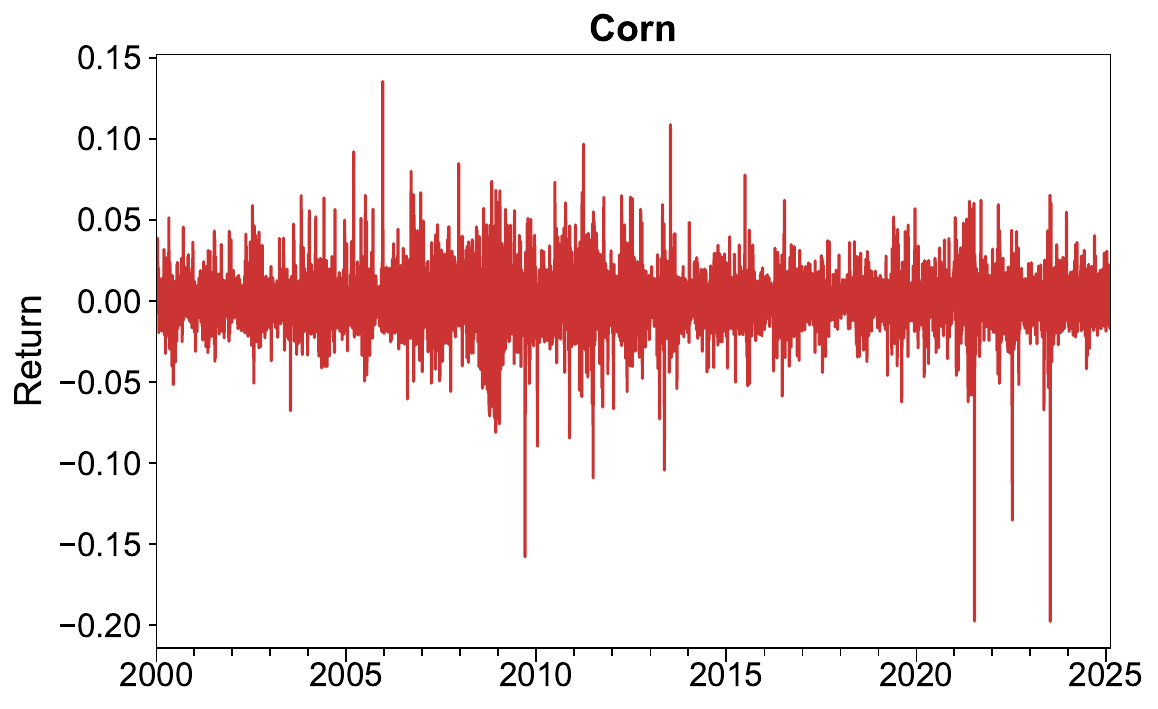}
  \includegraphics[width=0.245\linewidth]{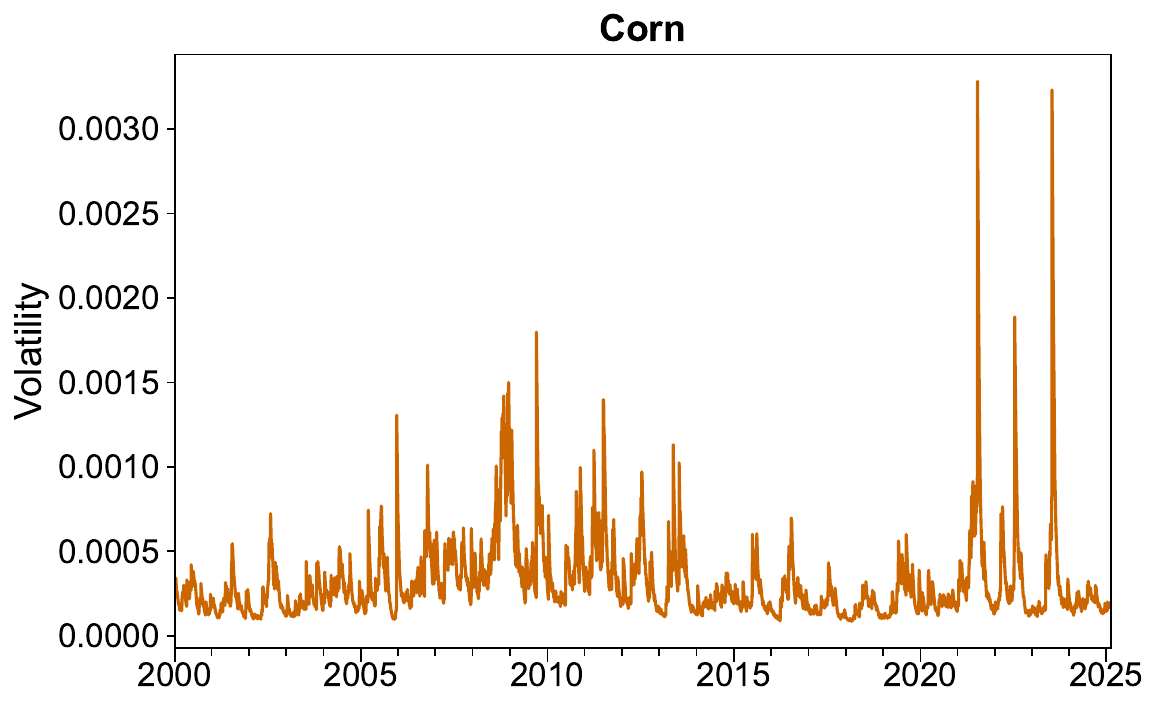}
  \includegraphics[width=0.245\linewidth]{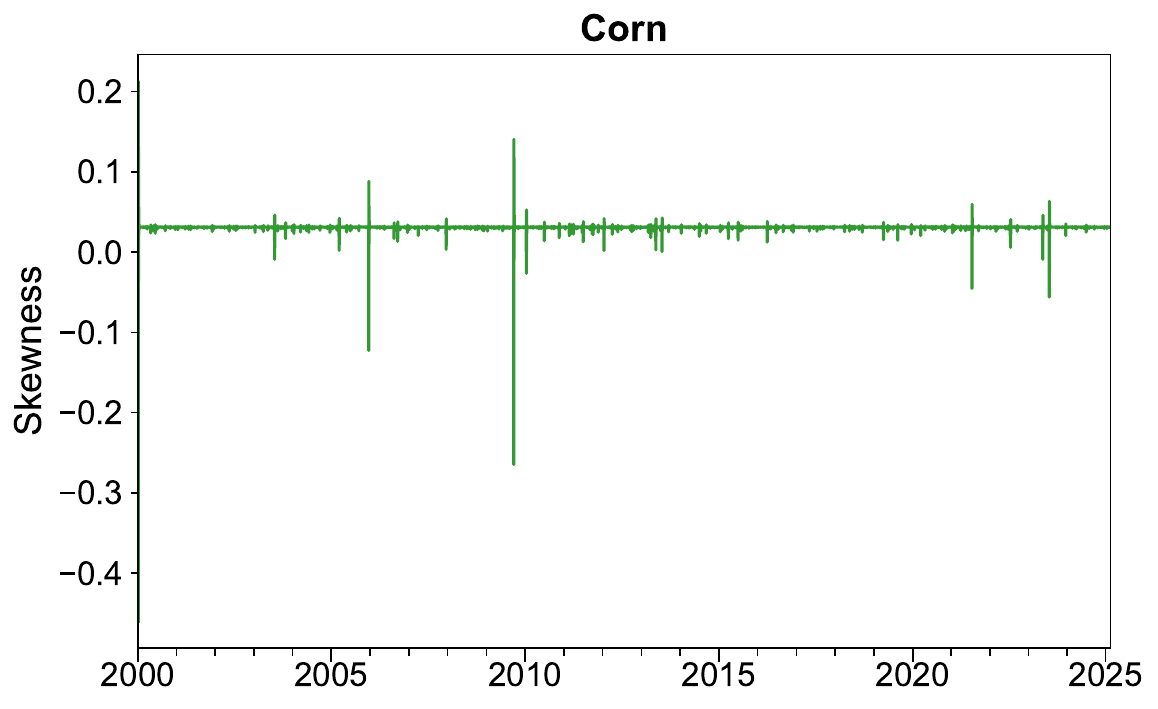}
  \includegraphics[width=0.245\linewidth]{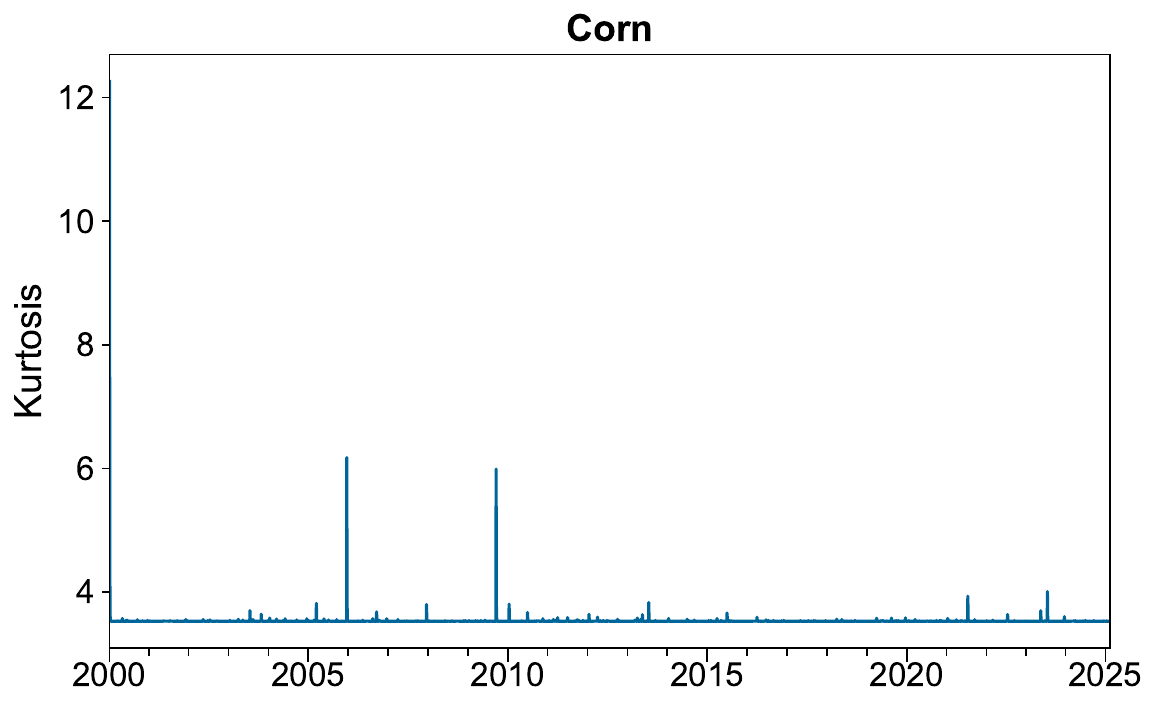}\\
  \includegraphics[width=0.245\linewidth]{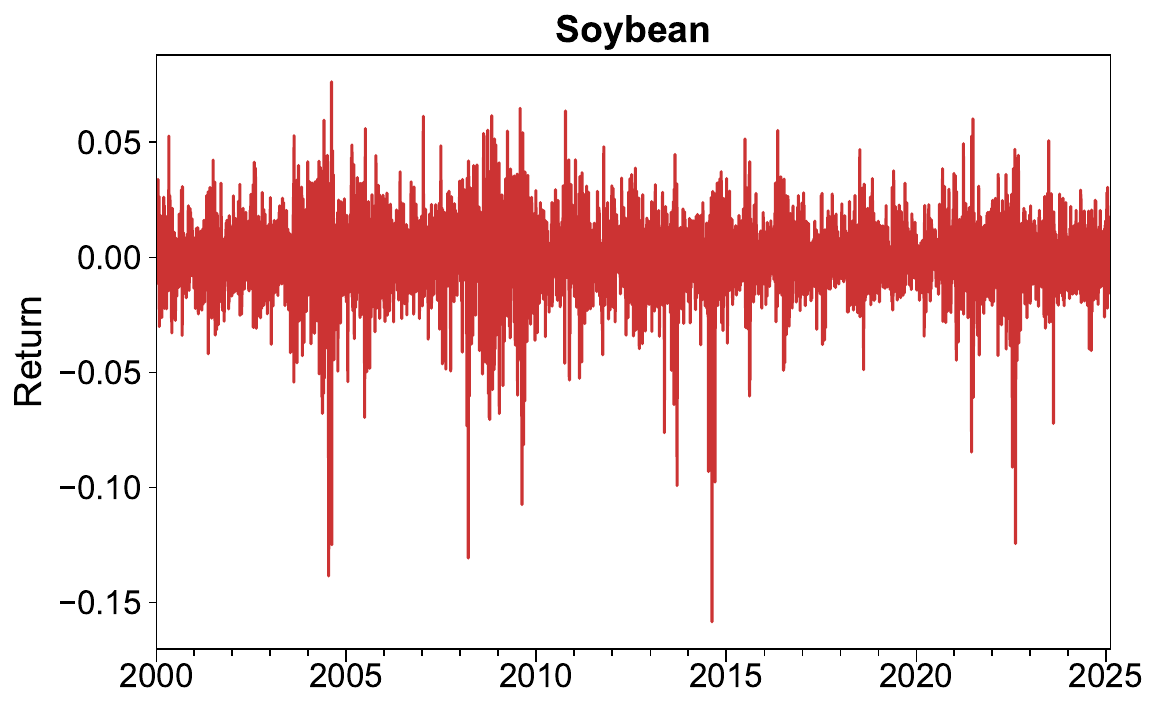}
  \includegraphics[width=0.245\linewidth]{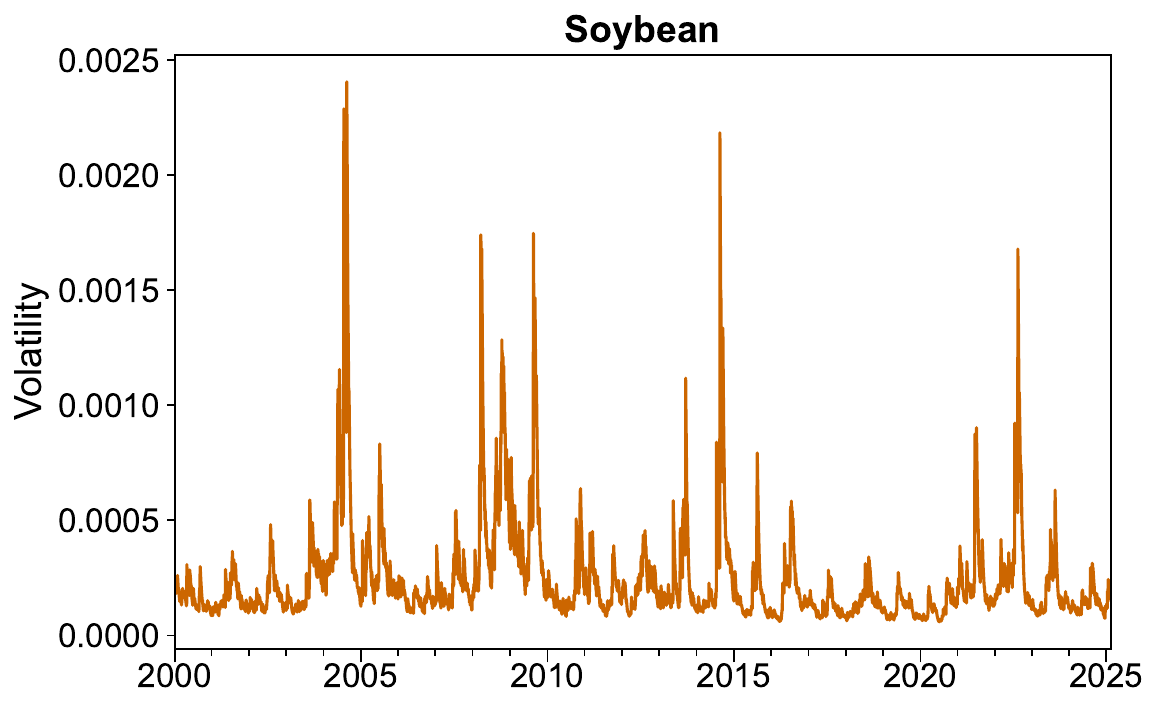}
  \includegraphics[width=0.245\linewidth]{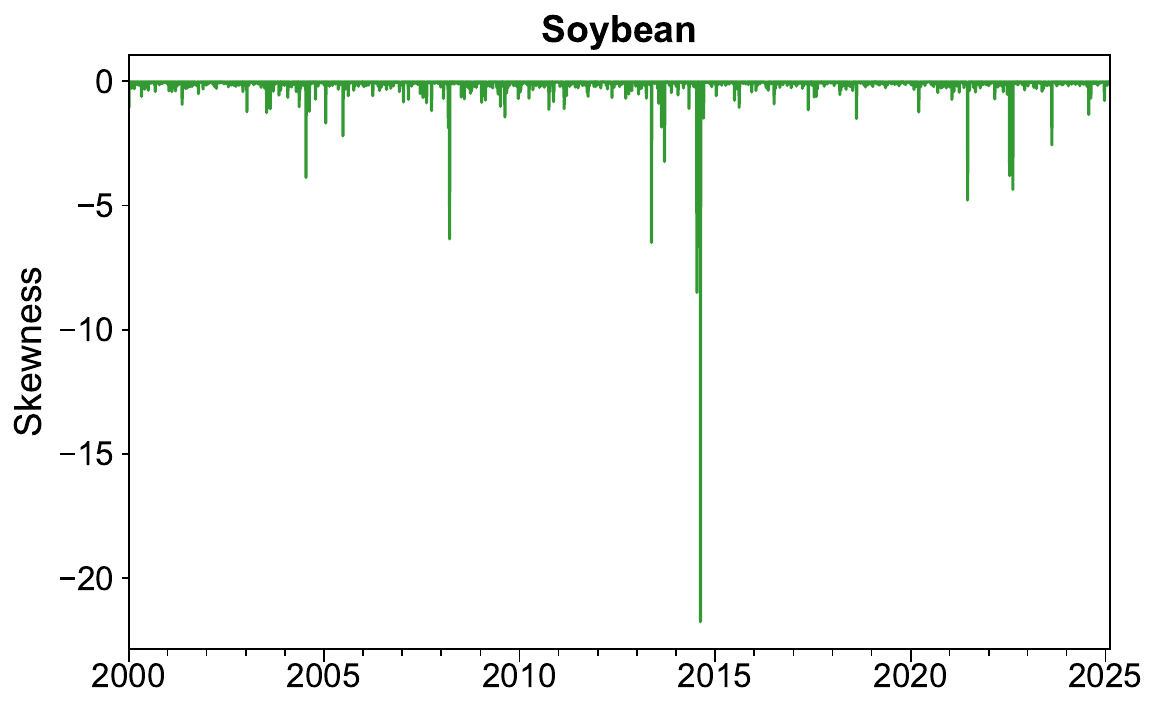}
  \includegraphics[width=0.245\linewidth]{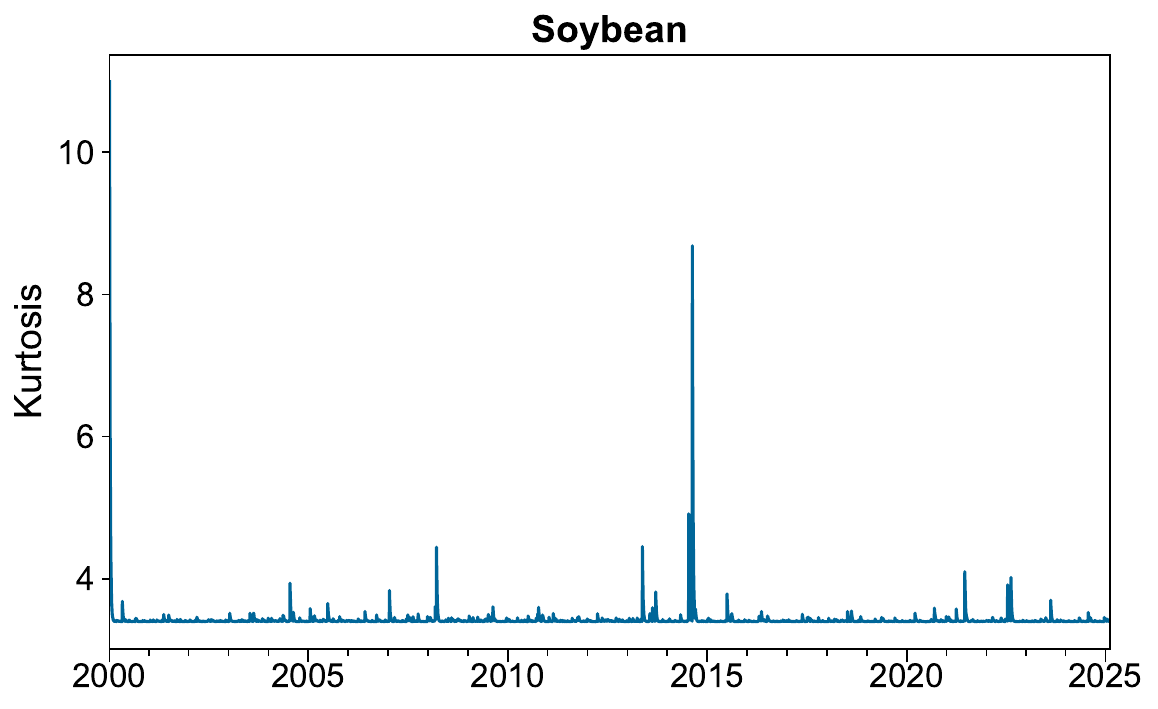}\\
  \includegraphics[width=0.245\linewidth]{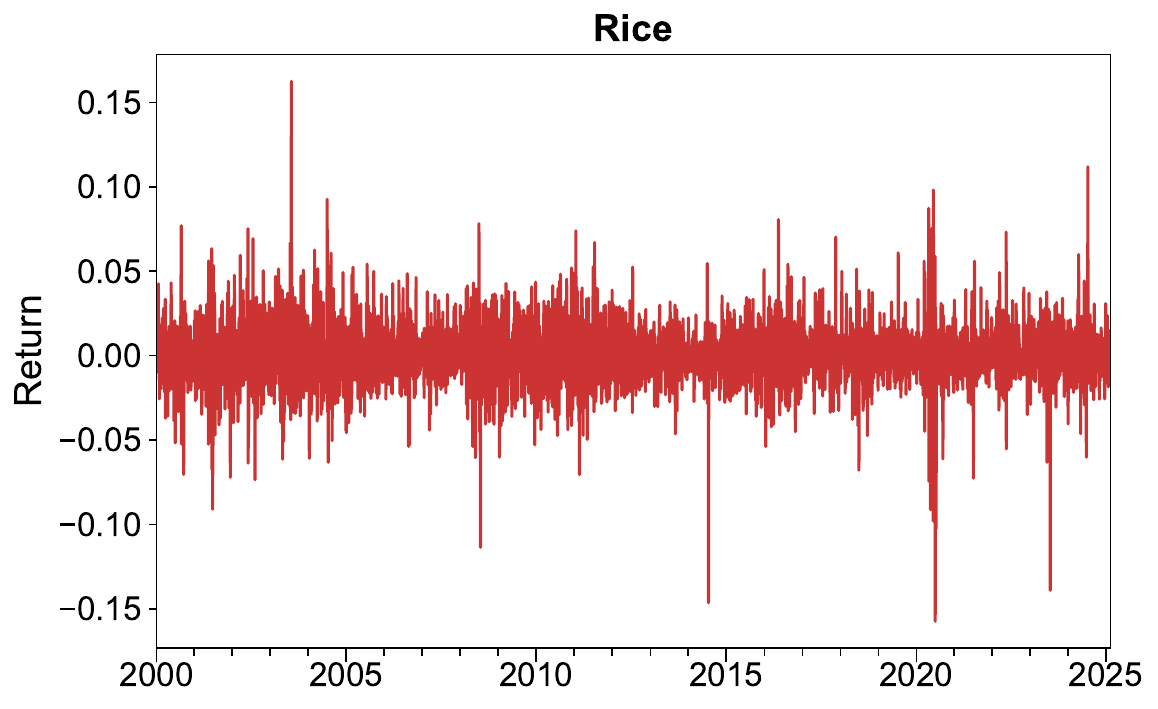}
  \includegraphics[width=0.245\linewidth]{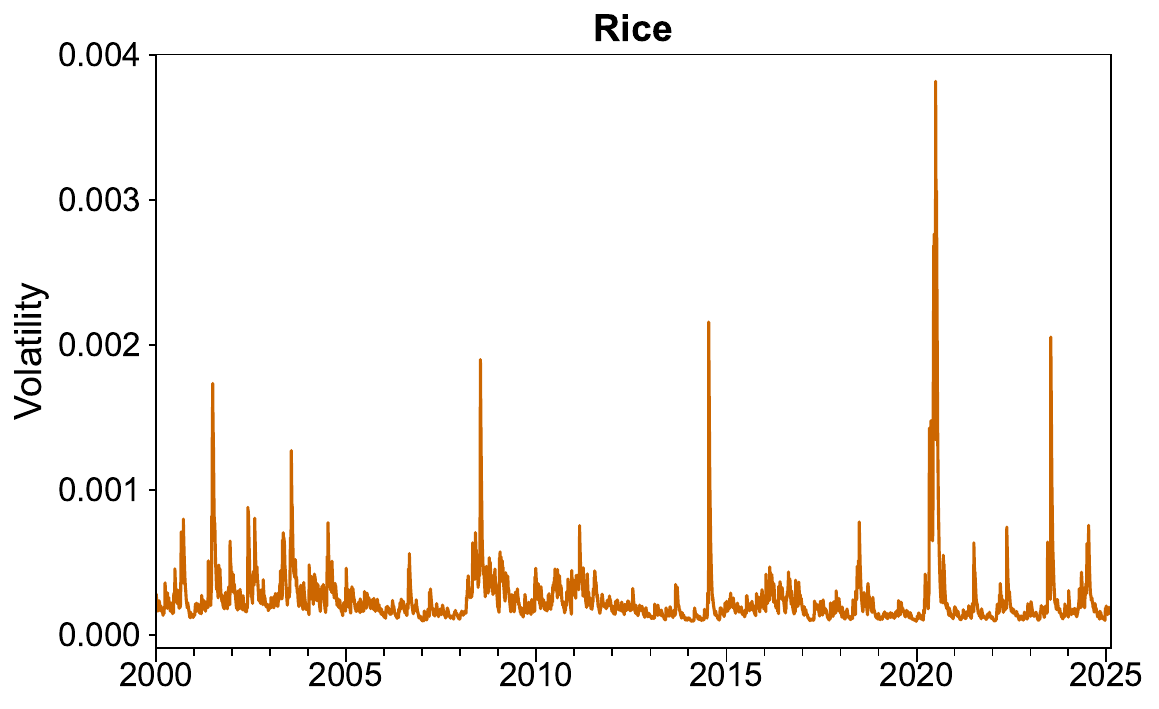}
  \includegraphics[width=0.245\linewidth]{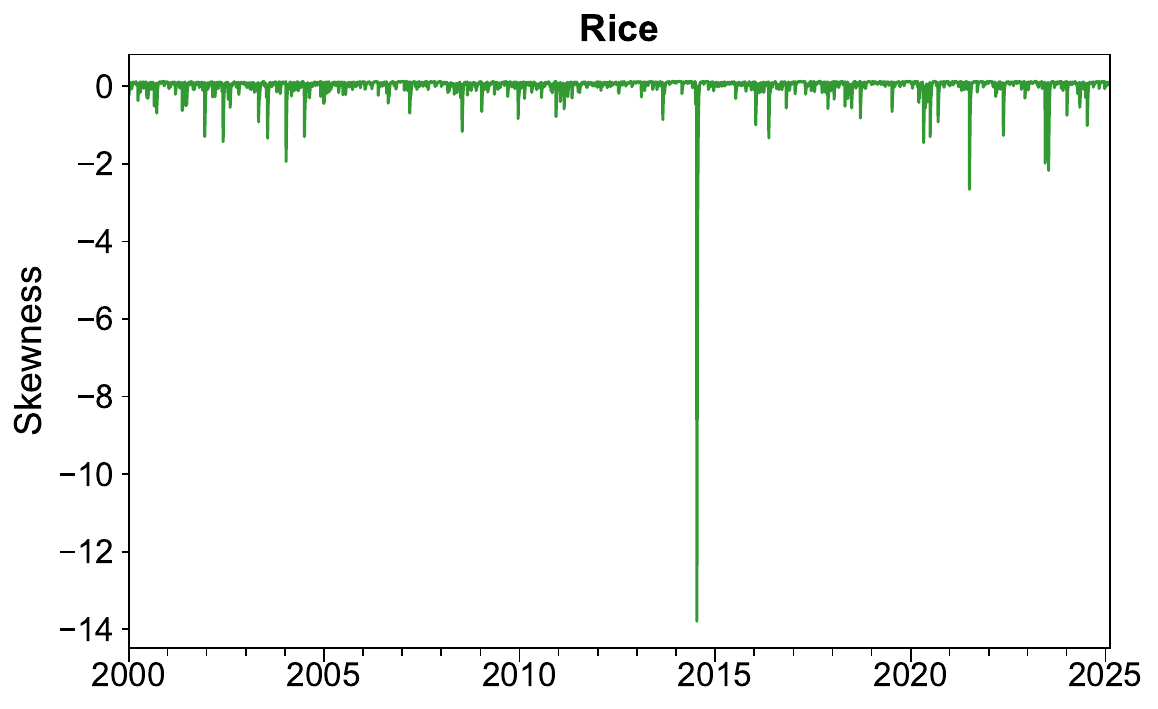}
  \includegraphics[width=0.245\linewidth]{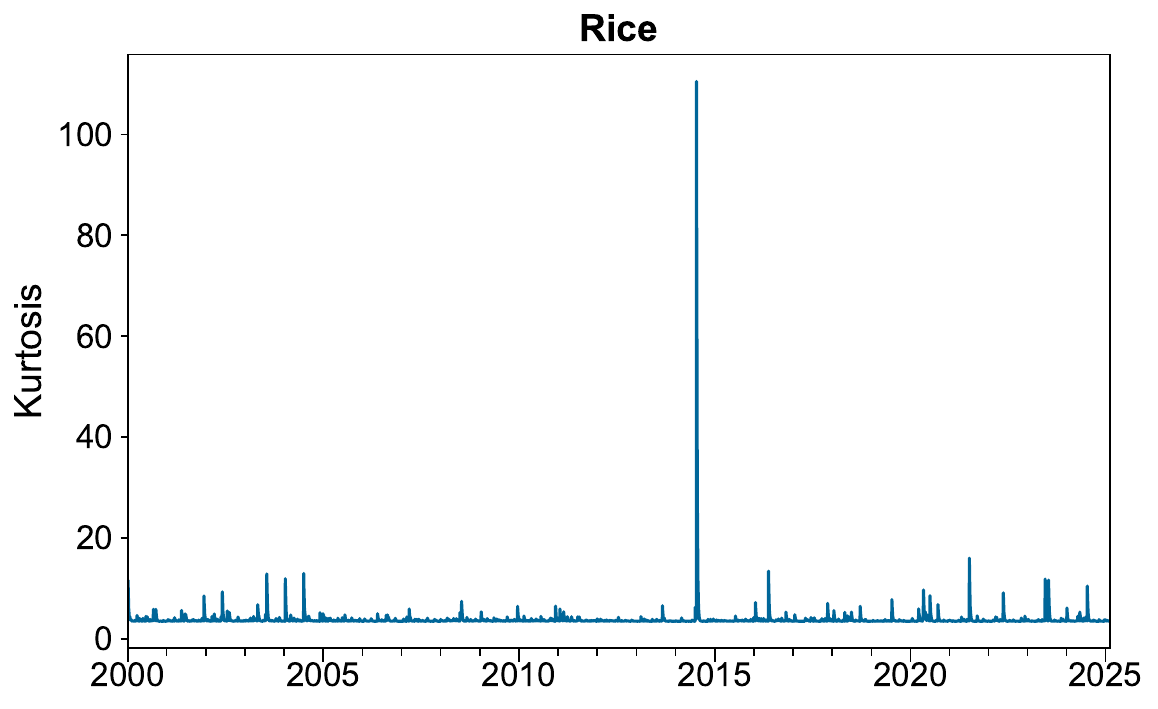}\\
  \includegraphics[width=0.245\linewidth]{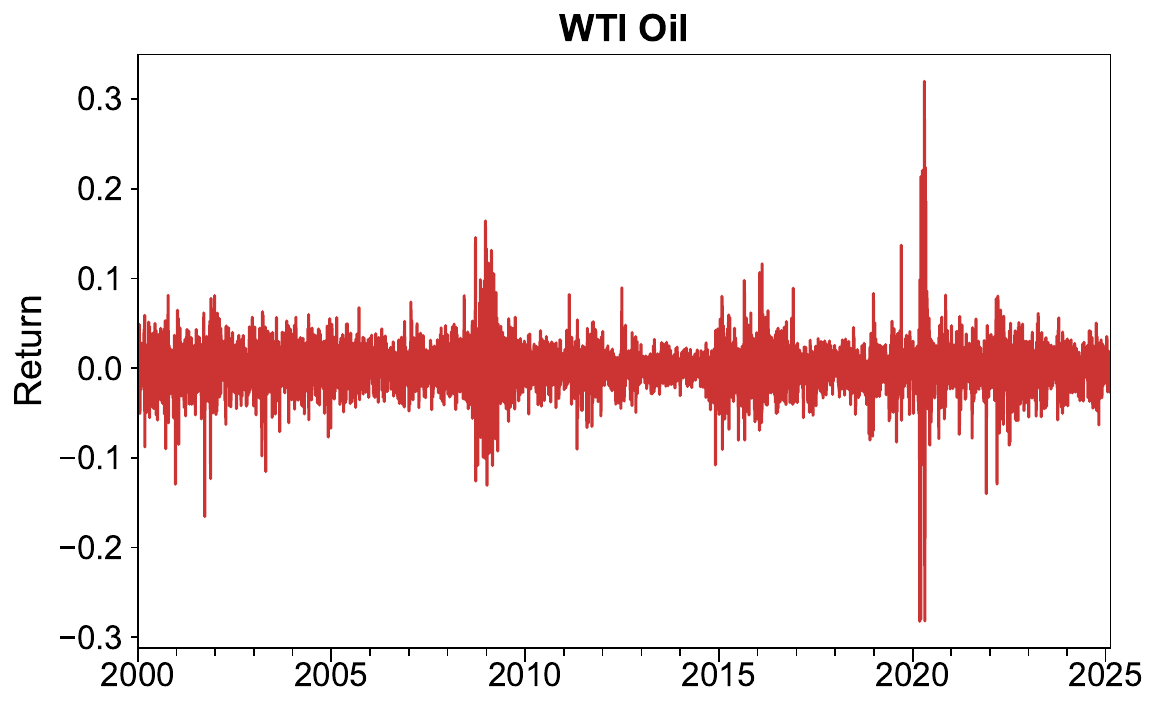}
  \includegraphics[width=0.245\linewidth]{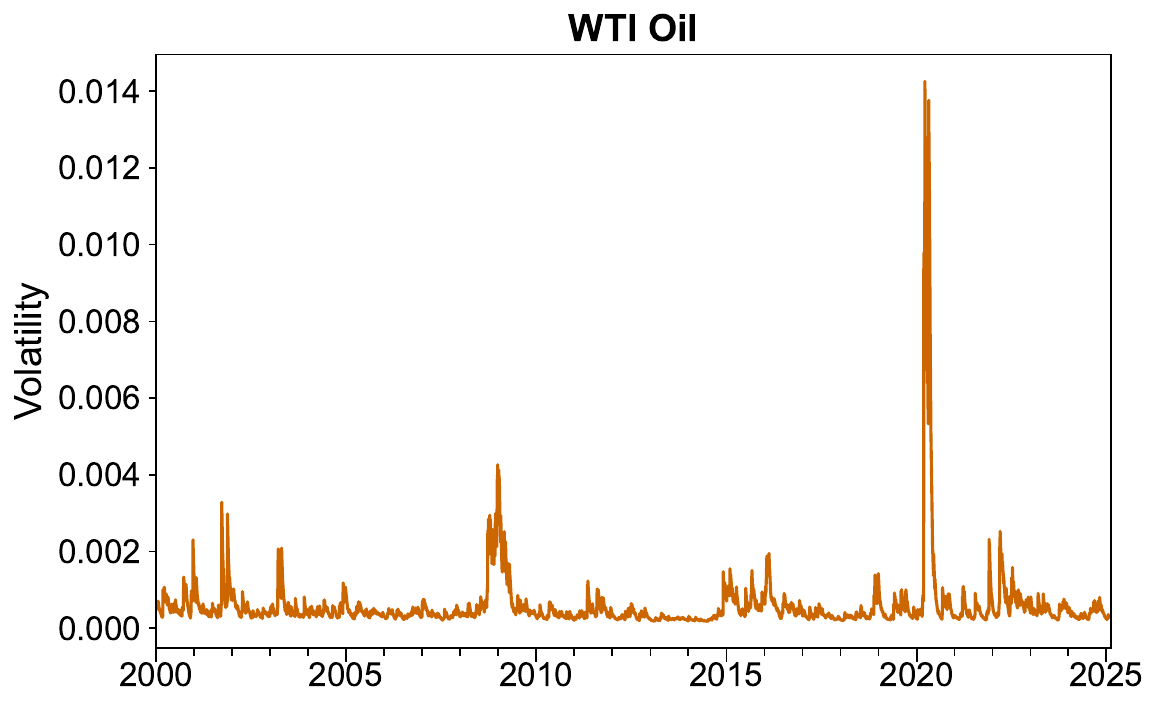}
  \includegraphics[width=0.245\linewidth]{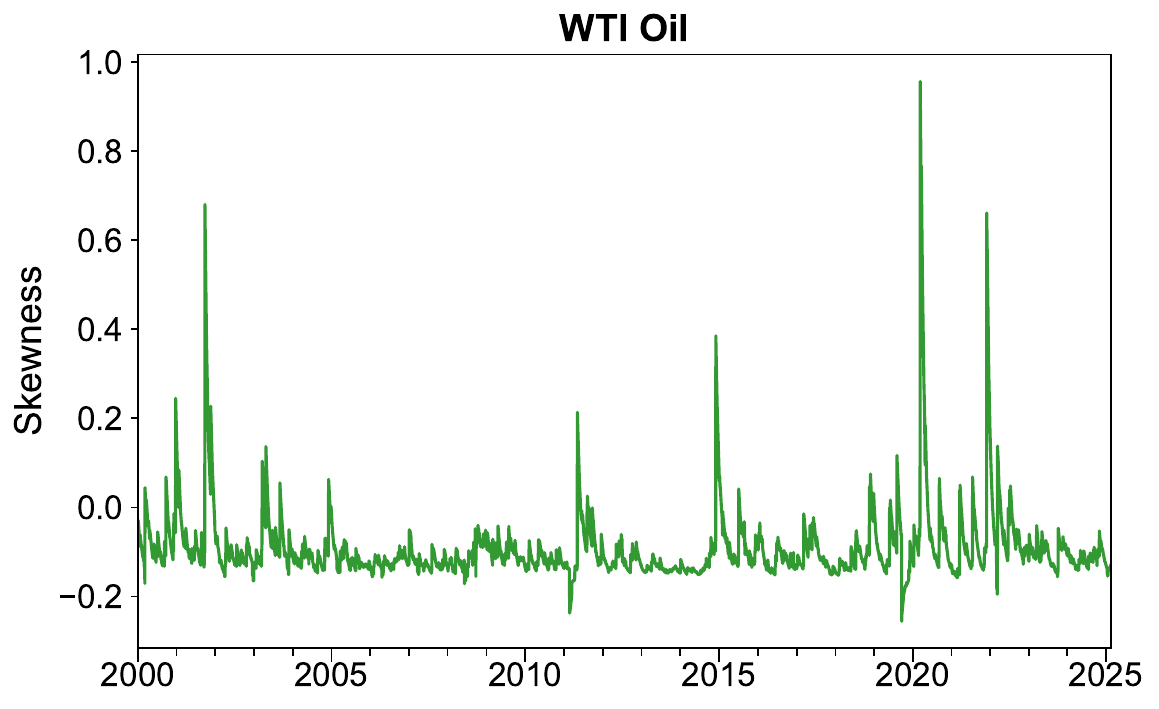}
  \includegraphics[width=0.245\linewidth]{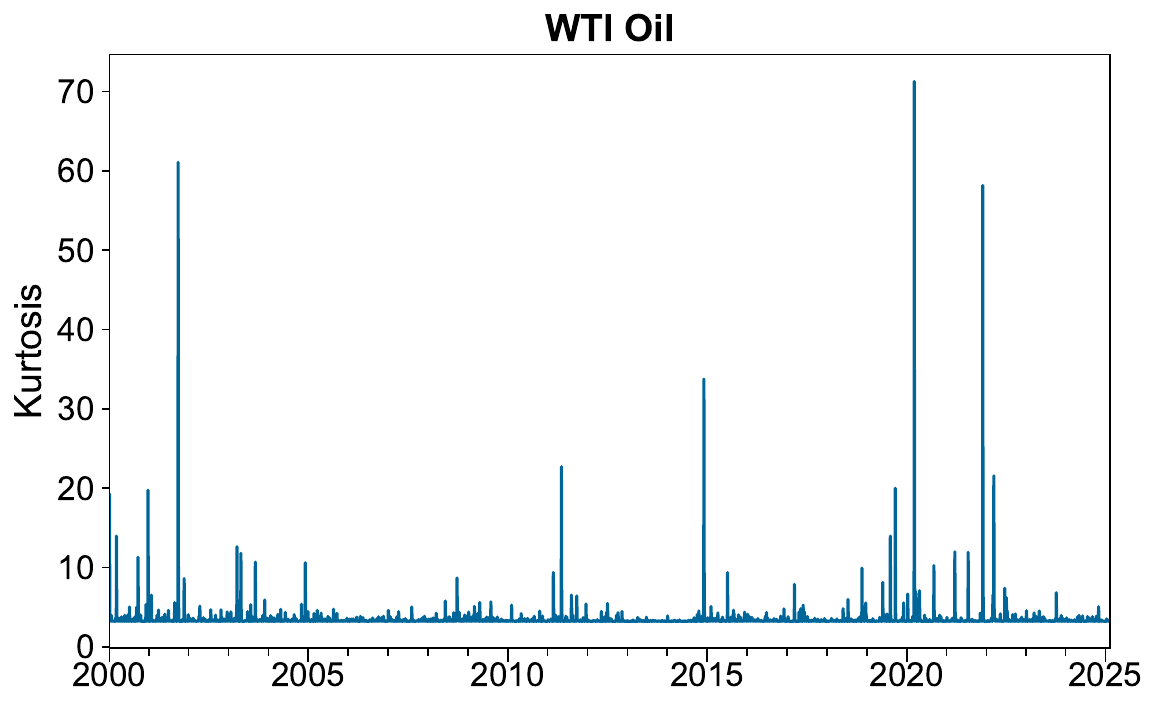}\\
  \includegraphics[width=0.245\linewidth]{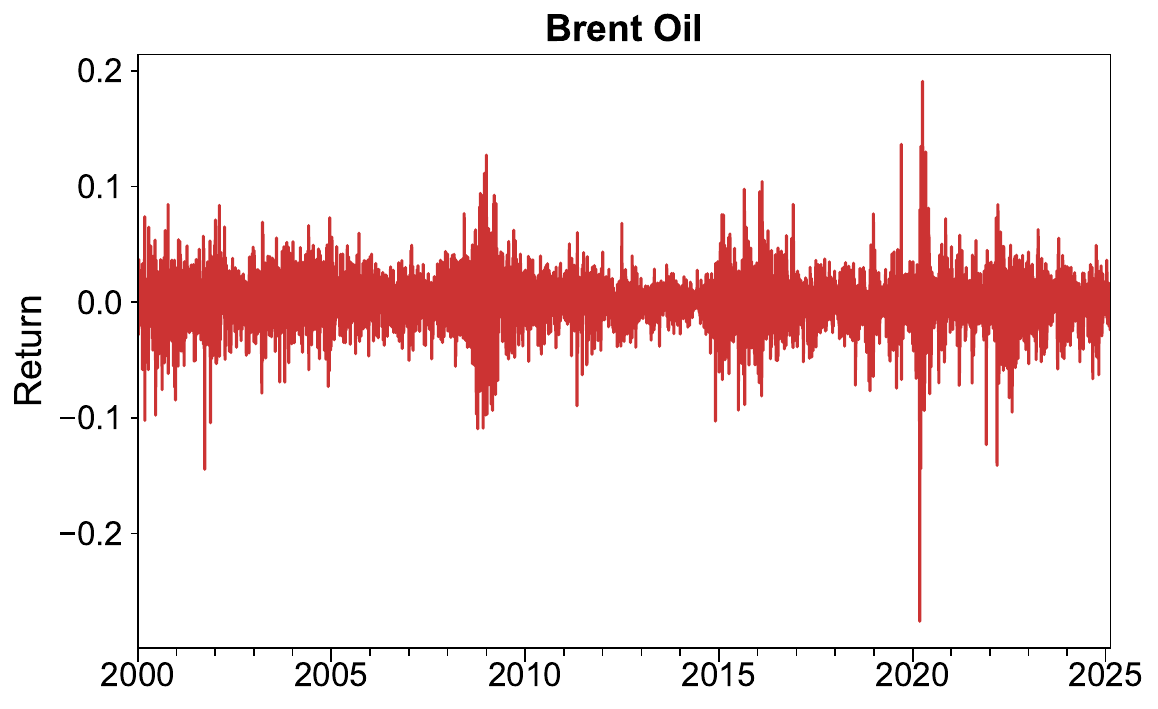}
  \includegraphics[width=0.245\linewidth]{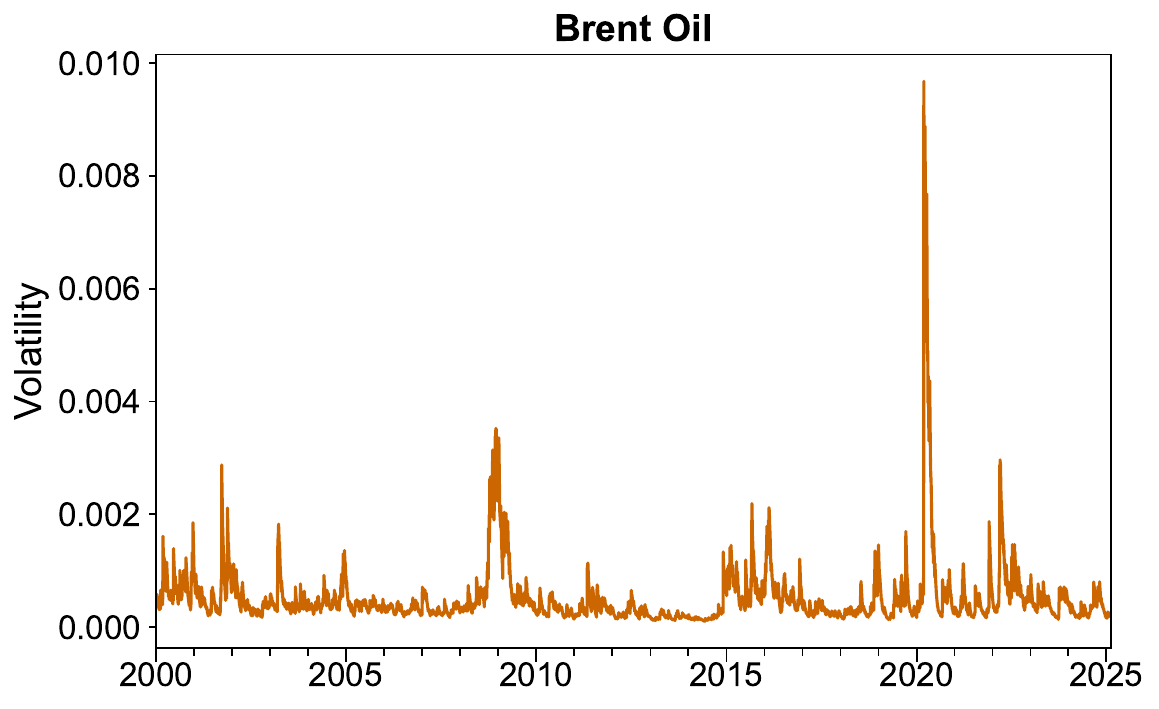}
  \includegraphics[width=0.245\linewidth]{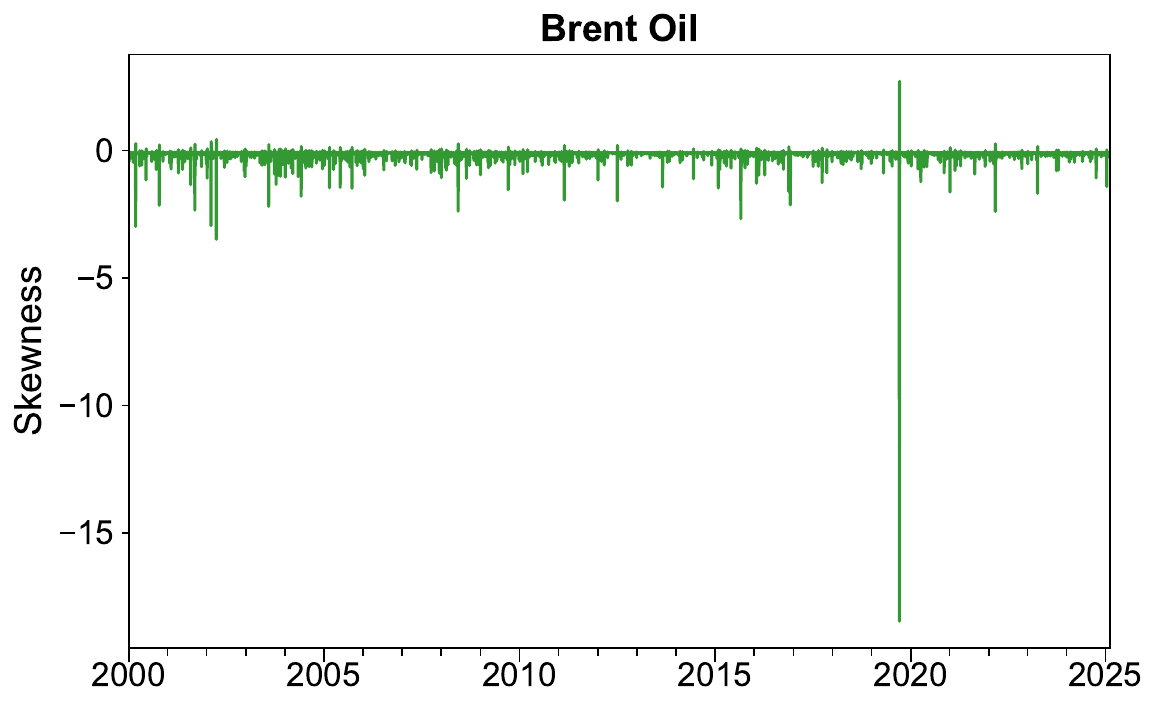}
  \includegraphics[width=0.245\linewidth]{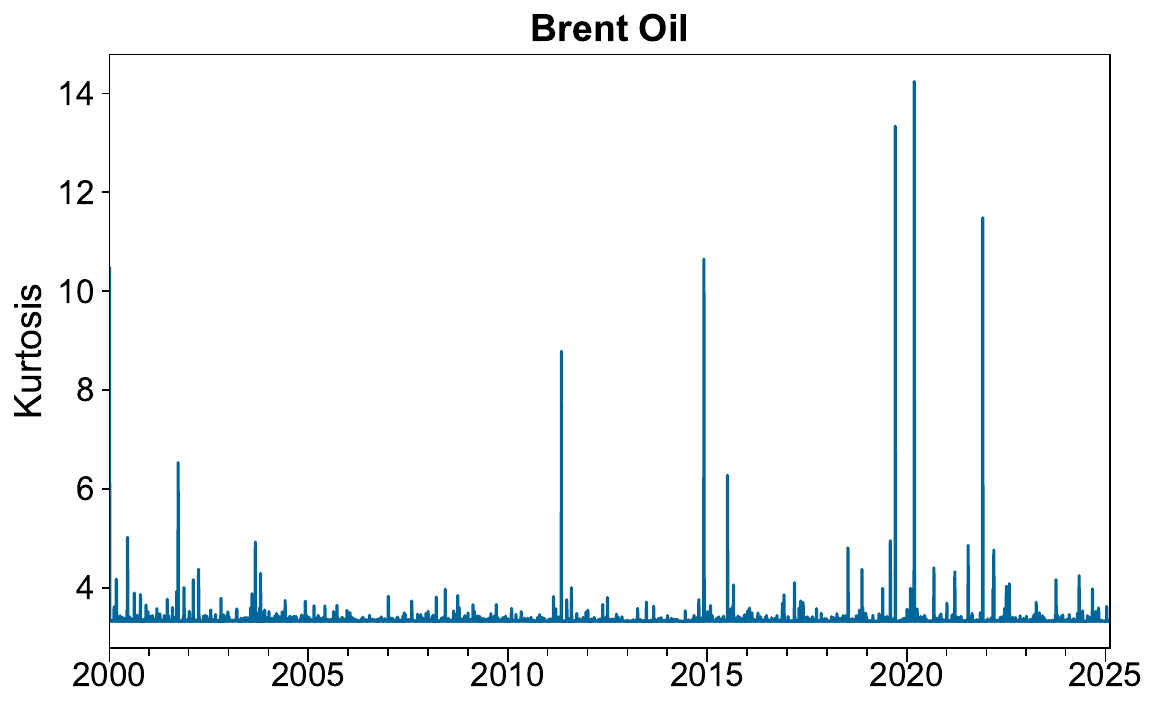}\\
  \includegraphics[width=0.245\linewidth]{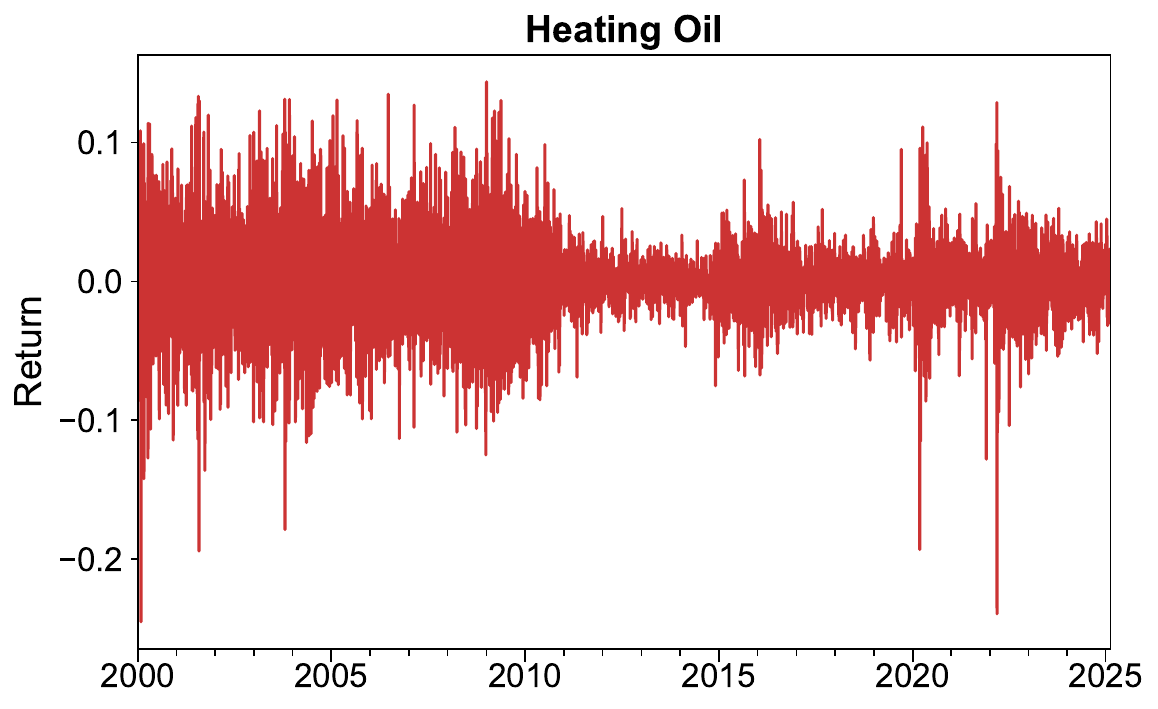}
  \includegraphics[width=0.245\linewidth]{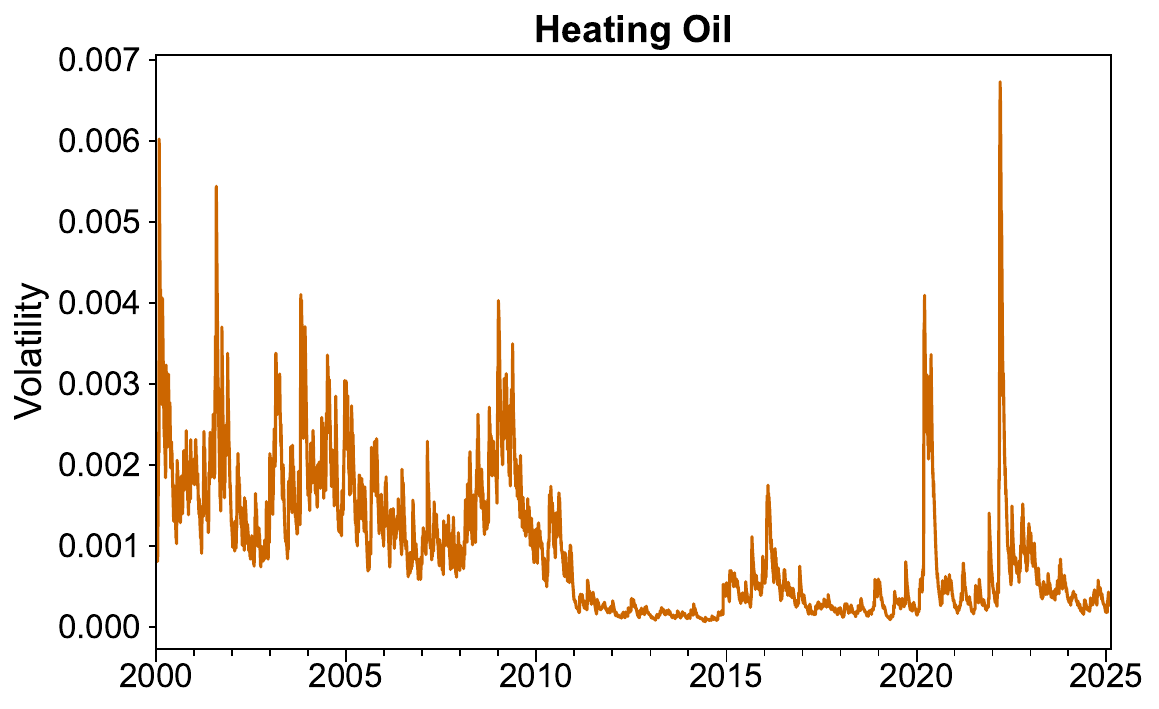}
  \includegraphics[width=0.245\linewidth]{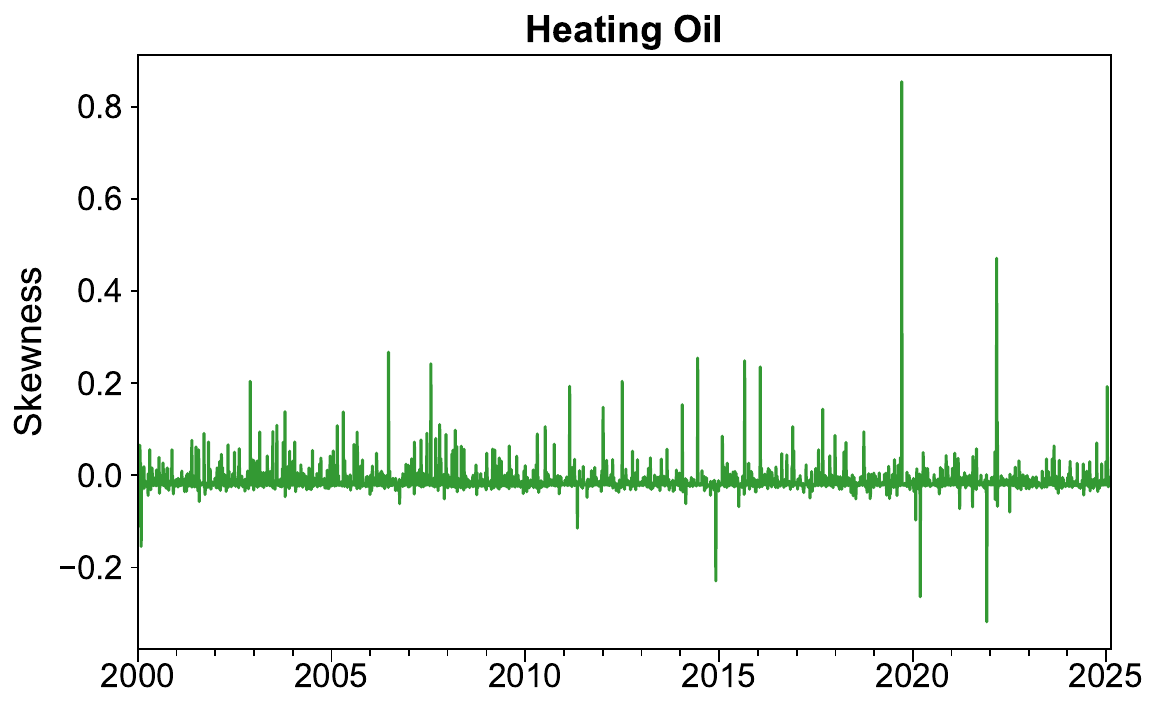}
  \includegraphics[width=0.245\linewidth]{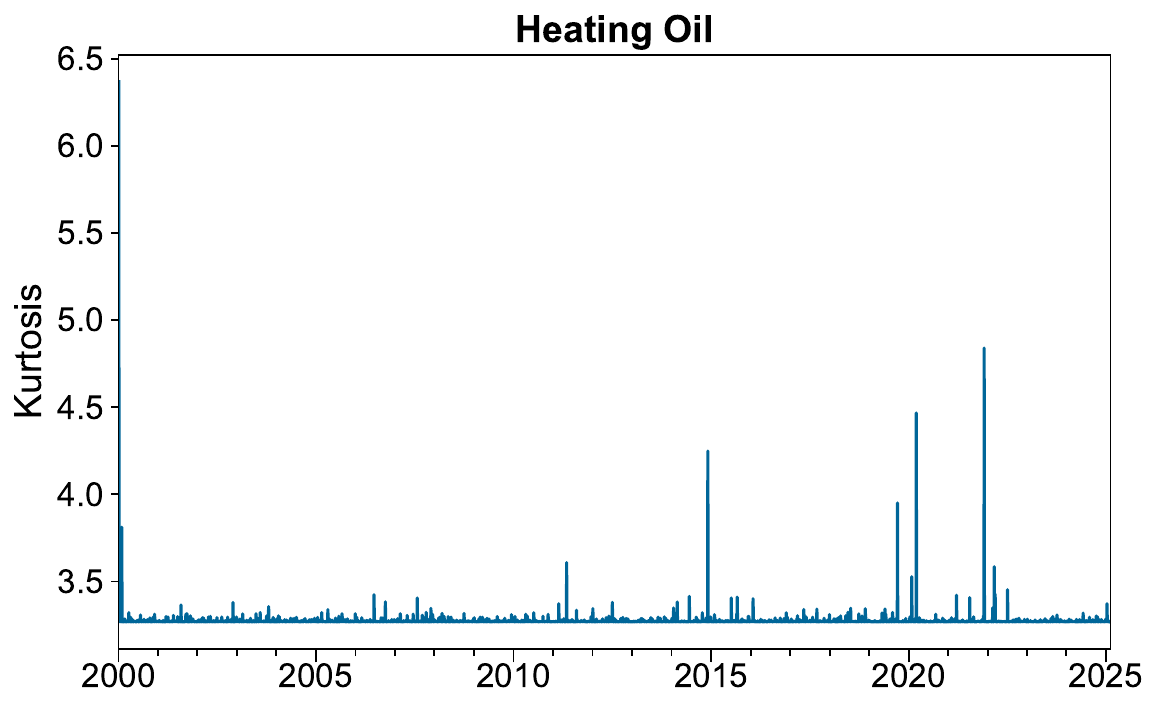}\\
  \includegraphics[width=0.245\linewidth]{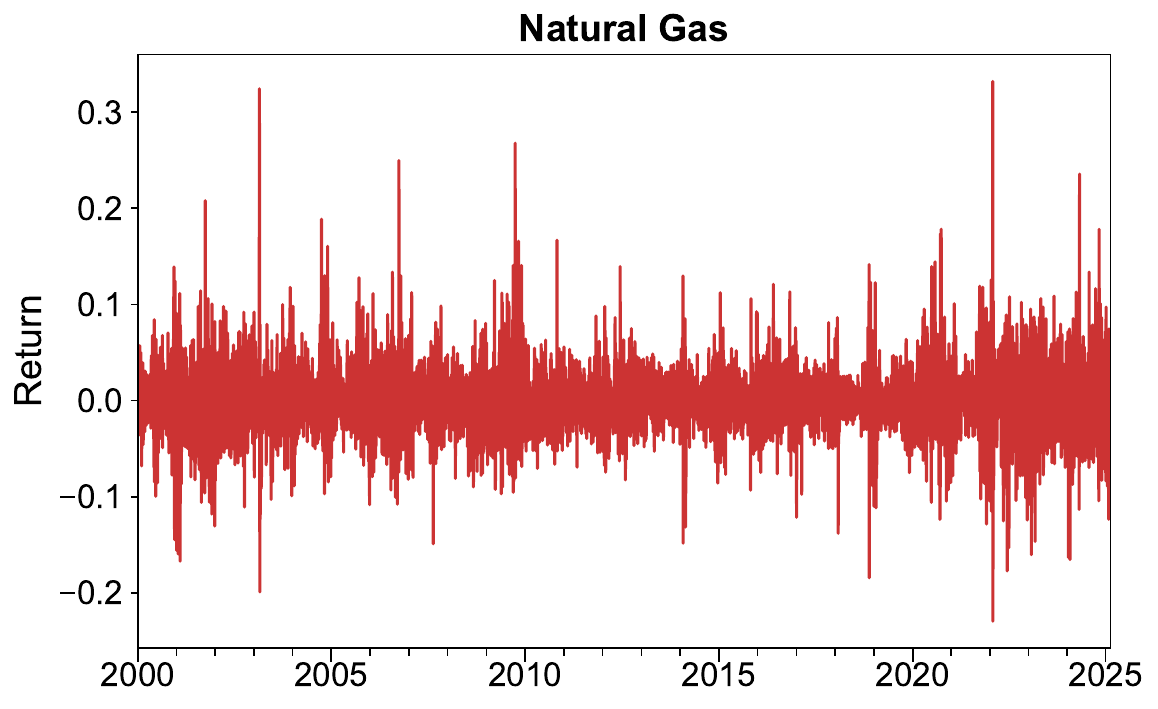}
  \includegraphics[width=0.245\linewidth]{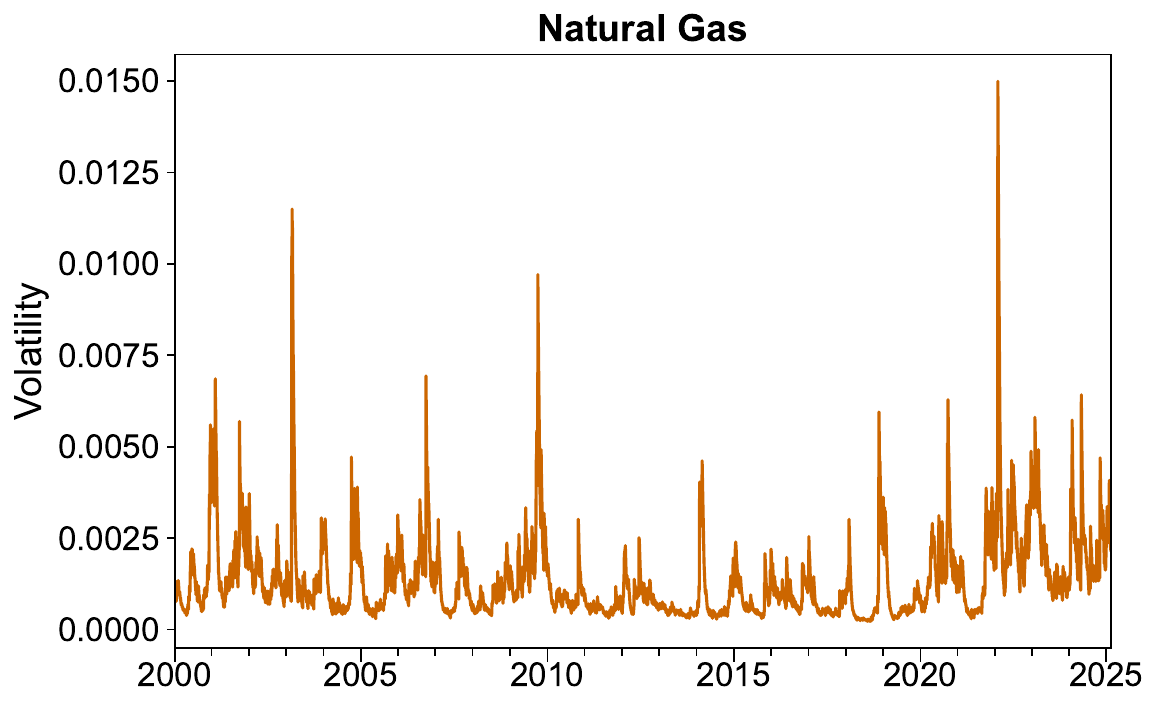}
  \includegraphics[width=0.245\linewidth]{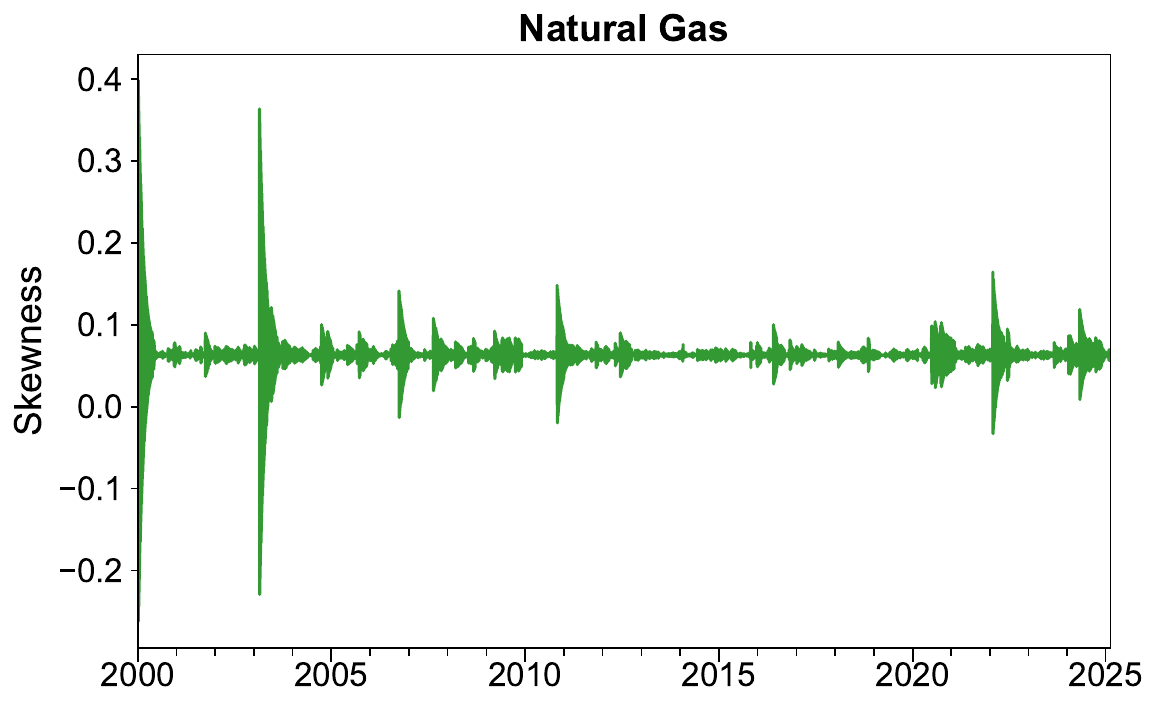}
  \includegraphics[width=0.245\linewidth]{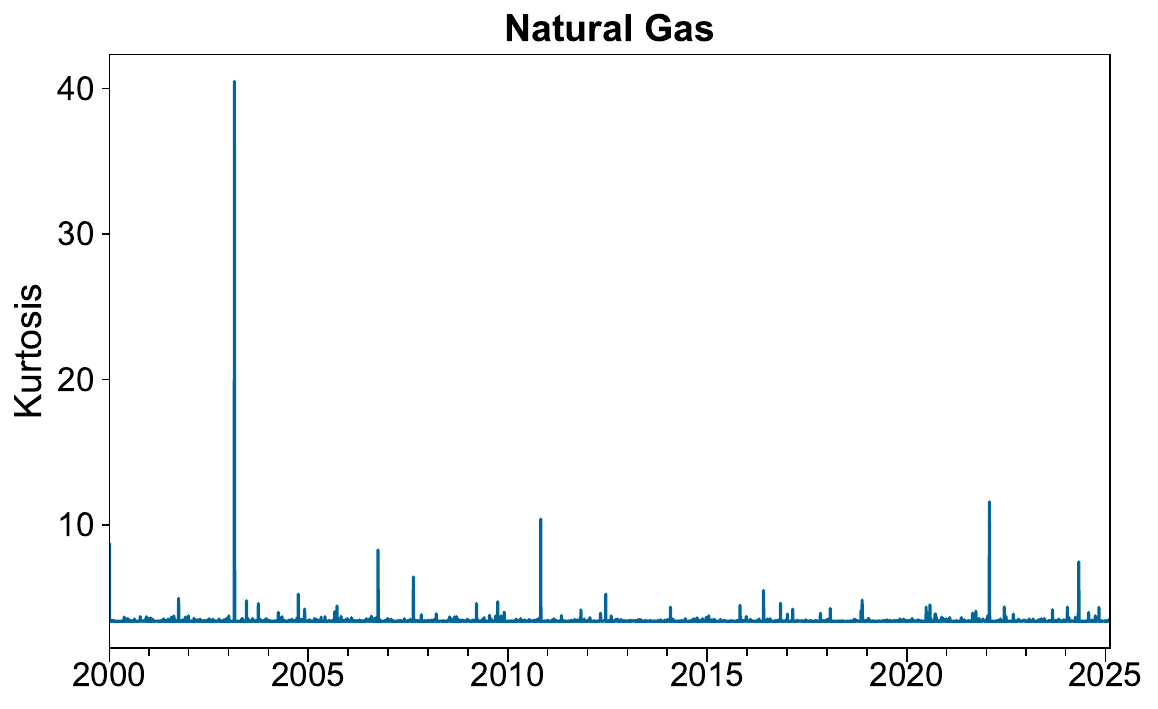}
  \caption{Evolution of return, volatility, skewness, and kurtosis for staple food and energy futures.}
\label{Fig:Agro_HighMoment_evolution}
\end{figure}

Figure~\ref{Fig:Agro_HighMoment_evolution} depicts the dynamic evolution of daily returns, volatility, skewness, and kurtosis for staple food and energy futures. It is evident that all four moments in both food and energy markets exhibit pronounced time-varying characteristics and are highly responsive to major crisis events. For instance, during the 2008 global financial crisis, the 2010–2012 global food crisis, the 2015 commodity downturn, the COVID-19 pandemic in 2020, the European energy crisis in 2020–2021, and the Russia–Ukraine conflict in 2022, the returns experienced extreme fluctuations, accompanied by notable spikes in volatility. Skewness deviated significantly from zero during these periods, indicating sharp price surges or crashes, while kurtosis rose dramatically, reflecting an increased probability of extreme risk events. Moreover, fluctuations in return, volatility, skewness, and kurtosis are generally more pronounced in energy markets than in food markets, likely due to the greater sensitivity of energy markets to macroeconomic changes, geopolitical tensions, OPEC+ production decisions, and the influence of financialization and speculative trading.

\begin{figure}[!h]
  \centering
  \includegraphics[width=0.45\linewidth]{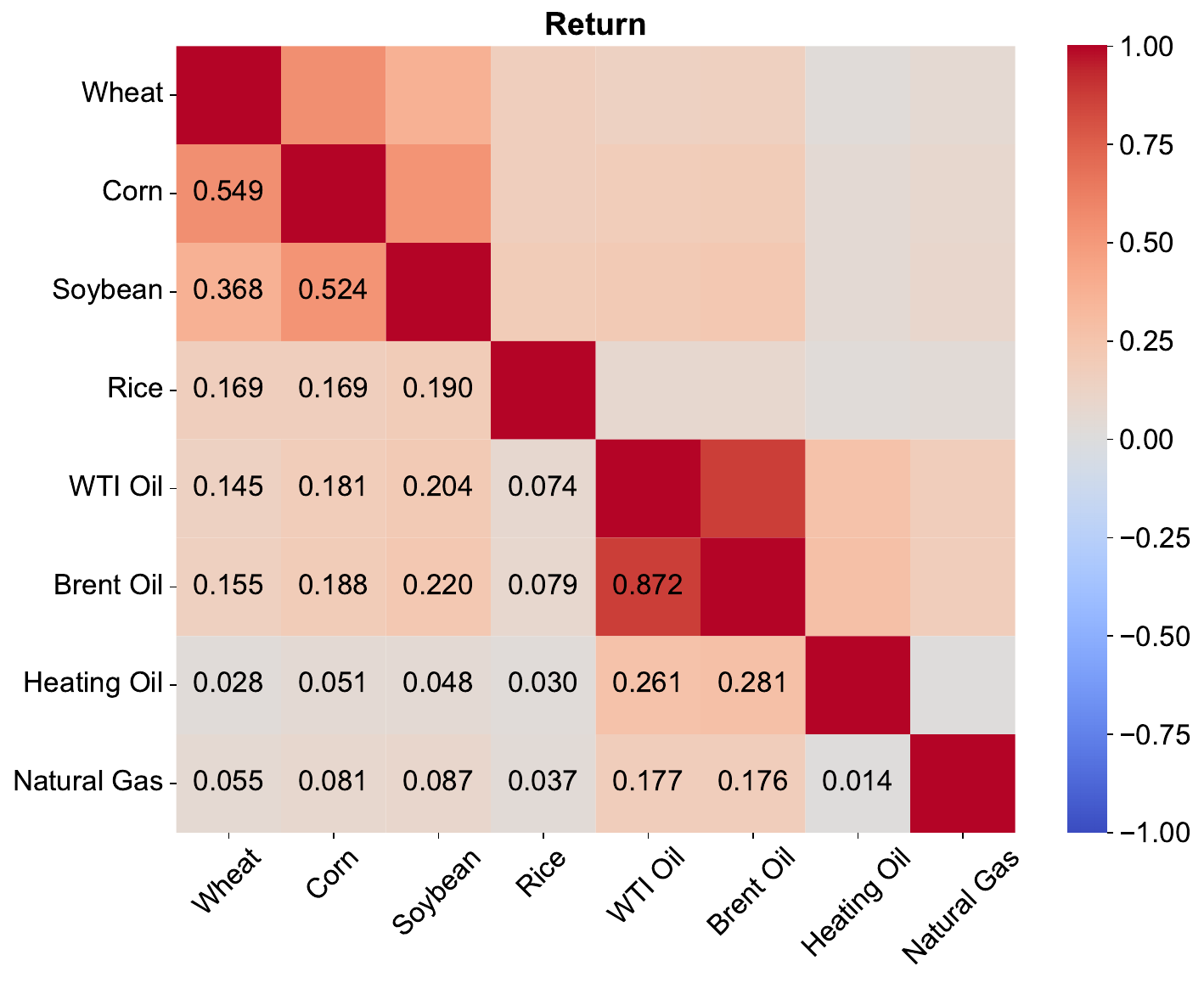}
  \includegraphics[width=0.45\linewidth]{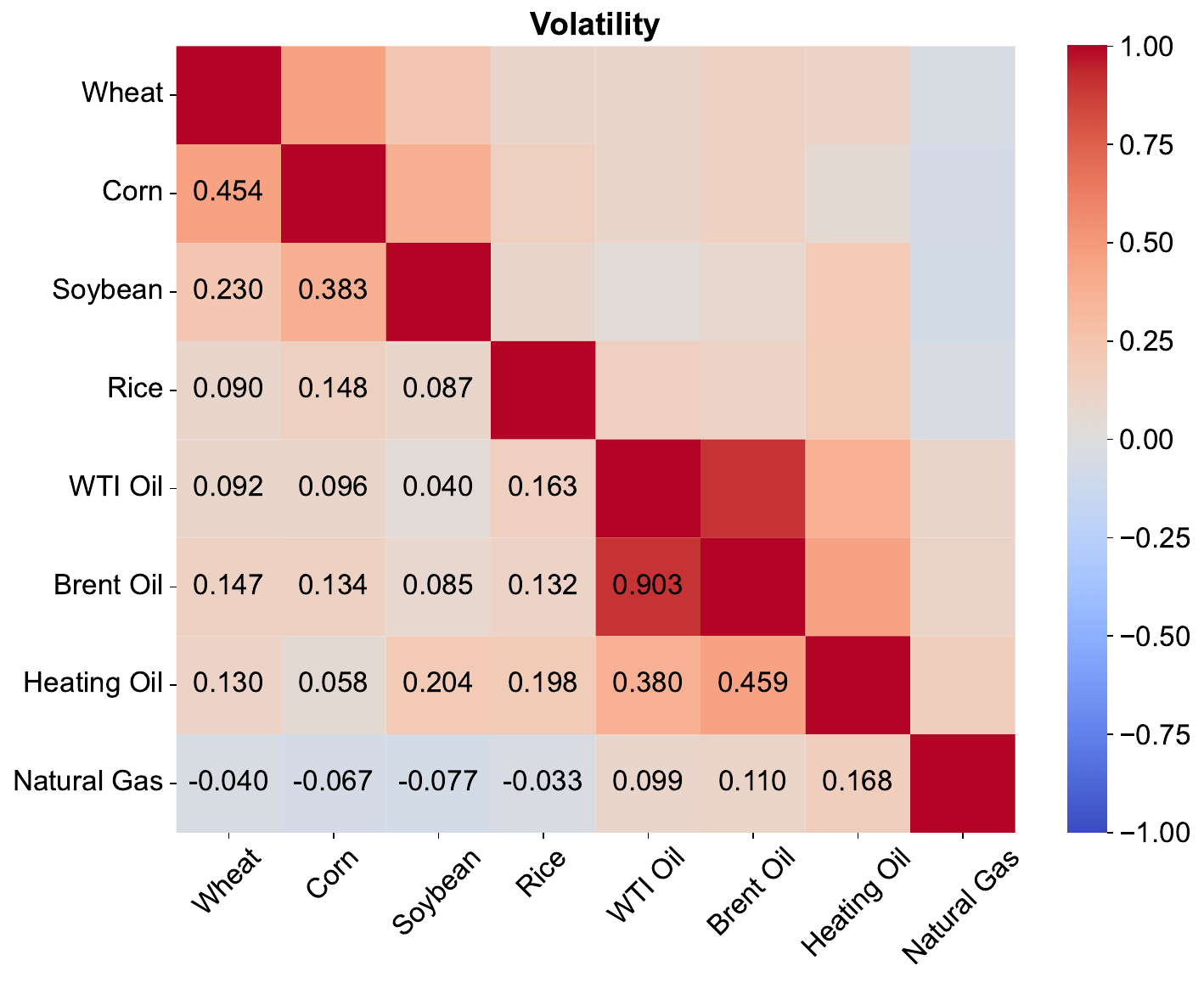}
  \includegraphics[width=0.45\linewidth]{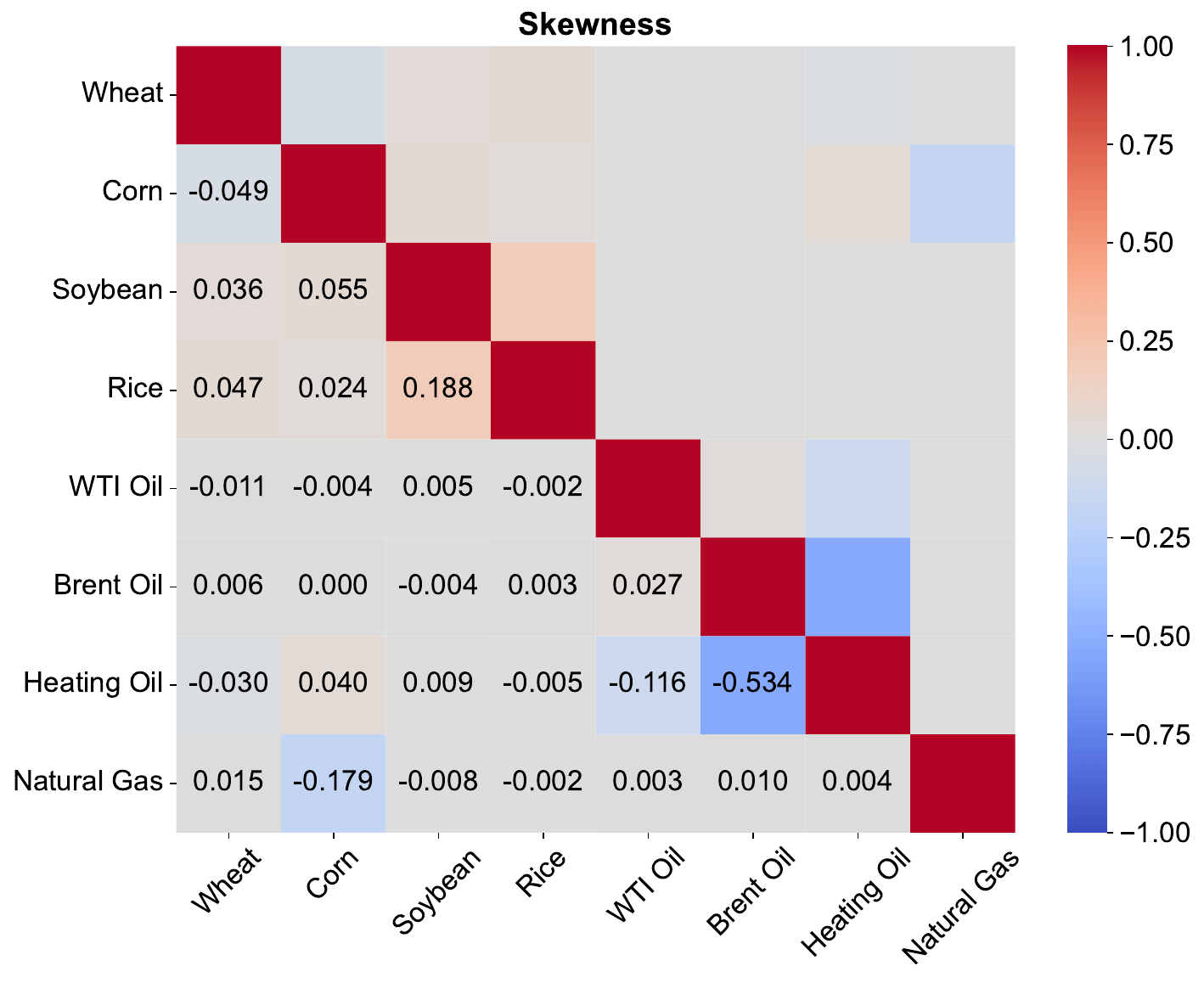}
  \includegraphics[width=0.45\linewidth]{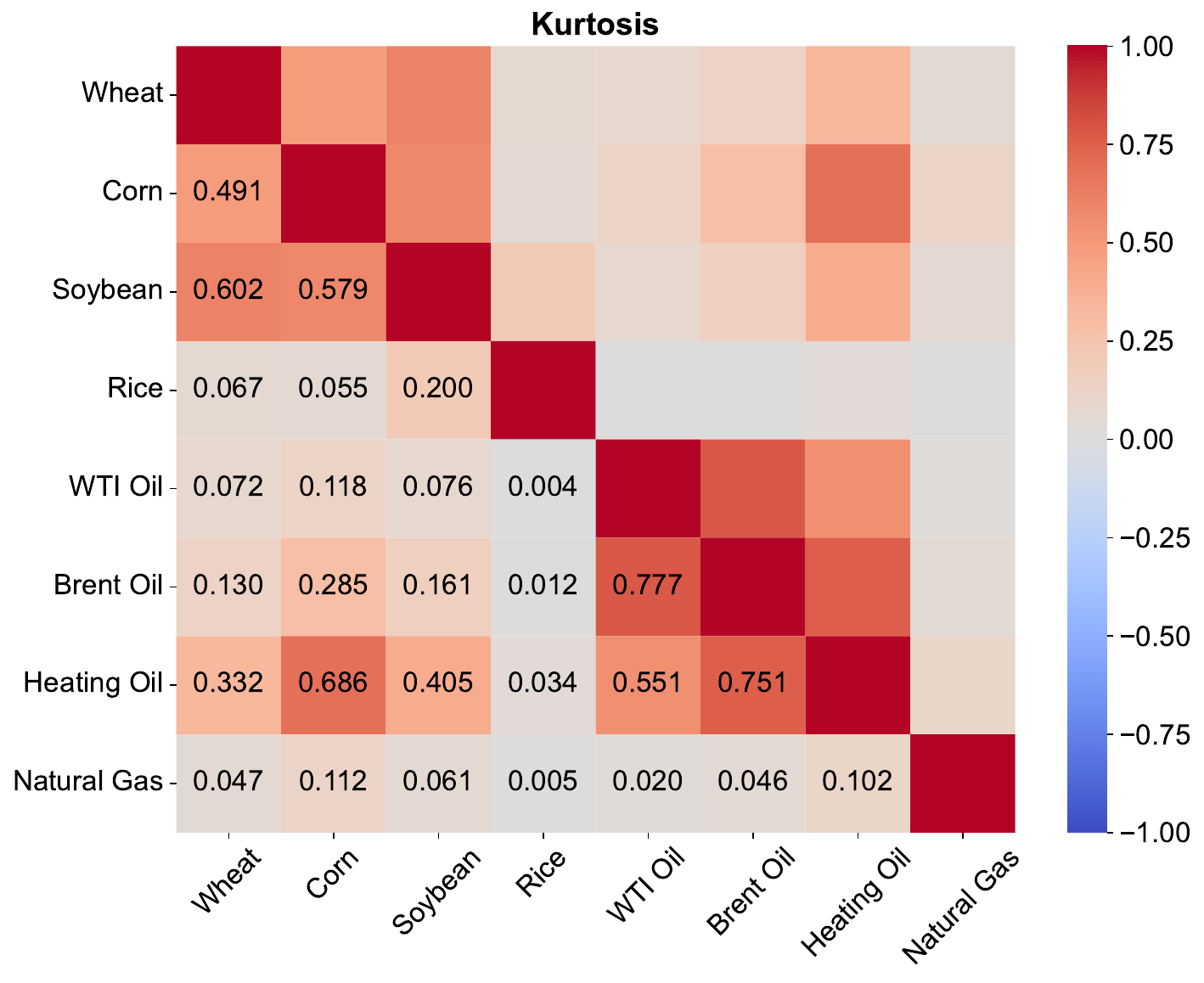}
  \caption{Heatmaps of correlation coefficients across return, volatility, skewness, and kurtosis for staple foods and energy.}  
\label{Fig:Heatmap_Correlation}
\end{figure}

Figure~\ref{Fig:Heatmap_Correlation} presents the heatmaps of correlation coefficients for return, volatility, skewness, and kurtosis across various food and energy futures. The correlations of returns and volatility within the food market and within the energy market are generally higher than those between food and energy markets, implying stronger internal linkages. For example, the return correlation between wheat and corn, as well as between corn and soybean, is relatively high. The former may be attributed to their substitutability in feed, while the latter likely stems from their shared role in biofuel production. Additionally, both pairs are subject to similar supply and demand shocks, such as weather conditions and agricultural policies, which further strengthen their interconnections. In the energy market, the return correlation between WTI and Brent oil reaches 0.872, and their volatility correlation is as high as 0.903. As both are benchmarks in the global crude oil market, their prices tend to move in tandem, especially during periods of market turbulence. Heating oil also shows strong correlations with crude oil in both returns and volatility, which can be explained by the fact that it is a refined product of crude oil and thus closely tied to crude oil price dynamics.

Skewness correlations are relatively weak. Within the food market, skewness correlations are close to zero, and even within the energy market, they remain low or sometimes negative, suggesting that the asymmetry in return distributions does not always shift in a synchronized manner across commodities. In contrast, kurtosis displays stronger correlations both within and between food and energy markets. This indicates that during extreme events, such as financial crises, the COVID-19 pandemic, and the Russia-Ukraine conflict, these markets are more likely to exhibit fat-tailed return distributions simultaneously, reflecting an increased probability of extreme price movements. Moreover, despite differences in market characteristics, the overall correlation between food and energy remains notable, which is largely related to their common classification as bulk commodities and the growing interconnections driven by biofuel-related demand linkages.

\subsection{Time-domain risk connectedness}

The novel TVP-VAR-DY connectedness approach is utilized to examine the time-domain moment connectedness among staple food and energy markets. Table~\ref{Tab:Average_time_connectedness} reports the average time-domain connectedness measures corresponding to return, volatility, skewness, and kurtosis. Diagonal values reflect the connectedness triggered by own-market shocks, while off-diagonal values capture the connectedness arising from cross-market interactions. The total connectedness indices for return, volatility, skewness, and kurtosis are 38.94\%, 44.59\%, 18.20\%, and 34.21\%, respectively, indicating substantial risk spillovers among energy and food markets in the time domain, especially in terms of volatility, return, and tail risks.

\begin{table}[!ht]
  \centering
  \setlength{\abovecaptionskip}{0pt}
  \setlength{\belowcaptionskip}{10pt}
  \caption{Average return, volatility, skewness, and kurtosis connectedness among staple food and energy markets}
  \setlength\tabcolsep{8pt}   \resizebox{\textwidth}{!}{ 
    \begin{tabular}{l r@{.}l r@{.}l r@{.}l r@{.}l r@{.}l r@{.}l r@{.}l r@{.}l r@{.}l}
    \toprule
    & \multicolumn{2}{c}{Wheat} & \multicolumn{2}{c}{Corn} & \multicolumn{2}{c}{Soybean} & \multicolumn{2}{c}{Rice} & \multicolumn{2}{c}{WTI Oil} & \multicolumn{2}{c}{Brent Oil} & \multicolumn{2}{c}{Heating Oil} & \multicolumn{2}{c}{Natural Gas} & \multicolumn{2}{c}{FROM}  \\
    \midrule
    \multicolumn{19}{l}{\textit{Panel A: Average return connectedness}} \\
    Wheat & 62&91 & 19&89 & 9&13 & 2&56 & 1&85 & 1&94 & 0&89 & 0&85 & 37&09 \\
    Corn & 18&10 & 57&18 & 16&13 & 2&15 & 2&14 & 2&40 & 1&02 & 0&88 & 42&82 \\
    Soybean & 8&80 & 17&40 & 61&79 & 2&69 & 3&23 & 3&49 & 1&73 & 0&87 & 38&21 \\
    Rice & 3&42 & 3&25 & 3&81 & 84&22 & 1&59 & 1&81 & 1&01 & 0&89 & 15&78 \\
    WTI Oil & 1&25 & 1&66 & 2&45 & 0&88 & 44&14 & 33&84 & 12&95 & 2&82 & 55&86 \\
    Brent Oil & 1&36 & 1&82 & 2&60 & 0&98 & 33&03 & 43&11 & 14&50 & 2&60 & 56&89 \\
    Heating Oil & 0&87 & 1&23 & 1&75 & 0&80 & 20&13 & 22&62 & 50&19 & 2&42 & 49&81 \\
    Natural Gas & 1&02 & 1&37 & 1&26 & 0&95 & 4&68 & 4&24 & 1&49 & 84&98 & 15&02 \\
    TO & 34&82 & 46&61 & 37&13 & 11&02 & 66&66 & 70&35 & 33&59 & 11&32 & \multicolumn{2}{c}{TCI} \\
    Net & $-$2&27 & 3&79 & $-$1&07 & $-$4&76 & 10&80 & 13&46 & $-$16&23 & $-$3&70 & 38&94  \vspace{2mm} \\
    \multicolumn{19}{l}{\textit{Panel B: Average volatility connectedness}} \\
    Wheat & 56&39 & 12&63 & 6&07 & 4&87 & 4&87 & 5&20 & 6&11 & 3&86 & 43&61 \\
    Corn & 12&82 & 57&30 & 8&39 & 4&58 & 4&01 & 4&35 & 4&65 & 3&89 & 42&70 \\
    Soybean & 6&54 & 13&16 & 59&30 & 3&97 & 4&52 & 4&84 & 4&62 & 3&05 & 40&70 \\
    Rice & 4&65 & 5&22 & 3&86 & 71&03 & 4&14 & 3&72 & 3&71 & 3&66 & 28&97 \\
    WTI Oil & 5&04 & 3&26 & 2&47 & 2&24 & 42&16 & 29&65 & 10&96 & 4&21 & 57&84 \\
    Brent Oil & 5&25 & 3&21 & 2&44 & 2&34 & 29&08 & 40&04 & 13&42 & 4&23 & 59&96 \\
    Heating Oil & 10&67 & 5&90 & 4&26 & 4&71 & 11&93 & 11&56 & 45&01 & 5&96 & 54&99 \\
    Natural Gas & 4&72 & 3&58 & 3&03 & 2&85 & 3&72 & 4&31 & 5&77 & 72&01 & 27&99 \\
    TO & 49&68 & 46&97 & 30&52 & 25&56 & 62&28 & 63&63 & 49&24 & 28&87 & \multicolumn{2}{c}{TCI} \\
    Net & 6&07 & 4&27 & $-$10&18 & $-$3&40 & 4&44 & 3&66 & $-$5&75 & 0&88 & 44&59  \vspace{2mm} \\
    \multicolumn{19}{l}{\textit{Panel C: Average skewness connectedness}} \\
    Wheat & 81&03 & 4&21 & 4&05 & 2&00 & 5&88 & 0&86 & 1&24 & 0&75 & 18&97 \\
    Corn & 4&51 & 84&16 & 6&18 & 2&37 & 0&74 & 0&59 & 0&31 & 1&13 & 15&84 \\
    Soybean & 3&87 & 6&06 & 84&40 & 2&66 & 1&40 & 0&57 & 0&57 & 0&47 & 15&60 \\
    Rice & 3&31 & 2&53 & 3&40 & 83&20 & 4&52 & 1&37 & 0&96 & 0&71 & 16&80 \\
    WTI Oil & 6&65 & 0&50 & 0&92 & 1&71 & 78&38 & 3&28 & 8&11 & 0&45 & 21&62 \\
    Brent Oil & 0&60 & 0&87 & 0&39 & 1&14 & 3&76 & 75&81 & 16&28 & 1&14 & 24&19 \\
    Heating Oil & 0&56 & 0&31 & 0&75 & 0&83 & 6&65 & 16&89 & 73&30 & 0&71 & 26&70 \\
    Natural Gas & 0&73 & 1&25 & 0&41 & 0&78 & 0&48 & 1&48 & 0&77 & 94&11 & 5&89 \\
    TO & 20&22 & 15&73 & 16&09 & 11&50 & 23&44 & 25&04 & 28&23 & 5&36 & \multicolumn{2}{c}{TCI} \\
    Net & 1&25 & $-$0&10 & 0&49 & $-$5&30 & 1&81 & 0&85 & 1&53 & $-$0&54 & 18&20  \vspace{2mm} \\
    \multicolumn{19}{l}{\textit{Panel D: Average kurtosis connectedness}} \\
    Wheat & 69&65 & 8&58 & 9&06 & 4&45 & 2&73 & 2&67 & 1&66 & 1&19 & 30&35 \\
    Corn & 8&03 & 71&47 & 9&14 & 4&11 & 1&93 & 2&35 & 1&47 & 1&50 & 28&53 \\
    Soybean & 7&52 & 9&35 & 69&15 & 6&15 & 2&74 & 2&64 & 1&85 & 0&59 & 30&85 \\
    Rice & 4&60 & 4&18 & 5&36 & 77&02 & 3&29 & 2&75 & 2&12 & 0&67 & 22&98 \\
    WTI Oil & 2&51 & 2&10 & 1&75 & 3&04 & 47&58 & 27&57 & 14&67 & 0&79 & 52&42 \\
    Brent Oil & 2&03 & 2&14 & 1&14 & 2&15 & 28&16 & 47&38 & 16&11 & 0&90 & 52&62 \\
    Heating Oil & 1&49 & 2&27 & 1&61 & 2&47 & 16&60 & 18&93 & 55&60 & 1&04 & 44&40 \\
    Natural Gas & 1&54 & 1&95 & 1&20 & 1&25 & 2&31 & 1&95 & 1&36 & 88&43 & 11&57 \\
    TO & 27&72 & 30&57 & 29&27 & 23&60 & 57&75 & 58&85 & 39&25 & 6&69 & \multicolumn{2}{c}{TCI} \\
    Net & $-$2&63 & 2&05 & $-$1&58 & 0&63 & 5&34 & 6&23 & $-$5&15 & $-$4&88 & 34&21 \\
  \bottomrule
    \end{tabular}
    }%
  \begin{flushleft}
    \footnotesize
    \justifying Note: This table presents the average time-domain connectedness measures of staple food and energy markets, corresponding to return, volatility, skewness, and kurtosis, respectively.
  \end{flushleft}
  \label{Tab:Average_time_connectedness}%
\end{table}%

In general, risk transmission within the energy market and within the food market is stronger than that between the two, and futures with larger risk spillovers tend to receive greater spillovers. Notably, soybean, corn, and wheat, along with WTI oil, Brent oil, and heating oil, usually exhibit higher TO and FROM indices, whereas rice and natural gas display weaker moment connectedness. This disparity can primarily be attributed to differences in market structures and supply-demand characteristics. Rice has a limited international trade volume, with most production reserved for domestic consumption. Major rice producers, such as China, Thailand, and India, typically implement strong domestic market regulation and reserve policies. In addition, the low liquidity of rice futures also results in a low degree of internationalization. Unlike corn and soybean, rice is not directly involved in the biofuel industry, which further reduces its connectedness within the energy-food system. Similarly, the supply of natural gas is highly regionalized due to transportation infrastructure constraints, unlike crude oil, which is freely traded globally. Moreover, natural gas is primarily used for heating and electricity generation, with demand patterns that differ from those of other commodities. As a result, the risk spillover effects between natural gas and other markets remain relatively limited.

The sign of the NET index reflects each market's role in risk transmission. Upon comparison, we note that different commodities dominate in different moment connectedness. Regarding return connectedness, Brent and WTI oil act as primary transmitters, while heating oil and rice are main receivers. As core energy commodities, crude oil prices affect cost structures of sectors, such as agriculture and transportation. Consequently, risks from the crude oil market tend to spill over into other markets, emphasizing oil's dominant position in the energy-food nexus. For volatility connectedness, wheat, WTI oil, and corn are key transmitters, whereas soybean and heating oil serve as primarily receivers. As global agricultural staples, wheat and corn are susceptible to external shocks, including extreme weather, geopolitical tensions, and export restrictions, which may propagate volatility to related markets. Moreover, corn's role as a major input for ethanol production establishes direct linkages with energy markets. Due to the substitutability between corn and soybean in cultivation, sharp fluctuations in corn prices may lead to shifts in planting decisions, thereby increasing volatility in the soybean market.

In terms of skewness connectedness, rice is identified as a major risk receiver. Owing to its limited trading volume and high self-sufficiency, rice may passively absorb external skewness shocks, resulting in asymmetric price movements. Similar to the return connectedness, Brent and WTI oil are also the primary transmitters to kurtosis connectedness, while heating oil and natural gas emerge as the main receivers. Crude oil markets frequently exhibit high kurtosis in response to unexpected events, such as the OPEC+ price war in March 2020 and the Russia-Ukraine conflict in 2022. Demand for natural gas and heating oil is closely tied to crude oil markets and weather conditions, making them vulnerable to tail risk spillovers from other markets. These findings further underscore the complex risk transmission between energy and food markets and highlight the heterogeneous roles played by different commodities across higher-order connectedness.

\begin{figure}[!h]
  \centering
  \includegraphics[width=0.45\linewidth]{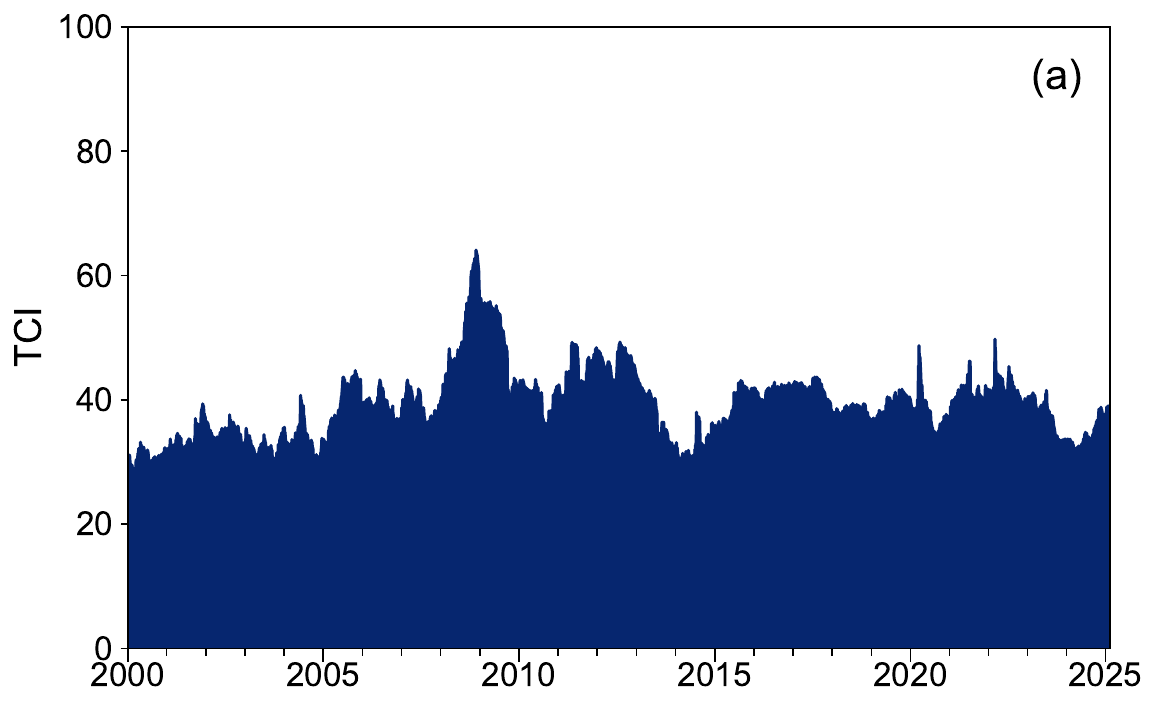}
  \includegraphics[width=0.45\linewidth]{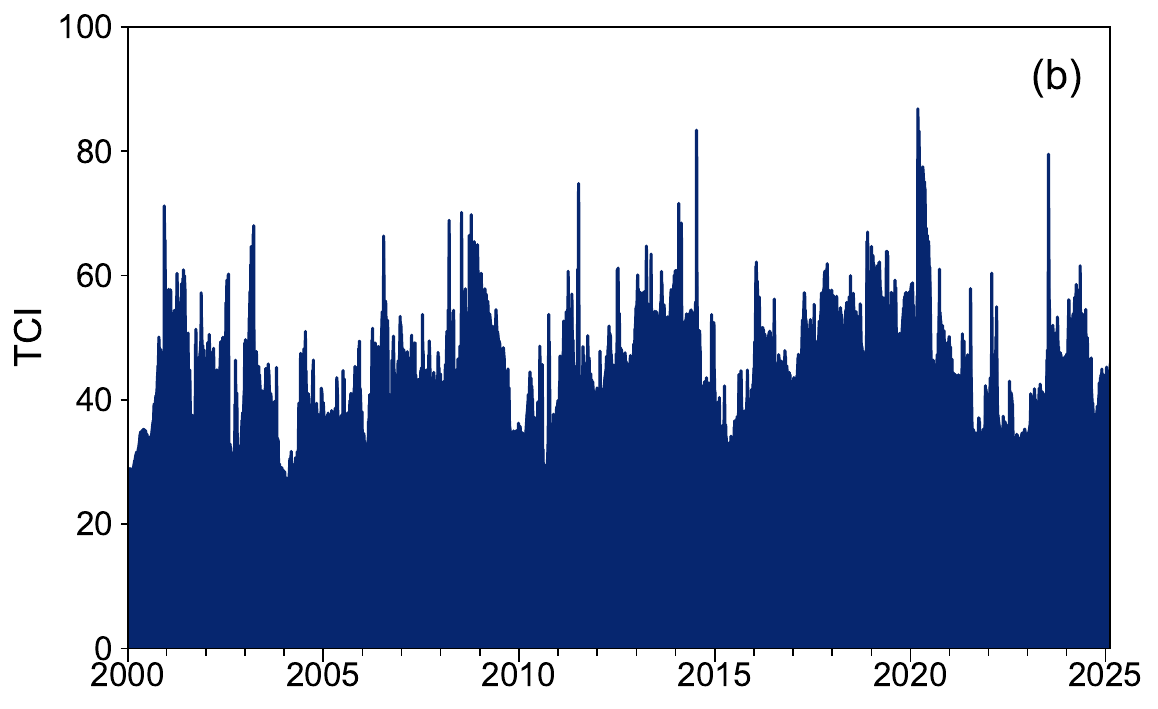}
  \includegraphics[width=0.45\linewidth]{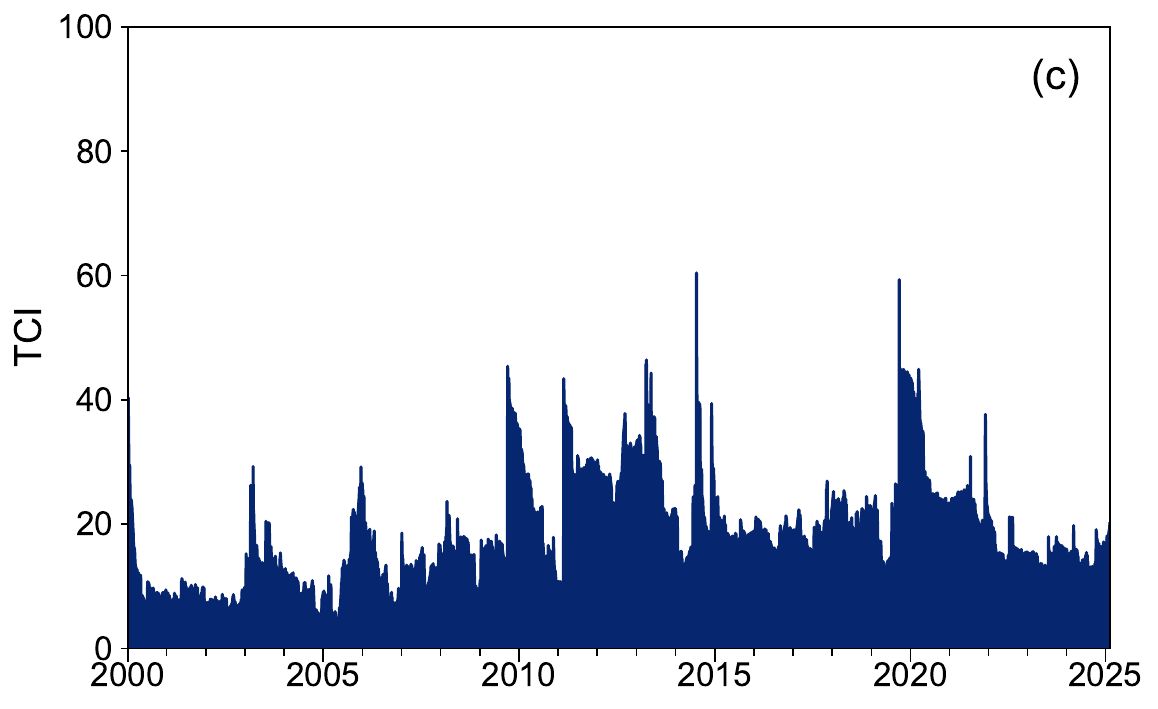}
  \includegraphics[width=0.45\linewidth]{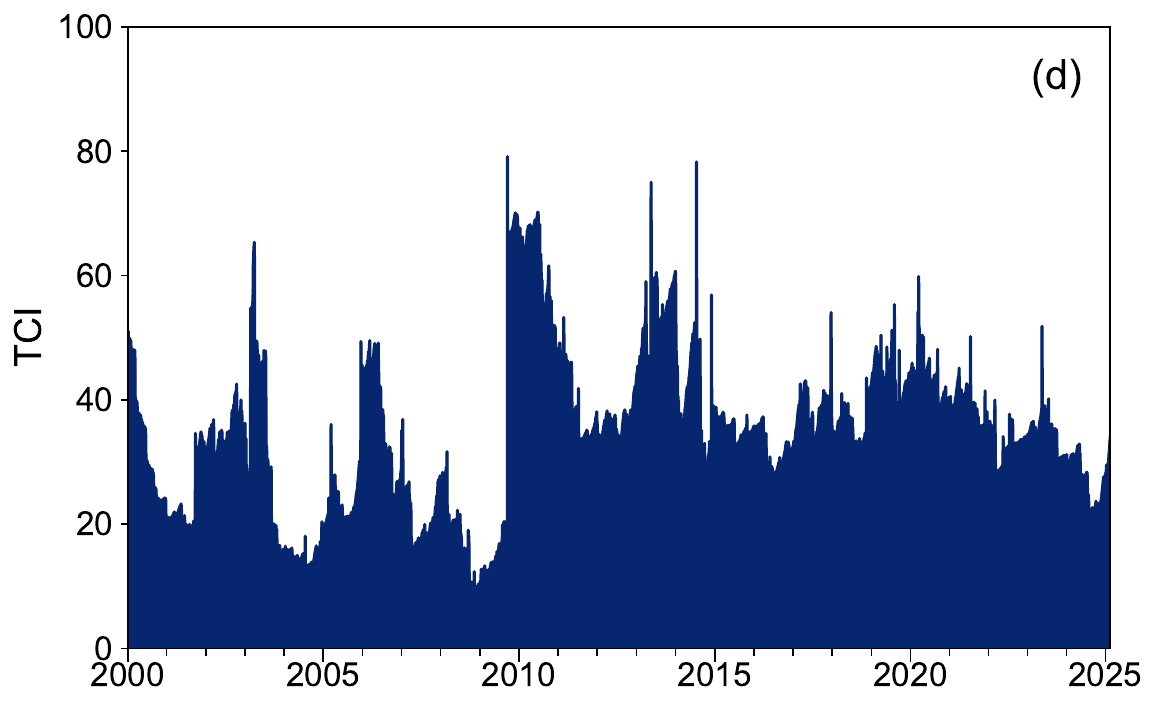}
  \caption{Dynamic total connectedness corresponding to return (a), volatility (b), skewness (c), and kurtosis (d) of staple food and energy markets.}
\label{Fig:Agro_Connectedness_time_TCI}
\end{figure}

Figure~\ref{Fig:Agro_Connectedness_time_TCI} illustrates the time-varying total connectedness indices for return (a), volatility (b), skewness (c), and kurtosis (d), offering a more intuitive view of the dynamic evolution of moment connectedness within the energy-food system. The return TCI consistently remains above 30\% and exhibits notable spikes during 2008–2009, 2020, and 2022. The global financial crisis in 2008–2009 triggered sharp fluctuations in both energy and food prices, amplifying systemic risk and thus increasing return connectedness. In early 2020, the outbreak of the COVID-19 pandemic led to a slowdown in global economic activity, a collapse in energy demand, and disruptions in food supply chains, which together intensified risk spillovers across markets. The Russia-Ukraine conflict that erupted in 2022 caused severe turmoil in global energy and food markets, with prices of crude oil, natural gas, wheat, and corn soaring, resulting in rising return connectedness.

The volatility TCI is markedly higher than the return TCI and fluctuates sharply, exceeding 80\% during crisis periods. In addition to the peaks in 2008, 2020, and 2022, the volatility TCI also shows phased spikes in 2001, 2011, 2014–2015, and 2023–2024. In 2001, the bursting of the dot-com bubble and the 9/11 terrorist attacks heightened global risk aversion, triggering shocks in the oil market and elevated volatility in food markets. The Arab Spring and the European debt crisis in 2011 pushed up energy prices, while extreme weather and protectionist trade policies fueled a food crisis. In 2014, OPEC's refusal to cut production led to a plunge in oil prices, and in 2015, the global economic slowdown coupled with China's exchange rate reform ushered in a commodity downturn. Frequent geopolitical conflicts during 2023–2024, including the Israel-Palestine conflict and tensions in the Middle East, along with the U.S. Federal Reserve's interest rate hikes, further exacerbated volatility spillovers in energy and food markets.

The skewness TCI remains relatively low overall but exhibits sharp increases during periods of market turmoil, indicating that return distributions become more asymmetric in the face of extreme events. Specifically, following the drastic decline and subsequent strong rebound during the 2008 financial crisis, skewness connectedness rose noticeably. In 2011, the Arab Spring, the European debt crisis, and the global food crisis heightened market risk aversion, leading to asymmetric shifts in return distributions of energy and food. The dramatic drop in oil prices during 2014–2015, followed by a modest recovery, also contributed to skewed return distributions in energy markets. In 2020, the COVID-19 pandemic caused a steep market downturn, which was rapidly followed by a strong rebound after the successful development of vaccines, resulting in substantial shifts in skewness spillovers. During the initial phase of the Russia-Ukraine conflict in 2022, the prices of wheat, corn, and crude oil soared before pulling back, further amplifying skewness connectedness.

The kurtosis TCI, by contrast, fluctuates dramatically and rises sharply in response to extreme event shocks. In addition to the peaks observed in 2008–2009, 2011, 2014–2015, 2020, and 2022–2023, kurtosis connectedness also rose abruptly around 2003, 2006, and 2018. The outbreak of the Iraq War in 2003 triggered significant turbulence in global energy markets, increasing the frequency of extreme returns and driving up kurtosis connectedness. In 2006, rising global oil demand and escalating geopolitical tensions in the Middle East pushed oil prices higher. Meanwhile, the introduction of the Renewable Fuel Standard significantly boosted ethanol demand, causing large price swings in corn and, through substitution effects, in wheat and soybeans. Concerns about future corn supply further fueled speculative activity, intensifying extreme risks in both food and energy markets. In 2018, the Sino-US trade war hit grain markets, while the continued interest rate hikes by the Federal Reserve, comprehensive sanctions on Iran, and uncertainty surrounding OPEC+ production policy led to energy market turmoil. As a result, tail risk spillovers between energy and food markets increased markedly. 

These findings underscore the profound impact that major crisis events exert on risk transmission within the energy-food system. In other words, the risk spillovers across these markets change significantly during turbulent periods. Therefore, policymakers and investors should closely monitor the dynamic evolution of moment connectedness to develop timely and effective risk management strategies.

\begin{figure}[!h]
  \centering
  \includegraphics[width=0.43\linewidth]{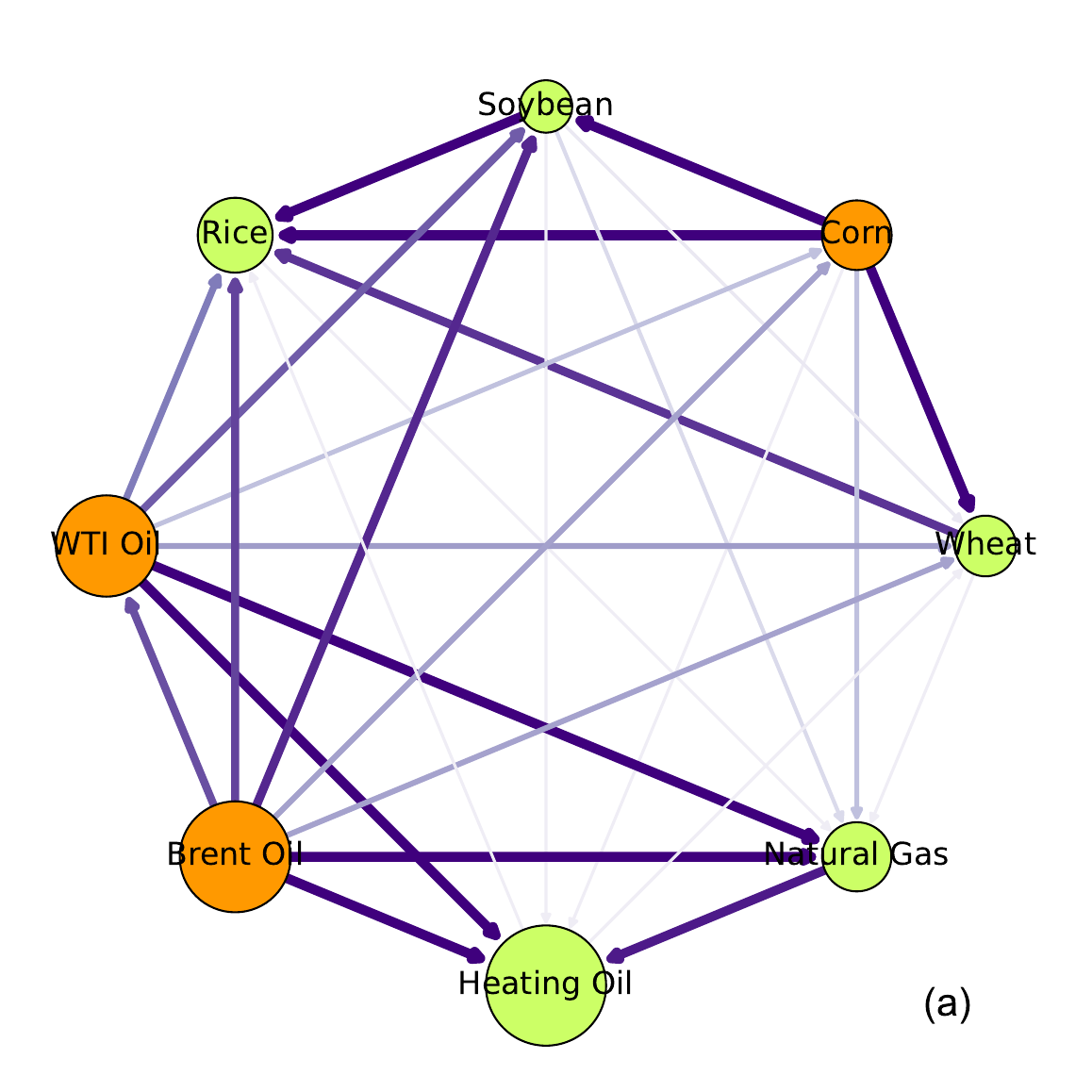}
  \includegraphics[width=0.43\linewidth]{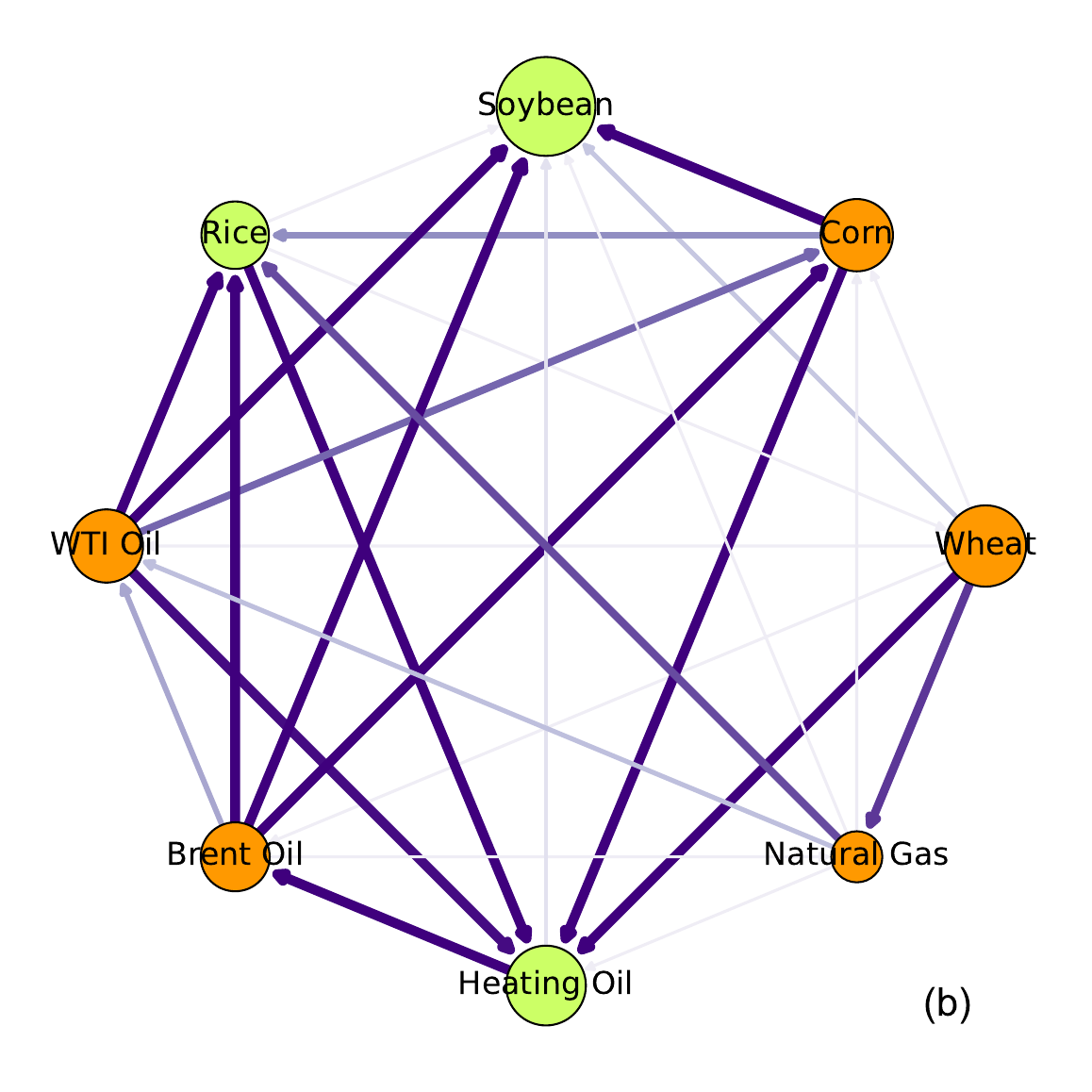}
  \includegraphics[width=0.43\linewidth]{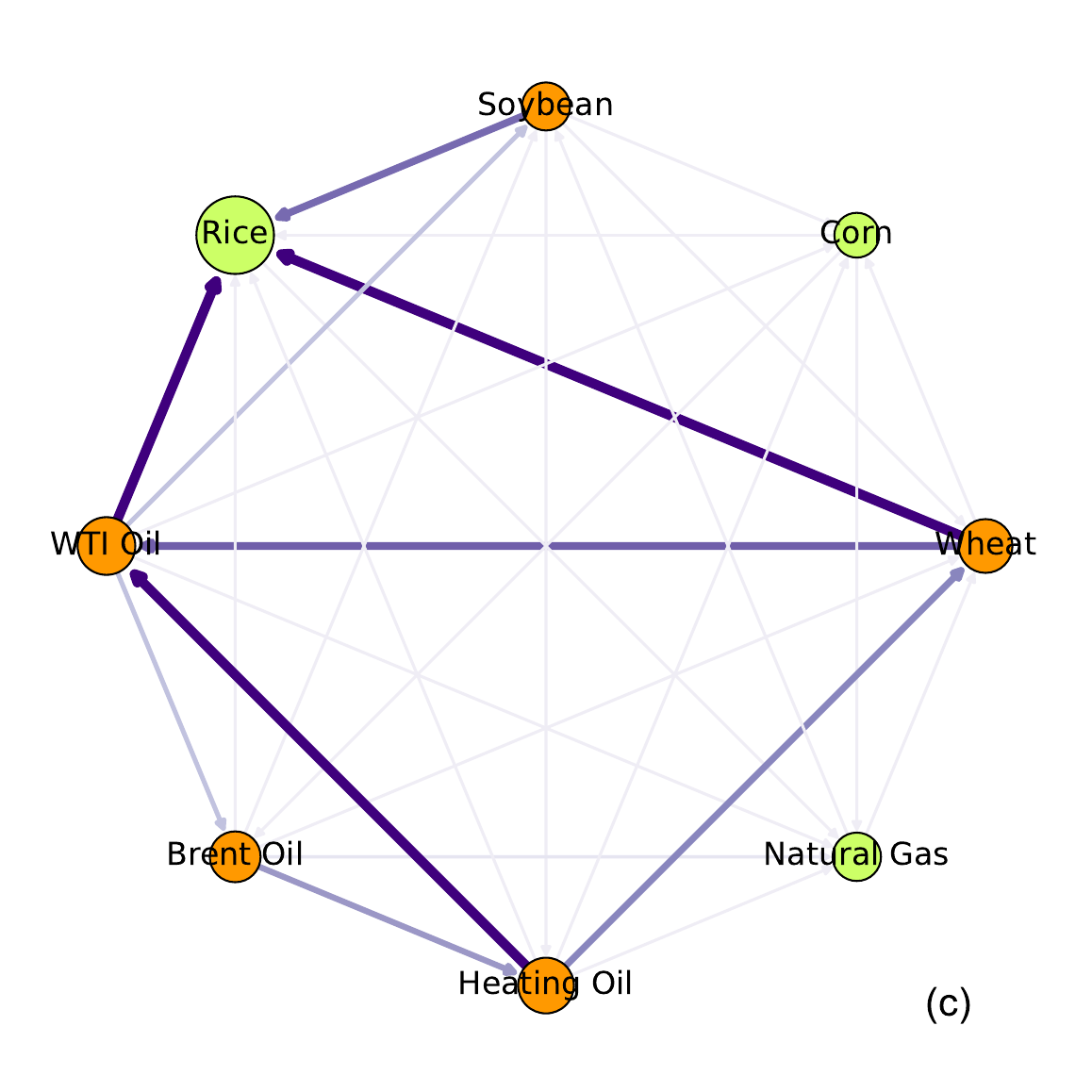}
  \includegraphics[width=0.43\linewidth]{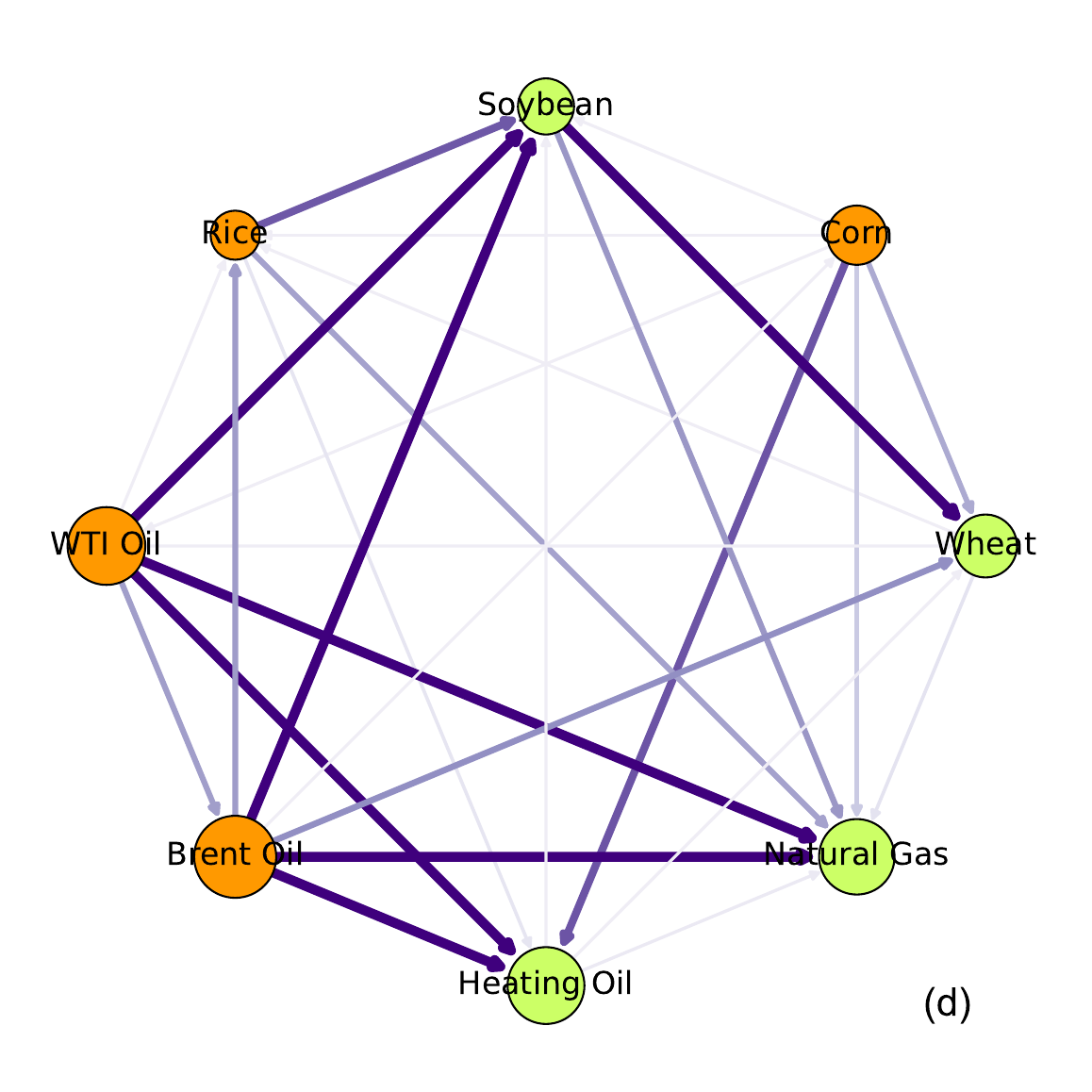}
  \caption{Net pairwise time-domain connectedness networks corresponding to return (a), volatility (b), skewness (c), and kurtosis (d) of staple food and energy markets. Node size and color indicate the out-strength and role of each node, where orange and yellow-green nodes refer to net transmitters and net receivers of spillovers. The direction of arrows shows the direction of net pairwise connectedness, and both the color gradient (from light to dark) and edge width (from thin to thick) correspond to the strength of net spillover effects, from weak to strong.} 
\label{Fig:Agro_Connectedness_time_Network}
\end{figure}

Net pairwise connectedness networks are further constructed to visualize the bilateral risk spillovers between individual energy and food markets in the time domain. The results for return, volatility, skewness, and kurtosis are presented in Figures~\ref{Fig:Agro_Connectedness_time_Network} (a–d), respectively. These are weighted directed networks, where edge weights represent the net pairwise directional connectedness indices. Node size and color indicate the out-strength of each node and its role within the network, respectively. Specifically, larger nodes reflect greater net connectedness, while orange and yellow-green nodes refer to net transmitters and net receivers of spillovers. In addition, the direction of the arrows shows the direction of net pairwise connectedness, and both the color gradient (from light to dark) and edge width (from thin to thick) correspond to the strength of net spillover effects, from weak to strong.

In terms of network structure, there exist significant differences in the characteristics of different moment connectedness. WTI and Brent oil emerge as the primary transmitters in the return connectedness network, further emphasizing the dominant role of crude oil within the energy-food system. Heating oil serves as the main receiver, experiencing the strongest spillovers from both Brent and WTI oil, which reflects its heavy dependence on the crude oil market. Other important return spillover pathways include WTI oil–natural gas, corn–wheat, Brent oil–natural gas, corn–soybean, soybean–rice, and corn–rice. These findings are consistent with Table~\ref{Tab:Average_time_connectedness}, highlighting that risk transmission is stronger within energy markets as well as within food markets than between energy and food markets.

In the volatility connectedness network, wheat, WTI oil, and corn act as major transmitters, indicating their greater influence on other markets in terms of volatility, while soybean and heating oil are the main receivers. Corn is a crucial input for ethanol production, and soybean oil is one of the major feedstocks for biodiesel, creating close links between these grains and energy. Moreover, fluctuations in oil prices may affect agricultural production costs through changes in fuel and fertilizer prices, thereby influencing food market volatility. Compared with return spillovers, cross-market volatility spillovers between energy and staple foods are significantly stronger, as reflected by the darker and thicker edges in the network, indicating closer volatility linkages between these markets. This finding is in line with the conclusions of \cite{Bouri-Lei-Jalkh-Xu-Zhang-2021-ResourPolicy} and \cite{Zhou-Wu-Liu-Rognone-2023-NatCommun}, which suggest that volatility spillovers tend to be more prominent in cross-market relationships.

In the skewness connectedness network, WTI and Brent oil remain the primary risk transmitters, which means that asymmetric risks in the crude oil market have substantial impacts on other markets. Rice acts as the main receiver of skewness spillovers, indicating that asymmetries in its return distribution are largely driven by external shocks, particularly from WTI oil and wheat. Similarly, Brent and WTI oil are also the dominant transmitters in the kurtosis connectedness network, implying that tail risks in the crude oil market, triggered by extreme events, can significantly affect other markets. Heating oil and natural gas are the main receivers of kurtosis risk, primarily absorbing spillovers from crude oil. Compared to energy markets, kurtosis connectedness is generally weaker among food markets, with only soybean exhibiting notable spillovers to wheat. In summary, the connectedness networks reveal heterogeneity in the interactions between energy and food markets, as well as in the roles of specific commodities, across diverse moments. These findings contribute to a deeper understanding of spillover pathways and transmission mechanisms of the energy-food nexus in the time domain.

\subsection{Frequency-domain risk connectedness}  

After examining the time-domain connectedness, we further apply the TVP-VAR-BK approach to investigate the frequency-domain connectedness between staple food and energy markets, which uncovers the risk transmission patterns at different timescales. Inspired by \cite{Barunik-Krehlik-2018-JFinancEconom} and \cite{Naeem-Qureshi-Farid-Tiwari-Elheddad-2024-AnnOperRes}, three frequency bands are considered, including the high frequency (short term) spanning from 1 day to 1 week, the medium frequency (medium term) from 1 week to 6 months, and the low frequency (long term) from 6 months to infinity. Tables~\ref{Tab:Average_frequency_connectedness_return}, \ref{Tab:Average_frequency_connectedness_volatility}, \ref{Tab:Average_frequency_connectedness_skewness}, and \ref{Tab:Average_frequency_connectedness_kurtosis} report the average frequency-domain connectedness measures for return, volatility, skewness, and kurtosis, respectively, where Panels A, B, and C correspond to the short-, medium-, and long-term connectedness.

\begin{table}[!ht]
  \centering
  \setlength{\abovecaptionskip}{0pt}
  \setlength{\belowcaptionskip}{10pt}
  \caption{High-, medium- and low-frequency return connectedness among staple food and energy markets}
  \setlength\tabcolsep{8pt}   \resizebox{\textwidth}{!}{ 
    \begin{tabular}{l r@{.}l r@{.}l r@{.}l r@{.}l r@{.}l r@{.}l r@{.}l r@{.}l r@{.}l}
    \toprule
    & \multicolumn{2}{c}{Wheat} & \multicolumn{2}{c}{Corn} & \multicolumn{2}{c}{Soybean} & \multicolumn{2}{c}{Rice} & \multicolumn{2}{c}{WTI Oil} & \multicolumn{2}{c}{Brent Oil} & \multicolumn{2}{c}{Heating Oil} & \multicolumn{2}{c}{Natural Gas} & \multicolumn{2}{c}{FROM}  \\
    \midrule
    \multicolumn{19}{l}{\textit{Panel A: High-frequency connectedness (1 day to 1 week)}} \\
    Wheat & 50&72 & 15&73 & 7&25 & 2&04 & 1&55 & 1&63 & 0&70 & 0&67 & 29&57 \\
    Corn & 14&64 & 45&81 & 12&92 & 1&74 & 1&81 & 2&04 & 0&82 & 0&66 & 34&63 \\
    Soybean & 7&27 & 14&14 & 50&15 & 2&20 & 2&72 & 2&94 & 1&39 & 0&67 & 31&35 \\
    Rice & 2&75 & 2&56 & 3&01 & 66&21 & 1&25 & 1&44 & 0&78 & 0&70 & 12&49 \\
    WTI Oil & 1&02 & 1&35 & 1&93 & 0&73 & 35&93 & 27&62 & 10&42 & 2&23 & 45&30 \\
    Brent Oil & 1&09 & 1&49 & 2&01 & 0&81 & 26&79 & 35&34 & 11&58 & 2&03 & 45&80 \\
    Heating Oil & 0&72 & 1&01 & 1&37 & 0&67 & 16&79 & 18&90 & 42&94 & 2&09 & 41&55 \\
    Natural Gas & 0&81 & 1&08 & 1&02 & 0&76 & 4&04 & 3&70 & 1&19 & 69&81 & 12&60 \\
    TO & 28&32 & 37&37 & 29&52 & 8&95 & 54&95 & 58&26 & 26&89 & 9&06 & \multicolumn{2}{c}{TCI} \\
    Net & $-$1&25 & 2&74 & $-$1&83 & $-$3&55 & 9&65 & 12&45 & $-$14&67 & $-$3&54 & 31&66 \vspace{2mm} \\
    \multicolumn{19}{l}{\textit{Panel B: Medium-frequency connectedness (1 week to 6 months)}} \\
    Wheat & 11&54 & 3&93 & 1&78 & 0&49 & 0&29 & 0&29 & 0&18 & 0&17 & 7&13 \\
    Corn & 3&27 & 10&76 & 3&05 & 0&39 & 0&32 & 0&34 & 0&18 & 0&20 & 7&76 \\
    Soybean & 1&45 & 3&08 & 11&03 & 0&46 & 0&48 & 0&53 & 0&32 & 0&18 & 6&49 \\
    Rice & 0&63 & 0&65 & 0&75 & 17&05 & 0&31 & 0&35 & 0&22 & 0&18 & 3&11 \\
    WTI Oil & 0&22 & 0&29 & 0&50 & 0&15 & 7&78 & 5&90 & 2&40 & 0&56 & 10&00 \\
    Brent Oil & 0&25 & 0&31 & 0&55 & 0&16 & 5&92 & 7&36 & 2&77 & 0&54 & 10&50 \\
    Heating Oil & 0&13 & 0&21 & 0&36 & 0&13 & 3&16 & 3&53 & 6&87 & 0&31 & 7&83 \\
    Natural Gas & 0&20 & 0&27 & 0&23 & 0&19 & 0&61 & 0&51 & 0&28 & 14&37 & 2&30 \\
    TO & 6&17 & 8&75 & 7&21 & 1&96 & 11&09 & 11&46 & 6&34 & 2&14 & \multicolumn{2}{c}{TCI} \\
    Net & $-$0&96 & 0&99 & 0&72 & $-$1&15 & 1&09 & 0&96 & $-$1&48 & $-$0&15 & 6&89 \vspace{2mm} \\
    \multicolumn{19}{l}{\textit{Panel C: Low-frequency connectedness (6 months to infinity)}} \\
    Wheat & 0&64 & 0&22 & 0&10 & 0&03 & 0&02 & 0&02 & 0&01 & 0&01 & 0&40 \\
    Corn & 0&18 & 0&60 & 0&17 & 0&02 & 0&02 & 0&02 & 0&01 & 0&01 & 0&43 \\
    Soybean & 0&08 & 0&17 & 0&61 & 0&03 & 0&03 & 0&03 & 0&02 & 0&01 & 0&36 \\
    Rice & 0&04 & 0&04 & 0&04 & 0&96 & 0&02 & 0&02 & 0&01 & 0&01 & 0&17 \\
    WTI Oil & 0&01 & 0&02 & 0&03 & 0&01 & 0&43 & 0&33 & 0&13 & 0&03 & 0&56 \\
    Brent Oil & 0&01 & 0&02 & 0&03 & 0&01 & 0&33 & 0&41 & 0&15 & 0&03 & 0&59 \\
    Heating Oil & 0&01 & 0&01 & 0&02 & 0&01 & 0&18 & 0&20 & 0&38 & 0&02 & 0&44 \\
    Natural Gas & 0&01 & 0&02 & 0&01 & 0&01 & 0&03 & 0&03 & 0&02 & 0&80 & 0&13 \\
    TO & 0&34 & 0&49 & 0&40 & 0&11 & 0&62 & 0&64 & 0&35 & 0&12 & \multicolumn{2}{c}{TCI} \\
    Net & $-$0&06 & 0&06 & 0&04 & $-$0&06 & 0&06 & 0&05 & $-$0&08 & $-$0&01 & 0&38 \\
  \bottomrule
    \end{tabular}
    }%
  \begin{flushleft}
    \footnotesize
    \justifying Note: This table reports the average frequency-domain measures of return connectedness between staple food and energy markets, where Panels A, B, and C correspond to the high-, medium- and low-frequency connectedness, respectively.
  \end{flushleft}
  \label{Tab:Average_frequency_connectedness_return}%
\end{table}%

\begin{table}[!ht]
  \centering
  \setlength{\abovecaptionskip}{0pt}
  \setlength{\belowcaptionskip}{10pt}
  \caption{High-, medium- and low-frequency volatility connectedness among staple food and energy markets}
  \setlength\tabcolsep{8pt}   \resizebox{\textwidth}{!}{ 
    \begin{tabular}{l r@{.}l r@{.}l r@{.}l r@{.}l r@{.}l r@{.}l r@{.}l r@{.}l r@{.}l}
    \toprule
    & \multicolumn{2}{c}{Wheat} & \multicolumn{2}{c}{Corn} & \multicolumn{2}{c}{Soybean} & \multicolumn{2}{c}{Rice} & \multicolumn{2}{c}{WTI Oil} & \multicolumn{2}{c}{Brent Oil} & \multicolumn{2}{c}{Heating Oil} & \multicolumn{2}{c}{Natural Gas} & \multicolumn{2}{c}{FROM}  \\
    \midrule
    \multicolumn{19}{l}{\textit{Panel A: High-frequency connectedness (1 day to 1 week)}} \\
    Wheat & 4&67 & 0&68 & 0&22 & 0&05 & 0&07 & 0&11 & 0&10 & 0&06 & 1&30 \\
    Corn & 0&81 & 5&46 & 0&60 & 0&17 & 0&06 & 0&08 & 0&06 & 0&05 & 1&83 \\
    Soybean & 0&25 & 0&62 & 6&02 & 0&13 & 0&07 & 0&09 & 0&16 & 0&08 & 1&41 \\
    Rice & 0&11 & 0&27 & 0&20 & 9&14 & 0&16 & 0&12 & 0&17 & 0&10 & 1&13 \\
    WTI Oil & 0&06 & 0&05 & 0&04 & 0&08 & 4&72 & 3&00 & 0&68 & 0&05 & 3&94 \\
    Brent Oil & 0&08 & 0&08 & 0&04 & 0&06 & 2&73 & 4&31 & 1&04 & 0&06 & 4&09 \\
    Heating Oil & 0&06 & 0&04 & 0&07 & 0&07 & 0&36 & 0&60 & 2&83 & 0&06 & 1&26 \\
    Natural Gas & 0&04 & 0&03 & 0&03 & 0&04 & 0&06 & 0&08 & 0&08 & 6&79 & 0&37 \\
    TO & 1&42 & 1&78 & 1&20 & 0&60 & 3&52 & 4&08 & 2&28 & 0&45 & \multicolumn{2}{c}{TCI} \\
    Net & 0&12 & $-$0&05 & $-$0&20 & $-$0&52 & $-$0&43 & $-$0&01 & 1&02 & 0&08 & 1&92 \vspace{2mm} \\
    \multicolumn{19}{l}{\textit{Panel B: Medium-frequency connectedness (1 week to 6 months)}} \\ 
    Wheat & 28&90 & 5&50 & 2&13 & 2&19 & 1&11 & 1&27 & 1&50 & 1&26 & 14&96 \\
    Corn & 5&55 & 32&01 & 3&76 & 2&21 & 1&06 & 1&17 & 1&15 & 1&30 & 16&18 \\
    Soybean & 2&44 & 5&64 & 33&85 & 1&52 & 1&15 & 1&19 & 1&13 & 0&87 & 13&93 \\
    Rice & 1&50 & 2&61 & 1&75 & 45&73 & 1&19 & 1&02 & 1&09 & 1&03 & 10&19 \\
    WTI Oil & 1&10 & 0&79 & 0&73 & 0&75 & 22&54 & 15&09 & 4&27 & 0&95 & 23&69 \\
    Brent Oil & 1&31 & 0&86 & 0&69 & 0&84 & 14&30 & 21&68 & 6&00 & 1&09 & 25&09 \\
    Heating Oil & 2&27 & 1&30 & 1&01 & 1&05 & 3&75 & 4&55 & 18&85 & 1&33 & 15&27 \\
    Natural Gas & 1&15 & 0&98 & 0&98 & 0&86 & 1&59 & 1&82 & 1&58 & 40&05 & 8&95 \\
    TO & 15&32 & 17&68 & 11&05 & 9&41 & 24&15 & 26&10 & 16&71 & 7&82 & \multicolumn{2}{c}{TCI} \\
    Net & 0&37 & 1&50 & $-$2&87 & $-$0&78 & 0&46 & 1&01 & 1&44 & $-$1&13 & 16&03 \vspace{2mm} \\
    \multicolumn{19}{l}{\textit{Panel C: Low-frequency connectedness (6 months to infinity)}} \\ 
    Wheat & 22&79 & 6&45 & 3&72 & 2&63 & 3&69 & 3&83 & 4&52 & 2&55 & 27&38 \\
    Corn & 6&46 & 19&80 & 4&03 & 2&20 & 2&90 & 3&11 & 3&46 & 2&55 & 24&71 \\
    Soybean & 3&84 & 6&90 & 19&40 & 2&33 & 3&31 & 3&57 & 3&35 & 2&10 & 25&40 \\
    Rice & 3&04 & 2&34 & 1&92 & 16&13 & 2&80 & 2&58 & 2&46 & 2&55 & 17&69 \\
    WTI Oil & 3&89 & 2&43 & 1&71 & 1&42 & 14&88 & 11&55 & 6&01 & 3&22 & 30&23 \\
    Brent Oil & 3&87 & 2&28 & 1&70 & 1&44 & 12&04 & 14&03 & 6&38 & 3&09 & 30&80 \\
    Heating Oil & 8&34 & 4&57 & 3&18 & 3&60 & 7&81 & 6&40 & 23&29 & 4&58 & 38&49 \\
    Natural Gas & 3&54 & 2&58 & 2&02 & 1&96 & 2&07 & 2&42 & 4&12 & 25&13 & 18&71 \\
    TO & 32&98 & 27&54 & 18&28 & 15&58 & 34&63 & 33&46 & 30&30 & 20&64 & \multicolumn{2}{c}{TCI} \\
    Net & 5&60 & 2&83 & $-$7&12 & $-$2&11 & 4&40 & 2&66 & $-$8&19 & 1&94 & 26&68 \\
  \bottomrule
    \end{tabular}
    }%
  \begin{flushleft}
    \footnotesize
    \justifying Note: This table reports the average frequency-domain measures of volatility connectedness between staple food and energy markets, where Panels A, B, and C correspond to the high-, medium- and low-frequency connectedness, respectively.
  \end{flushleft}
  \label{Tab:Average_frequency_connectedness_volatility}%
\end{table}%

\begin{table}[!ht]
  \centering
  \setlength{\abovecaptionskip}{0pt}
  \setlength{\belowcaptionskip}{10pt}
  \caption{High-, medium- and low-frequency skewness connectedness among staple food and energy markets}
  \setlength\tabcolsep{8pt}   \resizebox{\textwidth}{!}{ 
    \begin{tabular}{l r@{.}l r@{.}l r@{.}l r@{.}l r@{.}l r@{.}l r@{.}l r@{.}l r@{.}l}
    \toprule
    & \multicolumn{2}{c}{Wheat} & \multicolumn{2}{c}{Corn} & \multicolumn{2}{c}{Soybean} & \multicolumn{2}{c}{Rice} & \multicolumn{2}{c}{WTI Oil} & \multicolumn{2}{c}{Brent Oil} & \multicolumn{2}{c}{Heating Oil} & \multicolumn{2}{c}{Natural Gas} & \multicolumn{2}{c}{FROM}  \\
    \midrule
    \multicolumn{19}{l}{\textit{Panel A: High-frequency connectedness (1 day to 1 week)}} \\
    Wheat & 9&29 & 0&45 & 0&43 & 0&22 & 0&11 & 0&07 & 0&04 & 0&08 & 1&40 \\
    Corn & 3&52 & 75&88 & 5&40 & 2&00 & 0&18 & 0&50 & 0&22 & 1&03 & 12&84 \\
    Soybean & 2&68 & 4&53 & 64&35 & 1&79 & 0&36 & 0&39 & 0&34 & 0&36 & 10&45 \\
    Rice & 0&34 & 0&61 & 0&80 & 23&88 & 0&14 & 0&35 & 0&14 & 0&19 & 2&57 \\
    WTI Oil & 0&04 & 0&01 & 0&03 & 0&05 & 4&74 & 0&14 & 0&28 & 0&01 & 0&58 \\
    Brent Oil & 0&31 & 0&71 & 0&27 & 0&80 & 2&77 & 64&67 & 13&50 & 0&98 & 19&35 \\
    Heating Oil & 0&23 & 0&23 & 0&53 & 0&50 & 4&77 & 13&09 & 58&33 & 0&58 & 19&93 \\
    Natural Gas & 0&72 & 1&23 & 0&40 & 0&77 & 0&46 & 1&48 & 0&76 & 93&66 & 5&83 \\
    TO & 7&84 & 7&77 & 7&86 & 6&13 & 8&78 & 16&02 & 15&29 & 3&24 & \multicolumn{2}{c}{TCI} \\
    Net & 6&44 & $-$5&07 & $-$2&59 & 3&56 & 8&21 & $-$3&33 & $-$4&64 & $-$2&58 & 9&12 \vspace{2mm} \\
    \multicolumn{19}{l}{\textit{Panel B: Medium-frequency connectedness (1 week to 6 months)}} \\ 
    Wheat & 53&88 & 2&82 & 2&68 & 1&29 & 1&41 & 0&45 & 0&46 & 0&45 & 9&57 \\
    Corn & 0&70 & 7&78 & 0&71 & 0&33 & 0&12 & 0&07 & 0&04 & 0&09 & 2&06 \\
    Soybean & 0&94 & 1&43 & 18&87 & 0&77 & 0&33 & 0&14 & 0&14 & 0&10 & 3&86 \\
    Rice & 1&86 & 1&64 & 2&20 & 52&95 & 1&04 & 0&80 & 0&49 & 0&45 & 8&47 \\
    WTI Oil & 2&16 & 0&12 & 0&28 & 0&54 & 29&94 & 0&99 & 3&03 & 0&13 & 7&24 \\
    Brent Oil & 0&20 & 0&14 & 0&10 & 0&27 & 0&55 & 10&59 & 2&63 & 0&15 & 4&03 \\
    Heating Oil & 0&25 & 0&07 & 0&20 & 0&29 & 1&45 & 3&59 & 14&08 & 0&12 & 5&97 \\
    Natural Gas & 0&01 & 0&01 & 0&01 & 0&01 & 0&00 & 0&00 & 0&00 & 0&42 & 0&04 \\
    TO & 6&11 & 6&23 & 6&18 & 3&50 & 4&91 & 6&04 & 6&79 & 1&49 & \multicolumn{2}{c}{TCI} \\
    Net & $-$3&47 & 4&17 & 2&32 & $-$4&97 & $-$2&34 & 2&01 & 0&82 & 1&45 & 5&16 \vspace{2mm} \\
    \multicolumn{19}{l}{\textit{Panel C: Low-frequency connectedness (6 months to infinity)}} \\ 
    Wheat & 17&85 & 0&93 & 0&94 & 0&49 & 4&37 & 0&33 & 0&74 & 0&21 & 8&01 \\
    Corn & 0&29 & 0&50 & 0&07 & 0&05 & 0&45 & 0&03 & 0&05 & 0&01 & 0&94 \\
    Soybean & 0&24 & 0&11 & 1&18 & 0&09 & 0&71 & 0&04 & 0&08 & 0&01 & 1&30 \\
    Rice & 1&11 & 0&29 & 0&40 & 6&37 & 3&35 & 0&22 & 0&33 & 0&07 & 5&76 \\
    WTI Oil & 4&45 & 0&36 & 0&61 & 1&13 & 43&70 & 2&15 & 4&80 & 0&31 & 13&81 \\
    Brent Oil & 0&10 & 0&03 & 0&02 & 0&07 & 0&44 & 0&55 & 0&15 & 0&01 & 0&82 \\
    Heating Oil & 0&08 & 0&01 & 0&02 & 0&04 & 0&43 & 0&21 & 0&88 & 0&01 & 0&80 \\
    Natural Gas & 0&00 & 0&00 & 0&00 & 0&00 & 0&02 & 0&00 & 0&00 & 0&02 & 0&03 \\
    TO & 6&28 & 1&73 & 2&05 & 1&87 & 9&77 & 2&98 & 6&15 & 0&62 & \multicolumn{2}{c}{TCI} \\
    Net & $-$1&73 & 0&80 & 0&76 & $-$3&90 & $-$4&04 & 2&17 & 5&35 & 0&60 & 3&93 \\
  \bottomrule
    \end{tabular}
    }%
  \begin{flushleft}
    \footnotesize
    \justifying Note: This table reports the average frequency-domain measures of skewness connectedness between staple food and energy markets, where Panels A, B, and C correspond to the high-, medium- and low-frequency connectedness, respectively.
  \end{flushleft}
  \label{Tab:Average_frequency_connectedness_skewness}%
\end{table}%

\begin{table}[!ht]
  \centering
  \setlength{\abovecaptionskip}{0pt}
  \setlength{\belowcaptionskip}{10pt}
  \caption{High-, medium- and low-frequency kurtosis connectedness among staple food and energy markets}
  \setlength\tabcolsep{8pt}   \resizebox{\textwidth}{!}{ 
    \begin{tabular}{l r@{.}l r@{.}l r@{.}l r@{.}l r@{.}l r@{.}l r@{.}l r@{.}l r@{.}l}
    \toprule
    & \multicolumn{2}{c}{Wheat} & \multicolumn{2}{c}{Corn} & \multicolumn{2}{c}{Soybean} & \multicolumn{2}{c}{Rice} & \multicolumn{2}{c}{WTI Oil} & \multicolumn{2}{c}{Brent Oil} & \multicolumn{2}{c}{Heating Oil} & \multicolumn{2}{c}{Natural Gas} & \multicolumn{2}{c}{FROM}  \\
    \midrule
    \multicolumn{19}{l}{\textit{Panel A: High-frequency connectedness (1 day to 1 week)}} \\
    Wheat & 8&66 & 0&57 & 0&89 & 0&30 & 0&11 & 0&21 & 0&11 & 0&06 & 2&25 \\
    Corn & 2&31 & 37&24 & 2&17 & 1&23 & 0&47 & 0&85 & 0&53 & 0&45 & 8&01 \\
    Soybean & 1&13 & 0&86 & 14&05 & 0&67 & 0&19 & 0&23 & 0&16 & 0&04 & 3&29 \\
    Rice & 0&42 & 0&48 & 0&53 & 20&02 & 0&27 & 0&29 & 0&24 & 0&07 & 2&29 \\
    WTI Oil & 0&41 & 0&44 & 0&29 & 0&64 & 28&34 & 16&05 & 8&57 & 0&25 & 26&63 \\
    Brent Oil & 0&78 & 1&04 & 0&37 & 0&77 & 20&39 & 36&37 & 12&04 & 0&51 & 35&89 \\
    Heating Oil & 0&54 & 0&87 & 0&39 & 1&18 & 11&37 & 13&59 & 41&92 & 0&61 & 28&54 \\
    Natural Gas & 0&61 & 0&78 & 0&17 & 0&28 & 0&93 & 0&93 & 0&64 & 70&48 & 4&34 \\
    TO & 6&21 & 5&03 & 4&81 & 5&07 & 33&73 & 32&15 & 22&28 & 1&97 & \multicolumn{2}{c}{TCI} \\
    Net & 3&96 & $-$2&98 & 1&52 & 2&78 & 7&09 & $-$3&73 & $-$6&26 & $-$2&37 & 13&91 \vspace{2mm} \\
    \multicolumn{19}{l}{\textit{Panel B: Medium-frequency connectedness (1 week to 6 months)}} \\ 
    Wheat & 46&23 & 5&40 & 5&25 & 2&74 & 1&60 & 1&62 & 1&03 & 0&67 & 18&32 \\
    Corn & 3&88 & 29&65 & 4&42 & 1&84 & 1&00 & 1&17 & 0&76 & 0&82 & 13&89 \\
    Soybean & 4&37 & 5&95 & 44&54 & 3&40 & 1&78 & 1&77 & 1&24 & 0&36 & 18&86 \\
    Rice & 2&63 & 2&70 & 3&05 & 48&11 & 1&99 & 1&74 & 1&41 & 0&40 & 13&93 \\
    WTI Oil & 1&25 & 1&24 & 0&99 & 1&58 & 17&09 & 10&17 & 5&45 & 0&39 & 21&06 \\
    Brent Oil & 0&79 & 0&88 & 0&53 & 0&93 & 6&73 & 9&94 & 3&66 & 0&30 & 13&82 \\
    Heating Oil & 0&66 & 1&08 & 0&77 & 0&89 & 4&57 & 4&77 & 12&67 & 0&34 & 13&08 \\
    Natural Gas & 0&60 & 0&89 & 0&55 & 0&61 & 0&99 & 0&76 & 0&55 & 16&82 & 4&94 \\
    TO & 14&17 & 18&15 & 15&56 & 11&98 & 18&66 & 22&00 & 14&09 & 3&29 & \multicolumn{2}{c}{TCI} \\
    Net & $-$4&15 & 4&26 & $-$3&30 & $-$1&95 & $-$2&40 & 8&18 & 1&01 & $-$1&65 & 14&74 \vspace{2mm} \\
    \multicolumn{19}{l}{\textit{Panel C: Low-frequency connectedness (6 months to infinity)}} \\ 
    Wheat & 14&76 & 2&62 & 2&92 & 1&41 & 1&02 & 0&84 & 0&52 & 0&46 & 9&78 \\
    Corn & 1&83 & 4&58 & 2&56 & 1&03 & 0&46 & 0&33 & 0&19 & 0&23 & 6&63 \\
    Soybean & 2&02 & 2&53 & 10&56 & 2&08 & 0&77 & 0&64 & 0&46 & 0&20 & 8&69 \\
    Rice & 1&55 & 1&00 & 1&79 & 8&90 & 1&03 & 0&72 & 0&48 & 0&20 & 6&76 \\
    WTI Oil & 0&85 & 0&43 & 0&47 & 0&82 & 2&15 & 1&34 & 0&66 & 0&15 & 4&72 \\
    Brent Oil & 0&46 & 0&22 & 0&24 & 0&45 & 1&04 & 1&07 & 0&41 & 0&10 & 2&91 \\
    Heating Oil & 0&30 & 0&32 & 0&44 & 0&40 & 0&66 & 0&56 & 1&00 & 0&09 & 2&78 \\
    Natural Gas & 0&33 & 0&28 & 0&48 & 0&36 & 0&39 & 0&26 & 0&18 & 1&14 & 2&29 \\
    TO & 7&34 & 7&40 & 8&90 & 6&56 & 5&36 & 4&69 & 2&88 & 1&43 & \multicolumn{2}{c}{TCI} \\
    Net & $-$2&43 & 0&76 & 0&21 & $-$0&20 & 0&64 & 1&78 & 0&10 & $-$0&86 & 5&57 \\
  \bottomrule
    \end{tabular}
    }%
  \begin{flushleft}
    \footnotesize
    \justifying Note: This table reports the average frequency-domain measures of kurtosis connectedness between staple food and energy markets, where Panels A, B, and C correspond to the high-, medium- and low-frequency connectedness, respectively.
  \end{flushleft}
  \label{Tab:Average_frequency_connectedness_kurtosis}%
\end{table}%

As shown in Table~\ref{Tab:Average_frequency_connectedness_return}, the TCI of high-frequency connectedness reaches 31.66\%, significantly higher than that of medium-frequency (6.89\%) and low-frequency (0.38\%) connectedness. This suggests that short-term spillovers play an important role in return connectedness among the energy and food markets. The prominence of short-term return spillovers may be attributed to active short-run trading in futures markets, the rapid transmission of high-frequency information, and the large trading volumes and high liquidity of short-term contracts. Additionally, market sentiment fluctuates more sharply in the short term, and speculative activities are more frequent, further amplifying short-term return spillovers. Over longer time horizons, however, market prices gradually shift to be more influenced by fundamental factors such as macroeconomic conditions, policy adjustments, and structural changes, thereby reducing return spillovers in the medium and long term. Moreover, the high-frequency connectedness exhibits a modular structure similar to that observed in the time-domain analysis, where return spillovers are notably stronger within energy markets and within food markets than across them. Brent and WTI oil remain the primary transmitters, while heating oil continues to be the main receiver.

Based on Table~\ref{Tab:Average_frequency_connectedness_volatility}, long-term volatility spillovers are found to be dominant, with a corresponding TCI of 26.68\%, exceeding those of the medium-term (16.03\%) and short-term (1.92\%) connectedness. This pattern contrasts with the short-term dominance observed in return connectedness and aligns with the findings of \cite{He-Hamori-2021-JIntMoneyFinan} and \cite{Zhang-He-Hamori-2023-IntRevFinancAnal}. Volatility in futures markets often exhibits strong clustering, where periods of high or low volatility tend to persist. As a result, volatility spillovers are more pronounced over longer timescales. Furthermore, unlike return spillovers, which are largely influenced by short-term market sentiment and speculative behavior, volatility spillovers are more susceptible to macroeconomic conditions and policy changes, whose effects unfold over extended periods. Consequently, volatility tends to propagate more persistently across markets than returns, leading to stronger long-term volatility spillover effects. These findings highlight the critical importance of monitoring long-term volatility connectedness when designing risk management strategies and policy interventions aimed at mitigating the accumulation of systemic risk.

In Table~\ref{Tab:Average_frequency_connectedness_skewness}, the high-frequency skewness connectedness has the highest TCI (9.12\%), followed by the medium-frequency (5.16\%) and low-frequency (3.93\%) connectedness. This pattern suggests that the spillover of asymmetries in extreme returns between energy and food markets weakens over time. Due to heightened trading activity and stronger reactions to unexpected events, short-term skewness spillovers are more pronounced. This result is similar to the frequency-domain pattern of return connectedness, although the overall magnitude of skewness spillovers is smaller. Moreover, it is worth noting that the high-frequency FROM indices for wheat (1.40\%) and rice (2.57\%) are lower than their medium-frequency (9.57\% and 8.47\%) and low-frequency (8.01\% and 5.76\%) counterparts. Similarly, the high-frequency FROM index for WTI oil is 0.58\%, much lower than its medium-frequency (7.24\%) and low-frequency (13.81\%) indices. These results indicate that wheat, rice, and WTI oil are less exposed to external skewness spillovers in the short term but become more vulnerable over medium and long horizons, which differs from other food and energy markets.

According to Table~\ref{Tab:Average_frequency_connectedness_kurtosis}, the TCI indices for high- and medium-frequency kurtosis connectedness are 13.91\% and 14.74\%, respectively, while the low-frequency TCI index is considerably lower at 5.57\%. This implies that extreme risk spillovers within the energy-food system are more prominent in the short and medium term, whereas long-term spillovers are relatively weaker. Additionally, the high-frequency FROM and TO indices for WTI oil (26.63\% and 33.73\%), Brent oil (35.89\% and 32.15\%), and heating oil (28.54\% and 22.28\%) are substantially higher than those of other markets. This suggests that, in the short run, these energy markets act as both major receivers and transmitters of tail risk spillovers. In contrast, the grain markets display much lower high-frequency FROM and TO indices, indicating that short-term extreme risk spillovers are primarily driven by energy markets. However, the medium-frequency FROM and TO indices for wheat (18.32\% and 14.17\%), corn (13.89\% and 18.15\%), soybean (18.86\% and 15.56\%), and rice (13.93\% and 11.98\%) are notably higher than those at high frequency, which means that food markets become more susceptible to external tail risk shocks in the medium term and begin to transmit kurtosis risk to other markets. In the long term, the tail risk spillovers from food markets remain relatively strong, while those from energy markets diminish, highlighting a shift of extreme risk transmission over time.

\begin{figure}[!h]
  \centering
  \includegraphics[width=0.45\linewidth]{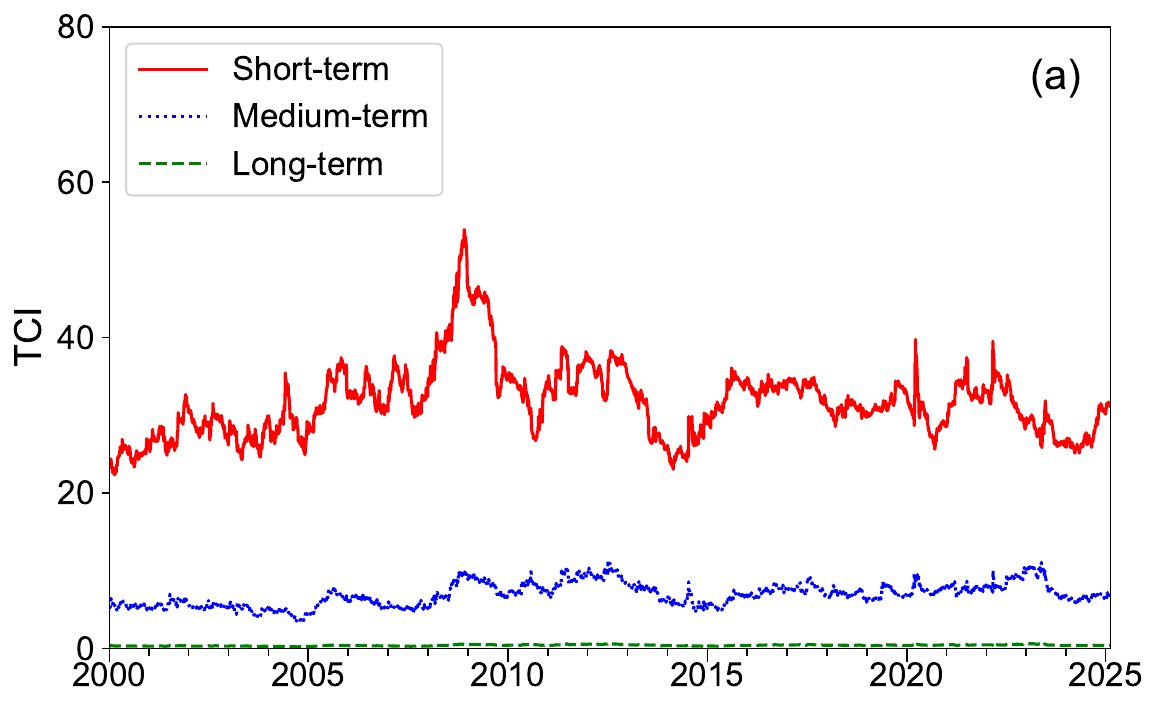}
  \includegraphics[width=0.45\linewidth]{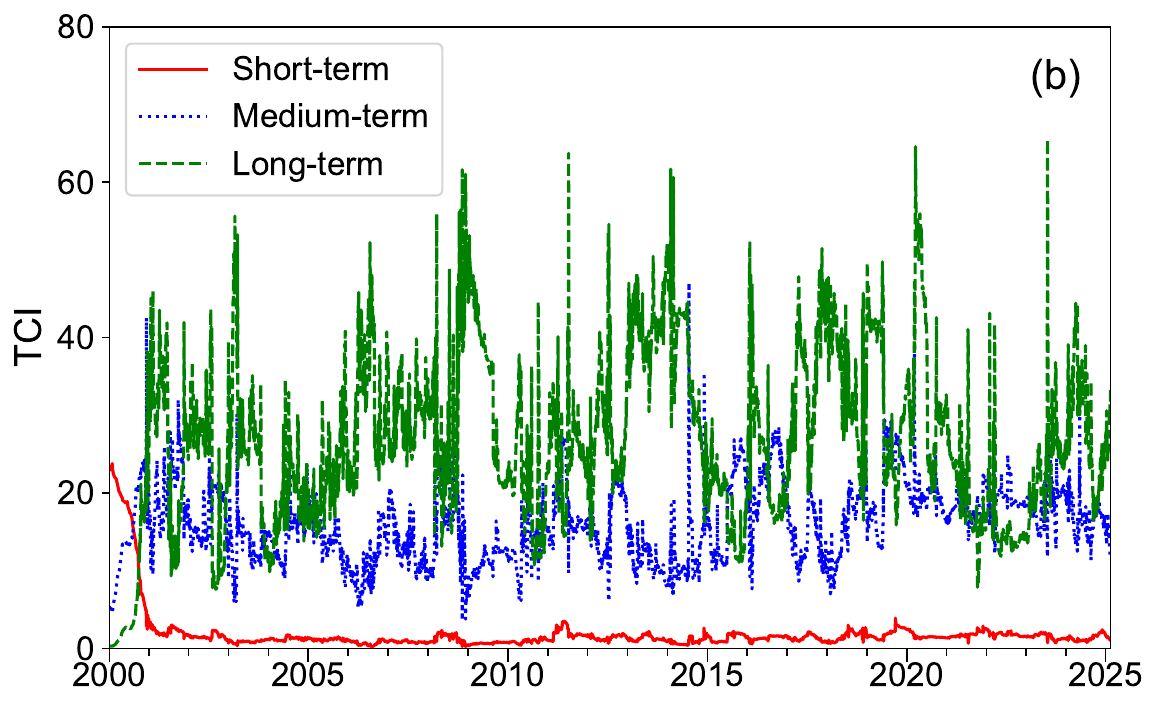}
  \includegraphics[width=0.45\linewidth]{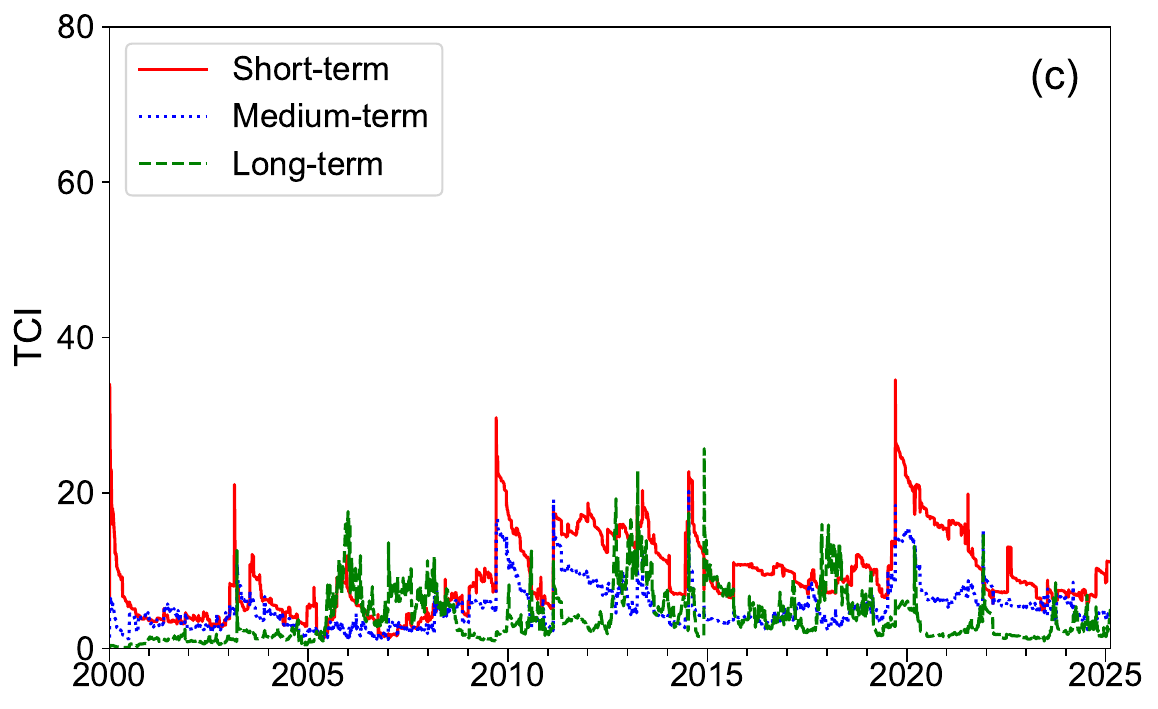}
  \includegraphics[width=0.45\linewidth]{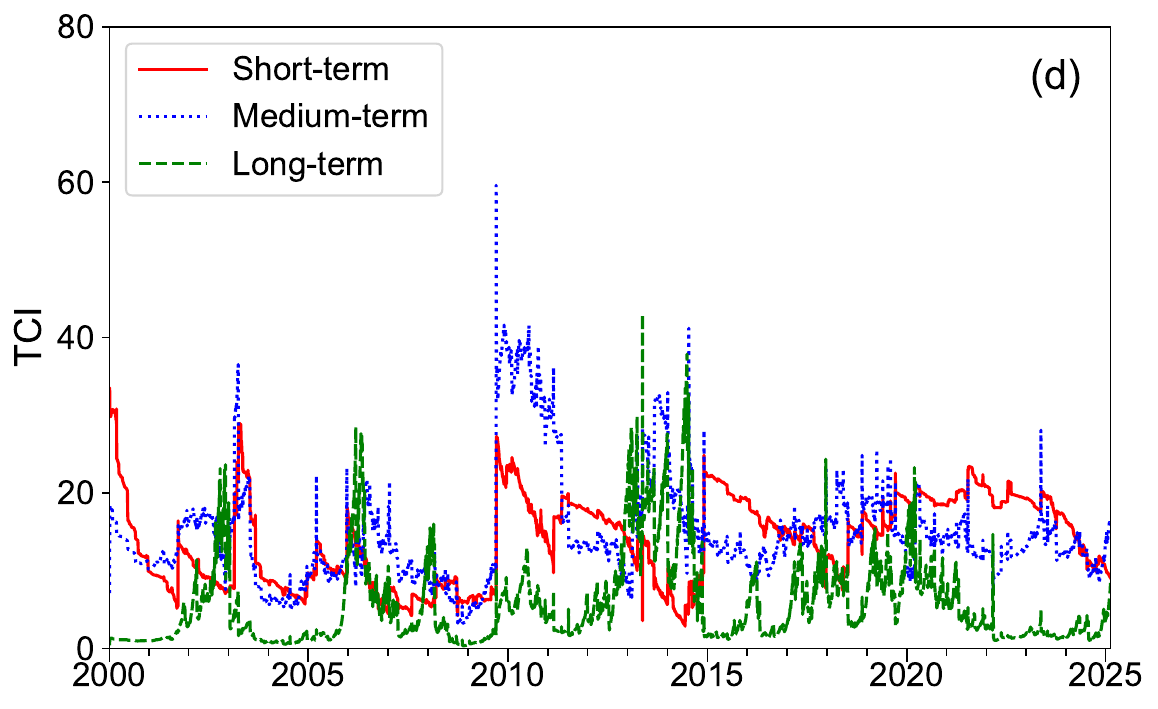}
  \caption{Short-, medium-, and long-term dynamic total connectedness corresponding to return (a), volatility (b), skewness (c), and kurtosis (d) of staple food and energy markets.}
\label{Fig:Agro_Connectedness_frequency_TCI}
\end{figure}

Figure~\ref{Fig:Agro_Connectedness_frequency_TCI} further illustrates the dynamic evolution of the total connectedness indices for short-, medium-, and long-term connectedness between energy and food markets, where the subfigures (a–d) correspond to return, volatility, skewness, and kurtosis, respectively. Overall, the moment connectedness at different timescales exhibits significant time variation and crisis sensitivity, and the frequency-domain connectedness varies notably across diverse moments.

For return connectedness, the short-term TCI is much higher than the medium- and long-term TCIs, with the long-term TCI being particularly low. This indicates that short-term return spillovers are most pronounced, while long-term spillovers are minimal. In other words, the frequency-domain return connectedness between energy and food markets is predominantly driven by short-term dynamics. Moreover, as shown in Figure~\ref{Fig:Agro_Connectedness_time_TCI}(a), the trajectory of the short-term TCI closely mirrors that of the time-domain return TCI, with noticeable spikes during the global financial crisis, the COVID-19 pandemic, and the Russia-Ukraine conflict. In contrast to return connectedness, the frequency-domain volatility connectedness exhibits a long-term dominant structure, as evidenced by the significantly higher long-term TCI than the medium- and short-term TCIs. This suggests that volatility spillovers within the energy-food system primarily unfold over longer horizons. Similar to the time-domain volatility TCI in Figure~\ref{Fig:Agro_Connectedness_time_TCI}(b), the long-term volatility TCI also experiences considerable fluctuations. Apart from the three aforementioned crises, it witnesses multiple local peaks during major events such as the Arab Spring and the European debt crisis in 2011, OPEC's policy shift and the commodity market downturn in 2014, and frequent geopolitical conflicts in 2023–2024.

Regarding skewness connectedness, the short-, medium-, and long-term TCIs do not exhibit a clear hierarchy over the sample period, but rather show intersecting trends. The short-term TCI dominates during several time intervals, particularly in turbulent periods such as 2000–2005, 2008–2012, and 2020–2024, indicating stronger short-term skewness spillovers during times of market stress. That is to say, asymmetries in return distributions tend to propagate rapidly in the short term during crises. The medium- and long-term TCIs surpass the short-term TCI in more tranquil periods, such as 2005–2008. Similar to skewness, the kurtosis TCIs at different timescales also demonstrate intersecting trajectories. The medium-term TCI stands out in several periods, especially between 2009 and 2011. The short-term TCI is relatively high during 2000–2005 and 2020–2023 but remains below the medium-term level for much of the sample. The long-term TCI exhibits greater fluctuations and occasionally exceeds the short-term TCI between 2013 and 2015, but its overall magnitude is less pronounced than the medium-term TCI. These findings suggest that extreme tail risk spillovers between the energy and food markets are most prominent in the medium term, followed by the short term, while long-term spillovers are relatively weaker but more volatile.

\begin{figure}[!h]
  \centering
  \includegraphics[width=0.3\linewidth]{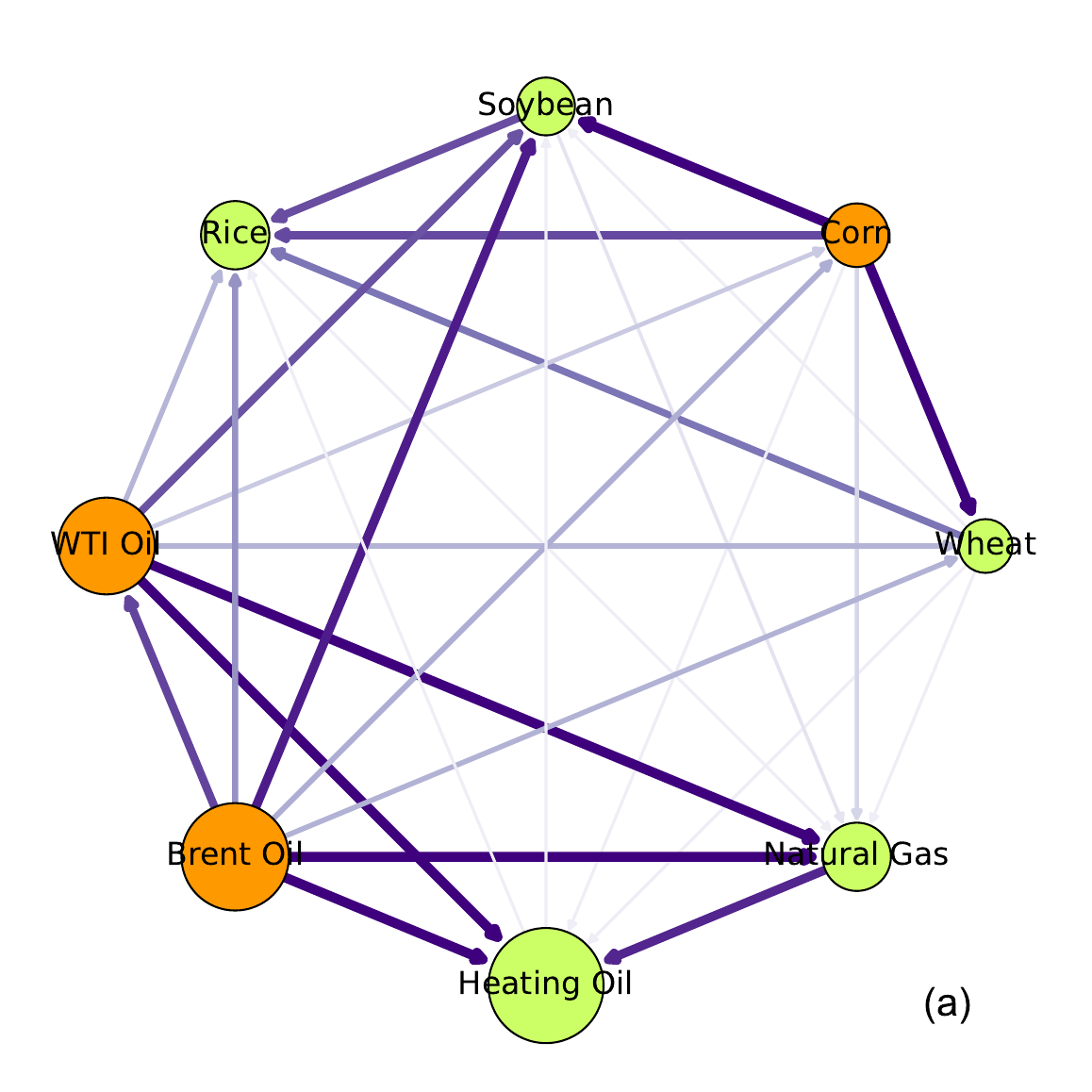}
  \includegraphics[width=0.3\linewidth]{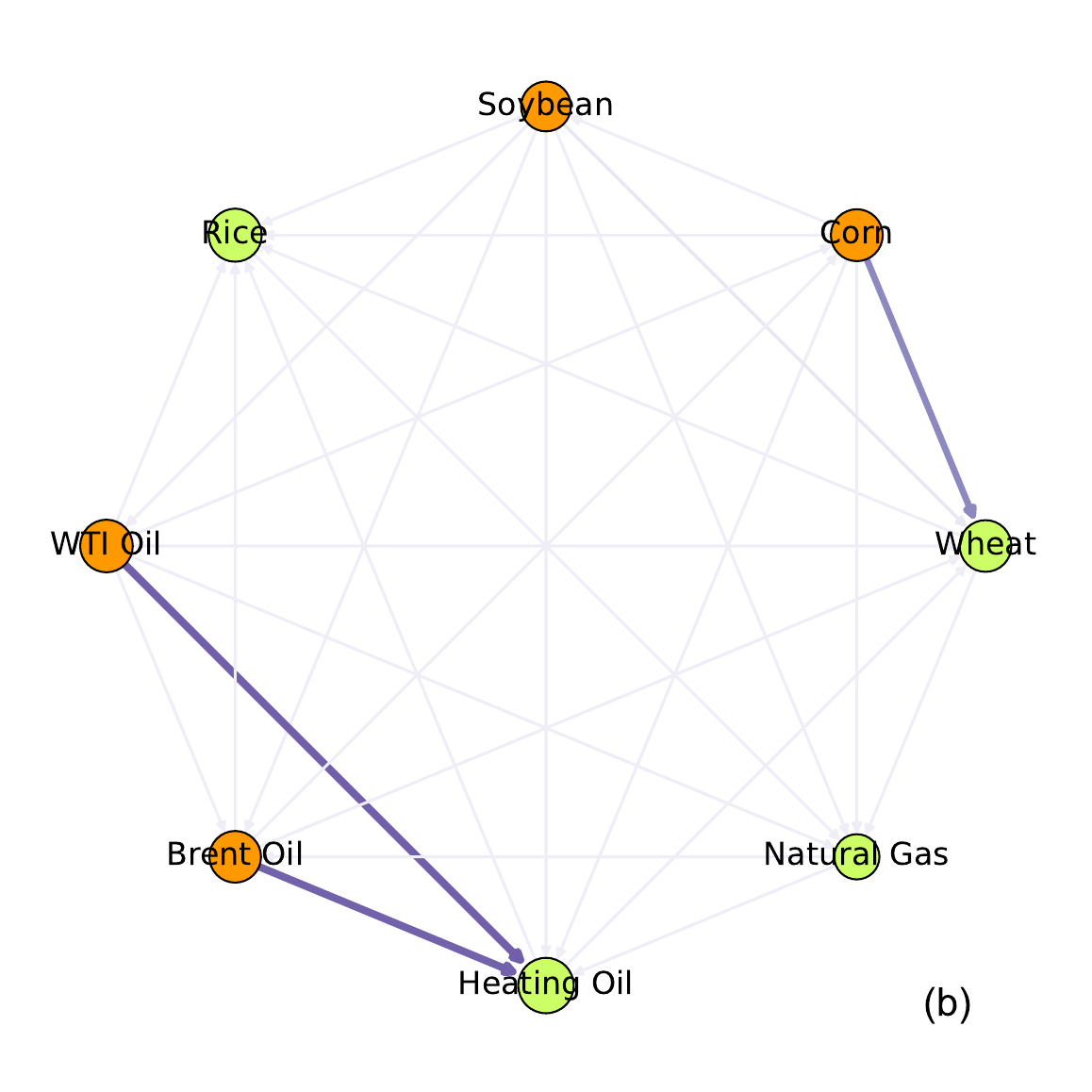}
  \includegraphics[width=0.3\linewidth]{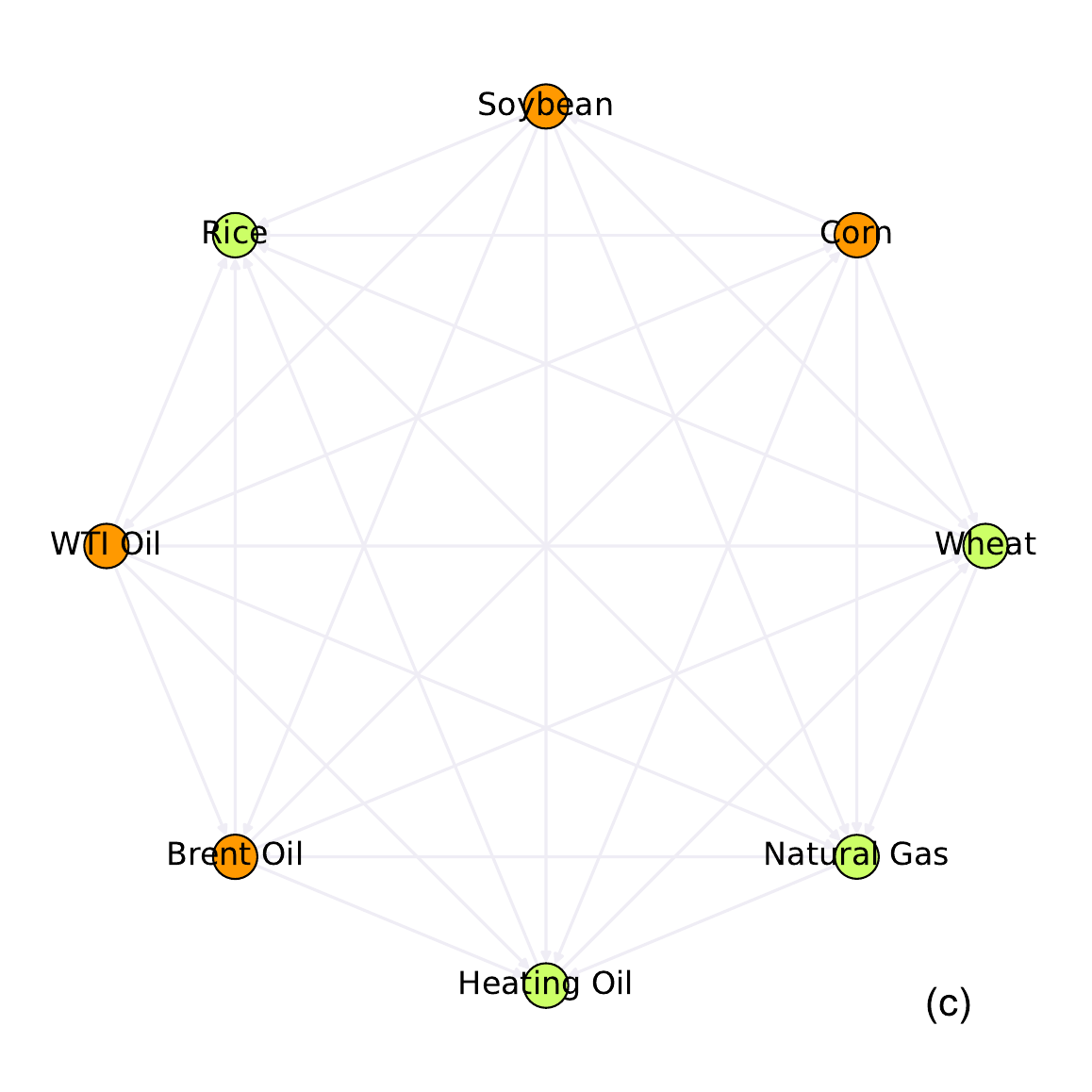}
  \includegraphics[width=0.3\linewidth]{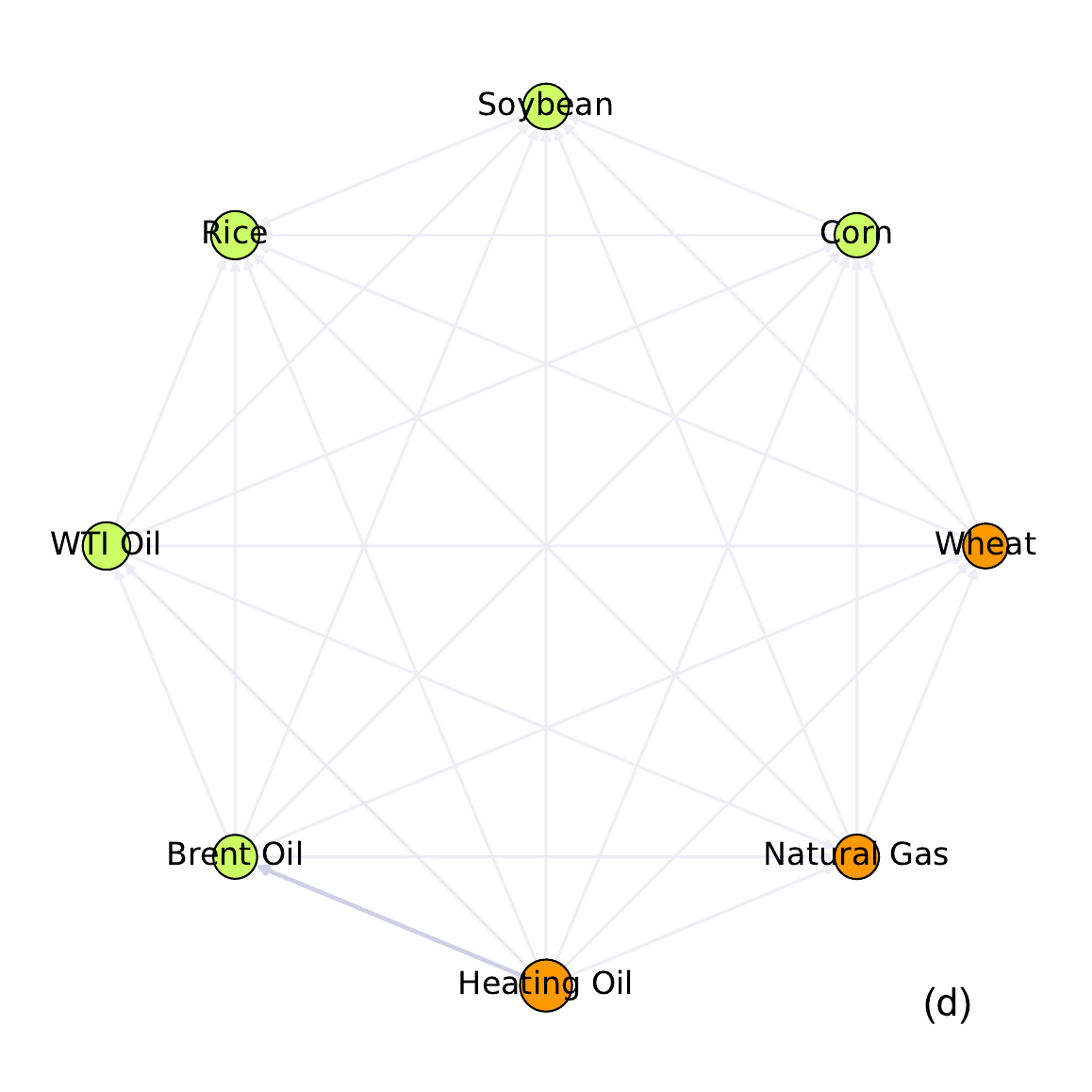}
  \includegraphics[width=0.3\linewidth]{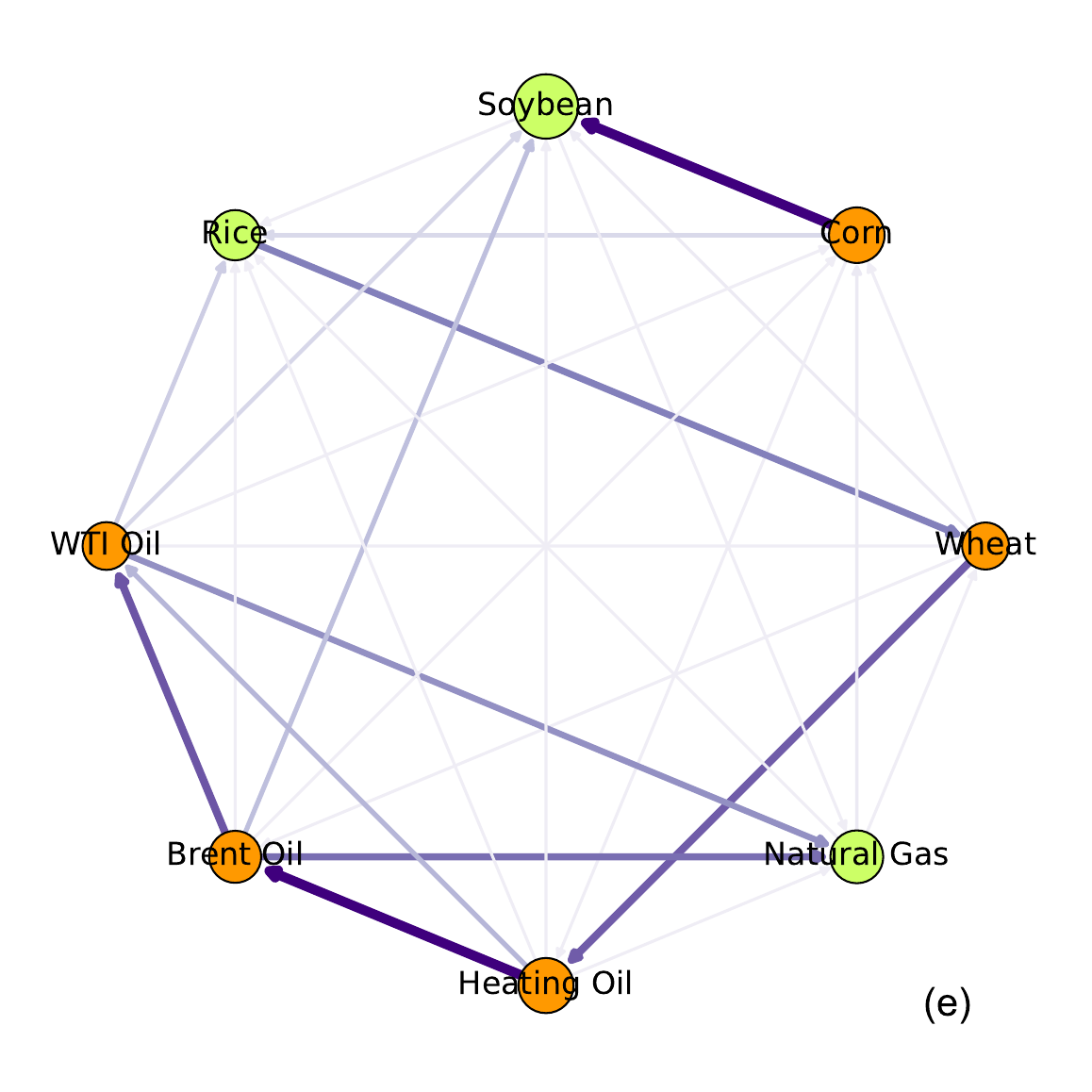}
  \includegraphics[width=0.3\linewidth]{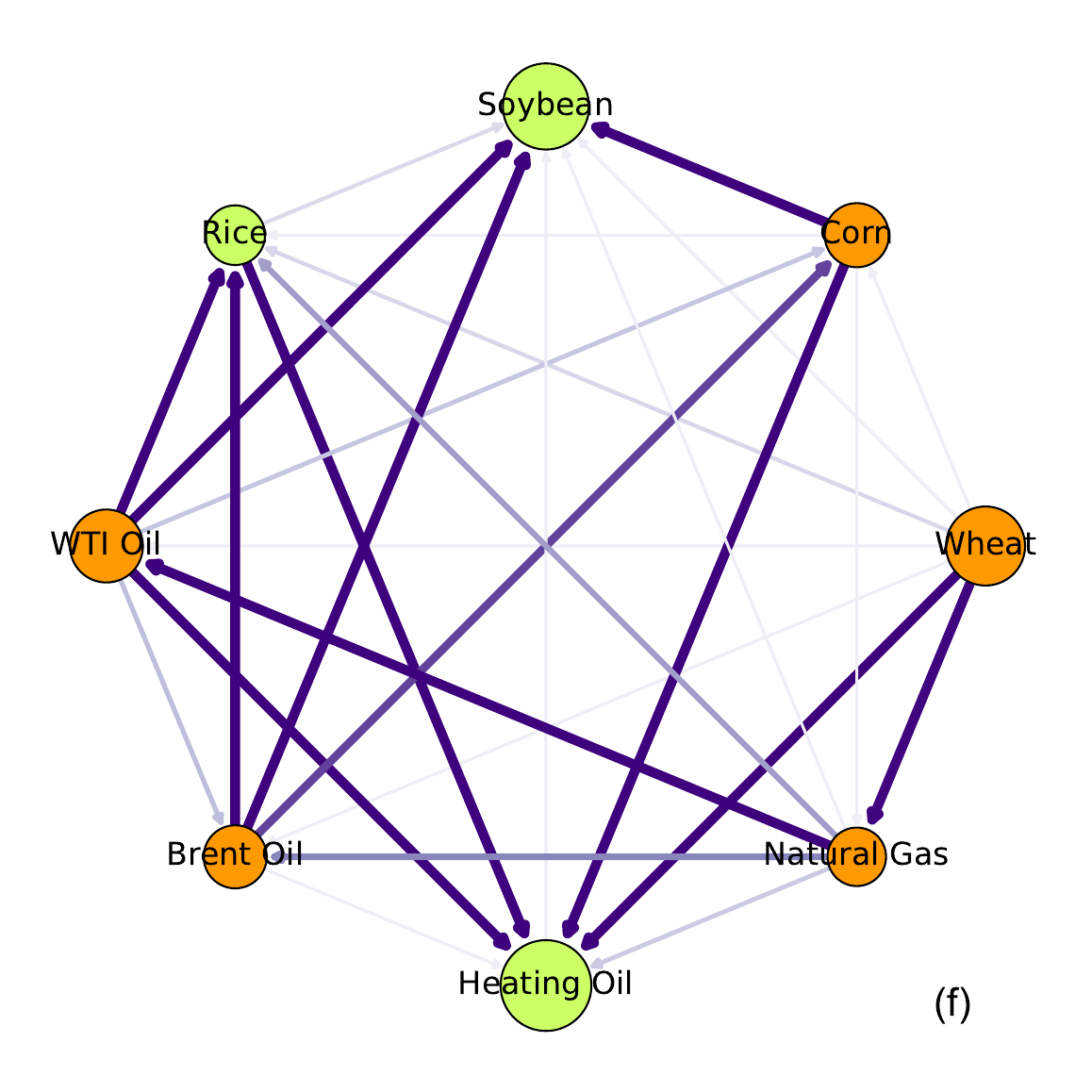}
  \includegraphics[width=0.3\linewidth]{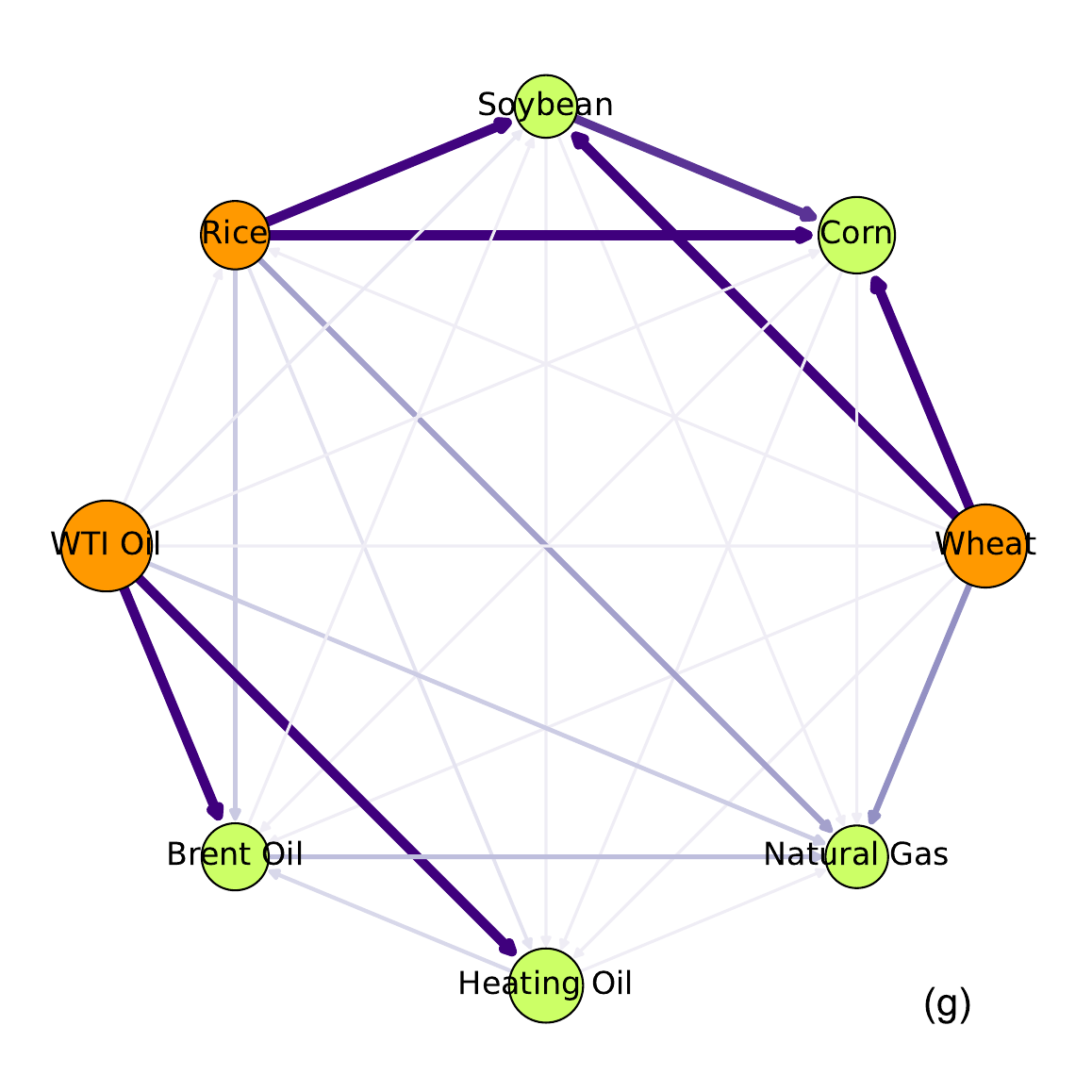}
  \includegraphics[width=0.3\linewidth]{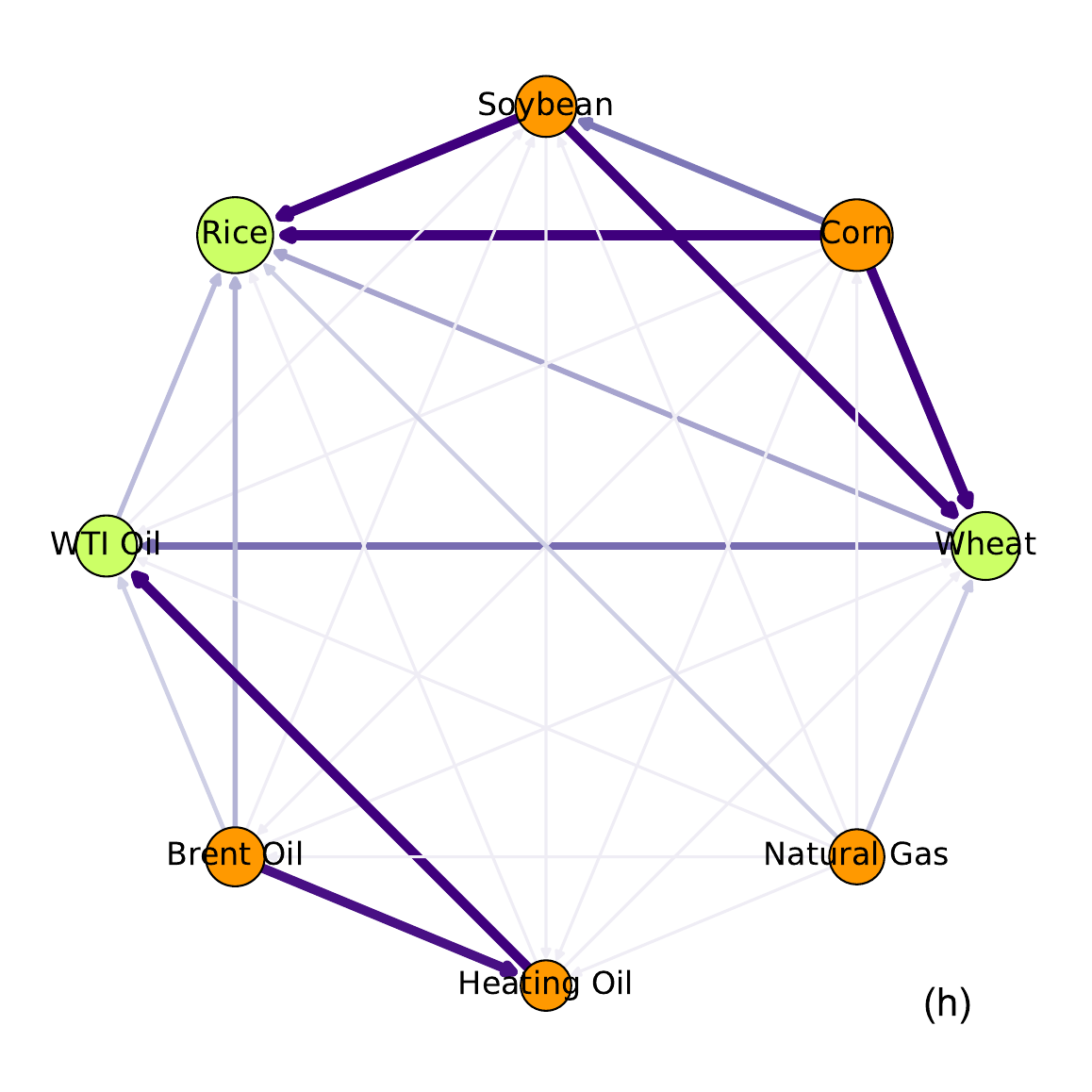}
  \includegraphics[width=0.3\linewidth]{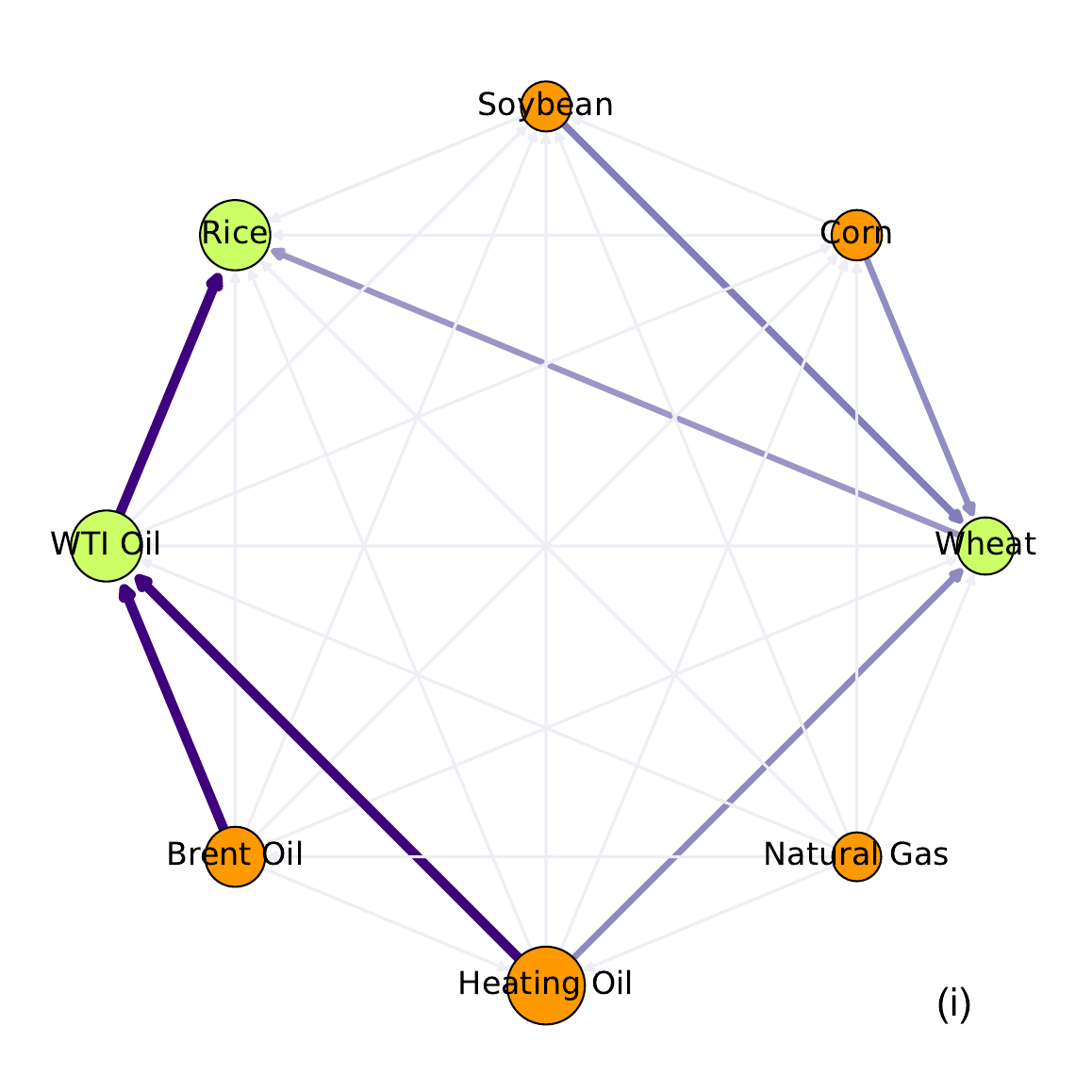}
  \includegraphics[width=0.3\linewidth]{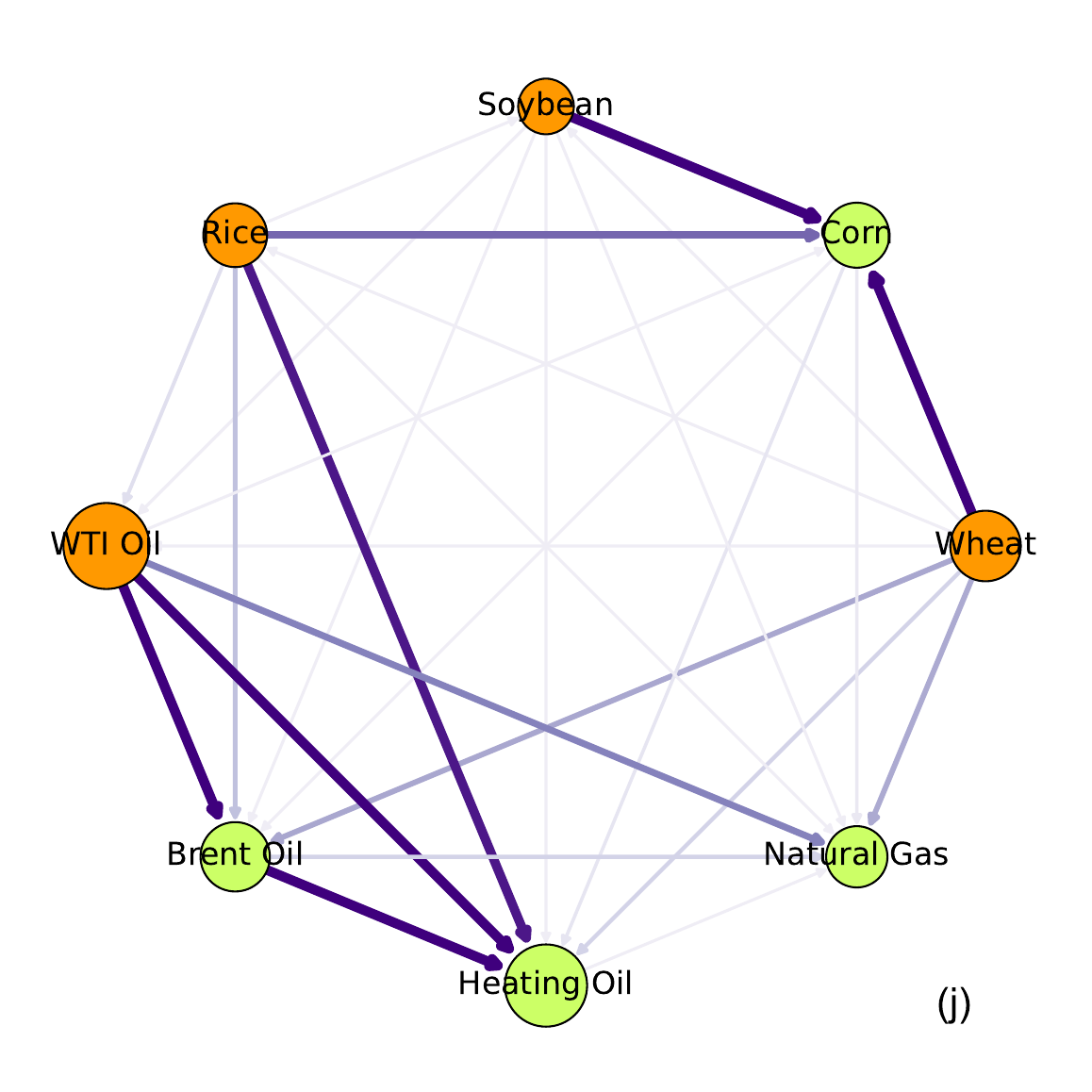}
  \includegraphics[width=0.3\linewidth]{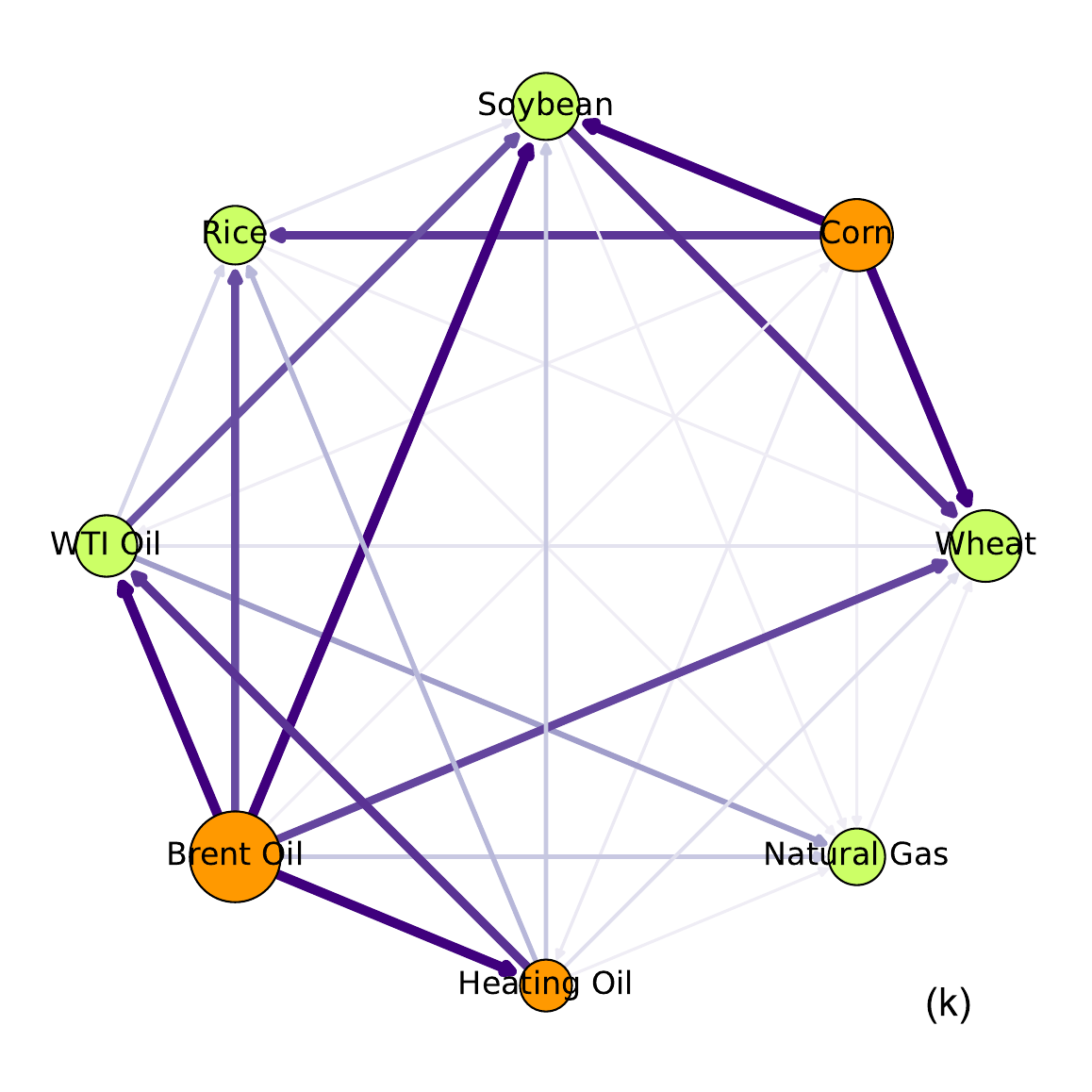}
  \includegraphics[width=0.3\linewidth]{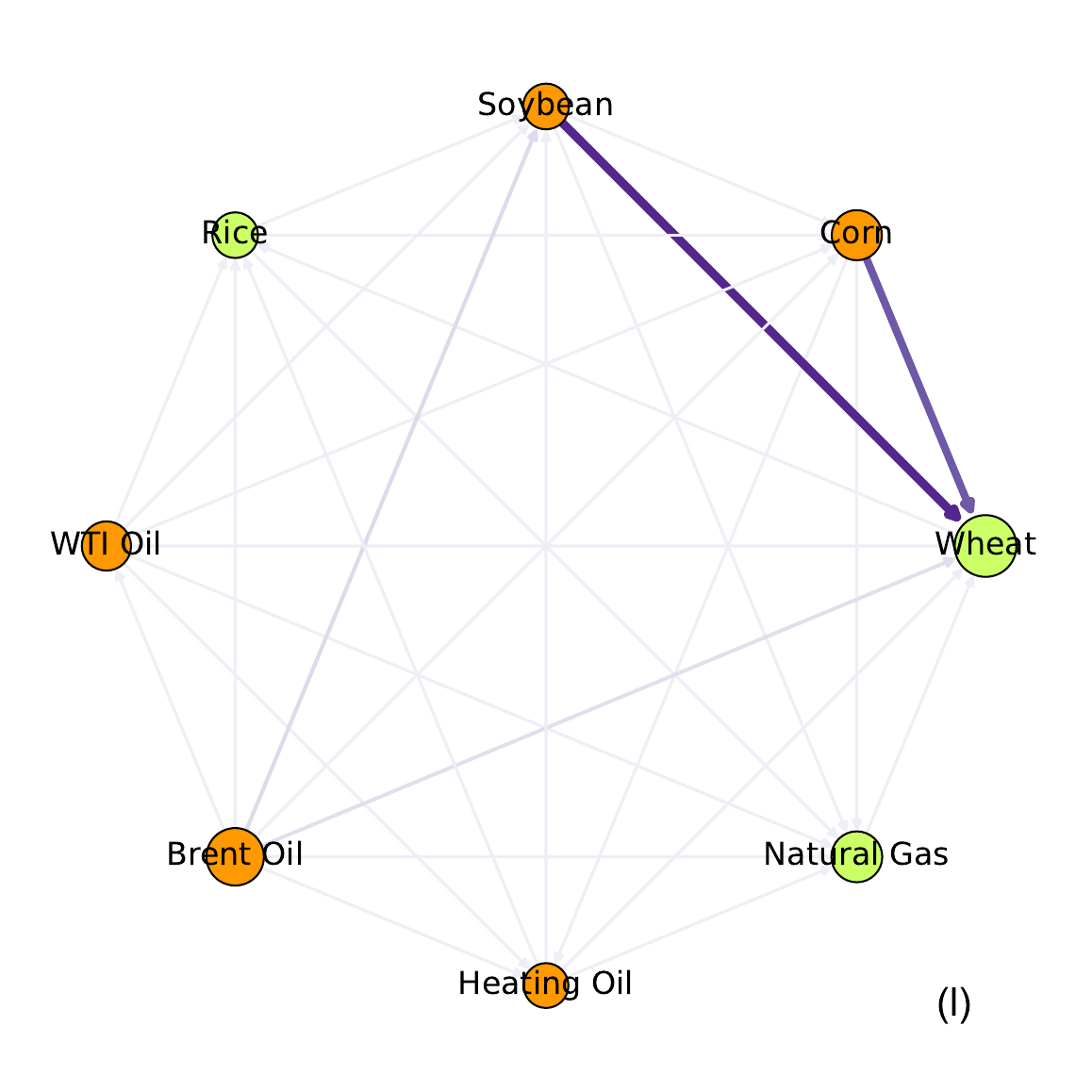}
  \caption{Networks of short-, medium-, and long-term net pairwise connectedness corresponding to return (a–c), volatility (d–f), skewness (g–i), and kurtosis (j–l) of staple food and energy markets.} 
\label{Fig:Agro_Connectedness_frequency_Network}
\end{figure}

Figure~\ref{Fig:Agro_Connectedness_frequency_Network} presents the short-, medium-, and long-term net pairwise connectedness networks of return (a–c), volatility (d–f), skewness (g–i), and kurtosis (j–l) for the staple food and energy markets.

The short-term return connectedness network exhibits a structure dominated by the energy market, where Brent and WTI oil serve as core transmitters, showing strong linkages with heating oil and natural gas. In contrast, return spillovers within the food market are relatively weak, with only corn transmitting large return shocks to soybean and wheat. Moreover, cross-market linkages between energy and food are limited, although weak connections, such as between Brent oil and soybean, are observed. In the medium term, return connectedness weakens overall, with most inter-market connections becoming less pronounced. The primary spillover pathways include WTI oil–heating oil, Brent oil–heating oil, and corn–wheat, all of which show a noticeable reduction in strength compared to the short term. The long-term return connectedness is the weakest, with the network becoming increasingly fragmented. These findings are consistent with the frequency-domain TCI results in Figure~\ref{Fig:Agro_Connectedness_frequency_TCI}(a), which demonstrate that return spillovers predominantly occur over short and medium horizons, with short-term effects being most significant.

The short-term volatility connectedness is generally low, with no evident risk transmission paths in the network. In the medium-term network, internal connections within the energy market strengthen, forming a relatively clear pathway of heating oil–Brent oil–WTI oil. Moderate medium-term volatility spillovers are also observed within the food market, particularly between corn and soybean. The long-term volatility connectedness increases significantly, especially within the energy sector and between energy and food markets, as indicated by darker edge colors in the network. Notably, in cross-market volatility spillovers, crude oil consistently acts as a risk transmitter to staple foods, while heating oil passively receives the volatility spillovers from staple foods. This reflects the complex risk spillover mechanism between energy and food markets. Furthermore, similar to the time-domain volatility network shown in Figure~\ref{Fig:Agro_Connectedness_time_Network}(b), wheat and WTI oil emerge as the primary transmitters of long-term volatility, whereas heating oil and soybean are the main receivers.

In the short-term skewness connectedness network, the food market exhibits stronger internal linkages than the energy market. Except for the wheat–rice pair, most nodes are involved in evident skewness spillovers, particularly along the wheat–corn and wheat–soybean paths. Within the energy sector, however, only WTI oil shows significant short-term skewness spillovers to heating oil and Brent oil. The medium-term skewness network presents similar transmission pathways with the short-term network, but the direction of spillovers reverses. Long-term skewness connectedness is notably weaker, with only a few visible pathways, such as heating oil–WTI, WTI–rice, and Brent–WTI. The comparison across timescales also reveals significant changes in the roles of individual commodities. Specifically, in the short-term network, wheat, rice, and WTI oil act as major transmitters of skewness risk, while the remaining markets serve as receivers. However, this pattern is reversed in the medium- and long-term networks, where wheat, rice, and WTI oil become net receivers, and the other commodities emerge as transmitters. These findings further highlight the temporal heterogeneity and complexity of skewness risk transmission within the energy-food nexus.

In the short-term kurtosis connectedness network, the energy sector shows stronger interlinkages, especially among WTI oil, Brent oil, and heating oil. Some food markets, such as wheat, corn, and soybean, also exhibit evidence of extreme tail risk spillovers. Medium-term kurtosis connectedness intensifies with increased cross-market interactions between energy and staple foods, suggesting that extreme risks are more prone to spread across markets over the medium term. In the long term, kurtosis connectedness decreases significantly, with only soybean, corn, and wheat showing kurtosis spillovers, implying that tail risk spillovers between energy and food markets are limited in the long run. Additionally, WTI and Brent oil hold central positions in the short- and medium-term kurtosis networks, and both also act as risk transmitters in the long-term network. These results further emphasize the pivotal role of crude oil within the energy-food system. That is to say, extreme risks in crude oil markets tend to spill over to other markets during major crisis events, potentially triggering systemic contagion effects. For example, the steep decline in international oil prices in the second half of 2014 is widely regarded as a major catalyst for the broad-based downturn in global commodity markets in 2015.

\subsection{Factors influencing risk connectedness}

Building on the identified time- and frequency-domain connectedness of return, volatility, skewness, and kurtosis between energy and food markets, we further explore the key drivers of these risk spillover effects using the random forest model. Following \cite{Kilian-Murphy-2014-JApplEconom}, \cite{Le-Pham-Do-2023-EnergyEcon}, \cite{Rao-Lucey-Kumar-2023-EnergyEcon}, and \cite{Zhou-Wu-Liu-Rognone-2023-NatCommun}, we select ten potential explanatory variables: DXY, TYR, VIX, EPU, COP, COS, EUA, CPU, GND, and GPR. These factors encompass a broad spectrum of influences, including macroeconomics and finance, policy uncertainty, crude oil supply and demand fundamentals, and climate and geopolitical shocks, all of which are closely associated with risk transmission dynamics within the energy-food system.

Table~\ref{Tab:RF_Evaluation_Metrics} reports the evaluation metrics for random forest models. It can be found that each time- and frequency-domain connectedness corresponds to $RAE$ and $RSE$ values below 1, indicating that all models outperform the benchmark. Specifically, for the time-domain connectedness, the $R^{2}$ values for the return TCI and skewness TCI models both exceed 0.8, suggesting that the ten selected variables have strong explanatory power for the return and skewness connectedness. These two models also show low absolute and relative errors, further validating their robustness and reliability. By contrast, the models for the kurtosis TCI and volatility TCI yield lower $R^{2}$ values, implying weaker explanatory power in capturing volatility and kurtosis spillovers. This may reflect the more complex transmission mechanisms of volatility and tail risks between energy and food markets.

Regarding frequency-domain connectedness, we observe that the models for short-term TCIs of return, volatility, skewness, and kurtosis consistently achieve better fits than those for the medium- and long-term TCIs, as reflected in their higher $R^{2}$ values. This indicates that short-term spillovers are generally more interpretable, possibly because they are more responsive to the selected explanatory variables, while medium- and long-term spillovers are often influenced by more intricate and persistent factors. In other words, the explanatory power of the models tends to decline with increasing timescales. Furthermore, the $MAE$ and $MSE$ values are relatively higher for the short-term return TCI and long-term volatility TCI, which implies that these two series exhibit more pronounced fluctuations, resulting in larger model errors. Conversely, the long-term return TCI and short-term volatility TCI display more stable trends, contributing to lower model errors. These findings are consistent with the results shown in Figure~\ref{Fig:Agro_Connectedness_frequency_TCI}.

\begin{table}[!ht]
  \centering
  \setlength{\abovecaptionskip}{0pt}
  \setlength{\belowcaptionskip}{10pt}
  \caption{Evaluation metrics for random forest models}
  \setlength\tabcolsep{25pt}   \resizebox{\textwidth}{!}{ 
    \begin{tabular}{@{\hspace{5pt}}l r@{.}l r@{.}l r@{.}l r@{.}l r@{.}l}
    \toprule
         & \multicolumn{2}{c}{$R^{2}$} & \multicolumn{2}{c}{$MAE$} & \multicolumn{2}{c}{$MSE$} & \multicolumn{2}{c}{$RAE$} & \multicolumn{2}{c}{$RSE$}  \\
        \midrule
    \multicolumn{11}{@{\hspace{5pt}}l}{\textit{Panel A: Time-domain connectedness}} \\
    Return TCI & 0&818  & 1&808  & 5&905  & 0&418  & 0&182 \\
    Volatility TCI & 0&434  & 5&206  & 42&521  & 0&747  & 0&566 \\
    Skewness TCI & 0&819  & 2&363  & 12&706  & 0&365  & 0&181 \\
    Kurtosis TCI & 0&546  & 6&041  & 68&078  & 0&653  & 0&454 \vspace{2mm} \\
    \multicolumn{11}{@{\hspace{5pt}}l}{\textit{Panel B: Frequency-domain return connectedness}} \\
    Short-term TCI & 0&806  & 1&546  & 4&305  & 0&439  & 0&194 \\
    Medium-term TCI & 0&665  & 0&642  & 0&733  & 0&532  & 0&335 \\
    Long-term TCI & 0&664  & 0&037  & 0&002  & 0&536  & 0&336 \vspace{2mm} \\
    \multicolumn{11}{@{\hspace{5pt}}l}{\textit{Panel C: Frequency-domain volatility connectedness}} \\    
    Short-term TCI & 0&871  & 1&396  & 4&274  & 0&353  & 0&129 \\
    Medium-term TCI & 0&400  & 2&957  & 14&377  & 0&743  & 0&600 \\
    Long-term TCI & 0&422  & 6&737  & 67&444  & 0&781  & 0&578 \vspace{2mm} \\
    \multicolumn{11}{@{\hspace{5pt}}l}{\textit{Panel D: Frequency-domain skewness connectedness}} \\
    Short-term TCI & 0&599  & 2&360  & 10&095  & 0&590  & 0&401 \\
    Medium-term TCI & 0&513  & 1&358  & 3&817  & 0&658  & 0&487 \\
    Long-term TCI & 0&523  & 1&463  & 4&848  & 0&594  & 0&477 \vspace{2mm} \\
    \multicolumn{11}{@{\hspace{5pt}}l}{\textit{Panel E: Frequency-domain kurtosis connectedness}} \\
    Short-term TCI & 0&544  & 2&855  & 14&156  & 0&592  & 0&456 \\
    Medium-term TCI & 0&514  & 3&388  & 23&803  & 0&716  & 0&486 \\
    Long-term TCI & 0&516  & 2&805  & 15&286  & 0&669  & 0&484 \\
  \bottomrule
    \end{tabular}
    }%
  \begin{flushleft}
    \footnotesize
\justifying Note: This table presents the evaluation metrics for random forest models of time- and frequency-domain moment connectedness between energy and food markets.
\end{flushleft} 
  \label{Tab:RF_Evaluation_Metrics}%
\end{table}%

\begin{figure}[!ht]
  \centering
  \includegraphics[width=0.45\linewidth]{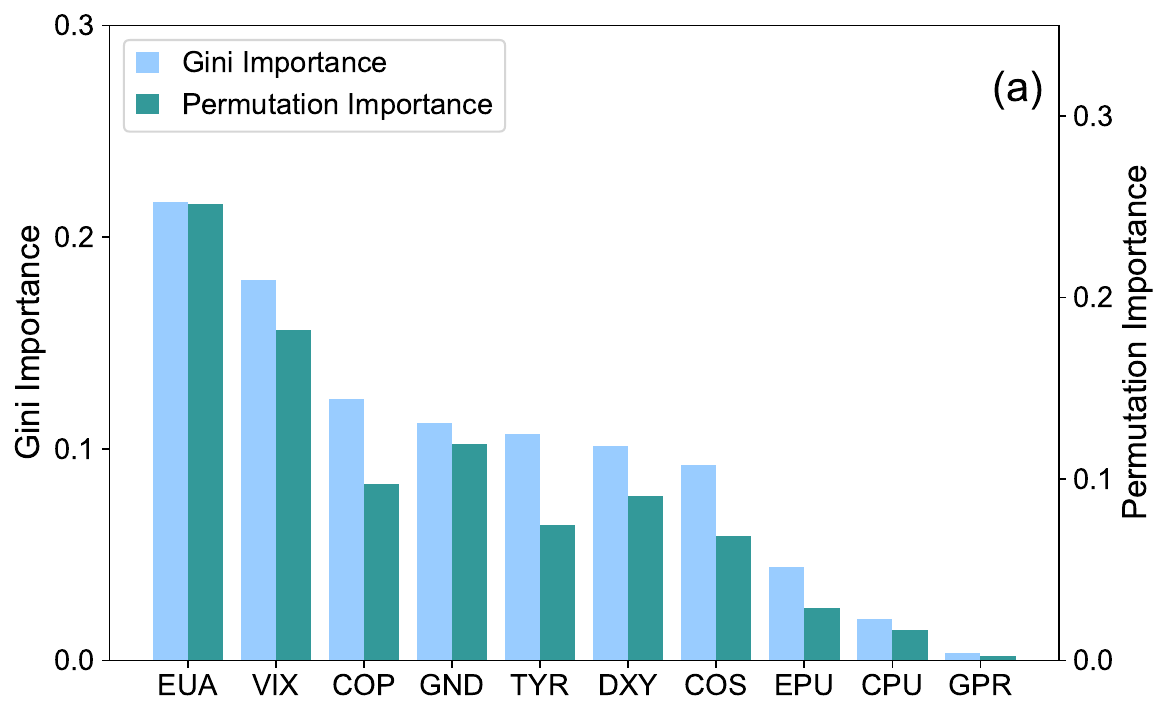}
  \includegraphics[width=0.45\linewidth]{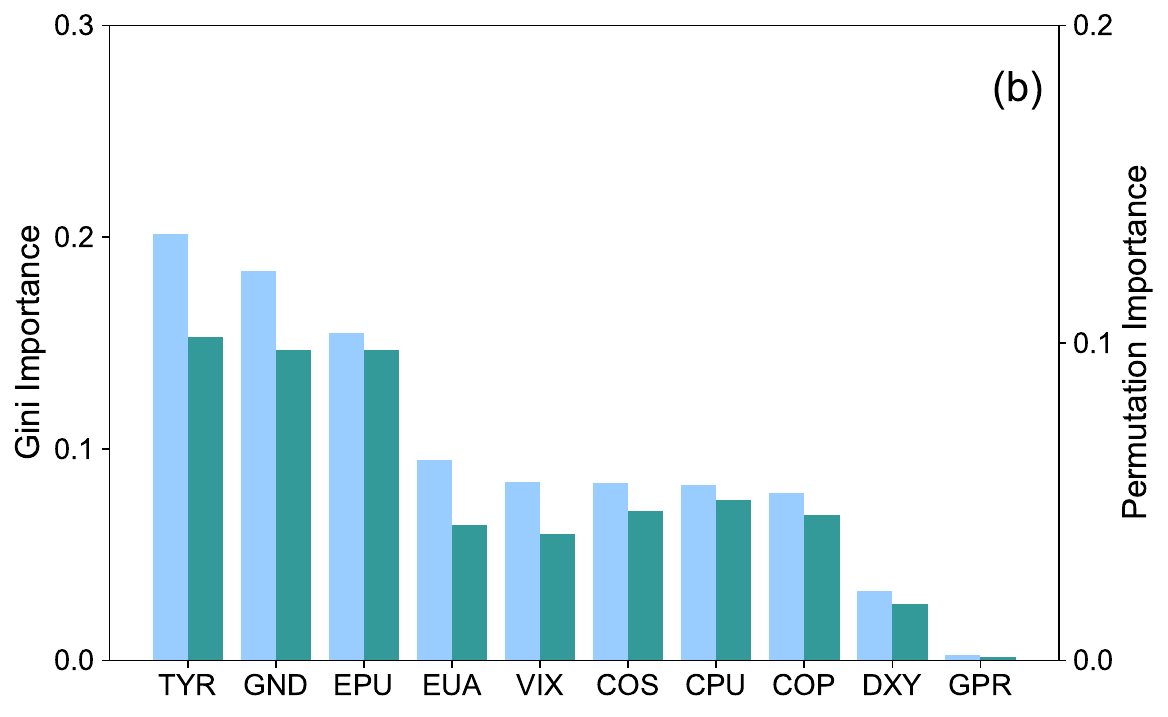}
  \includegraphics[width=0.45\linewidth]{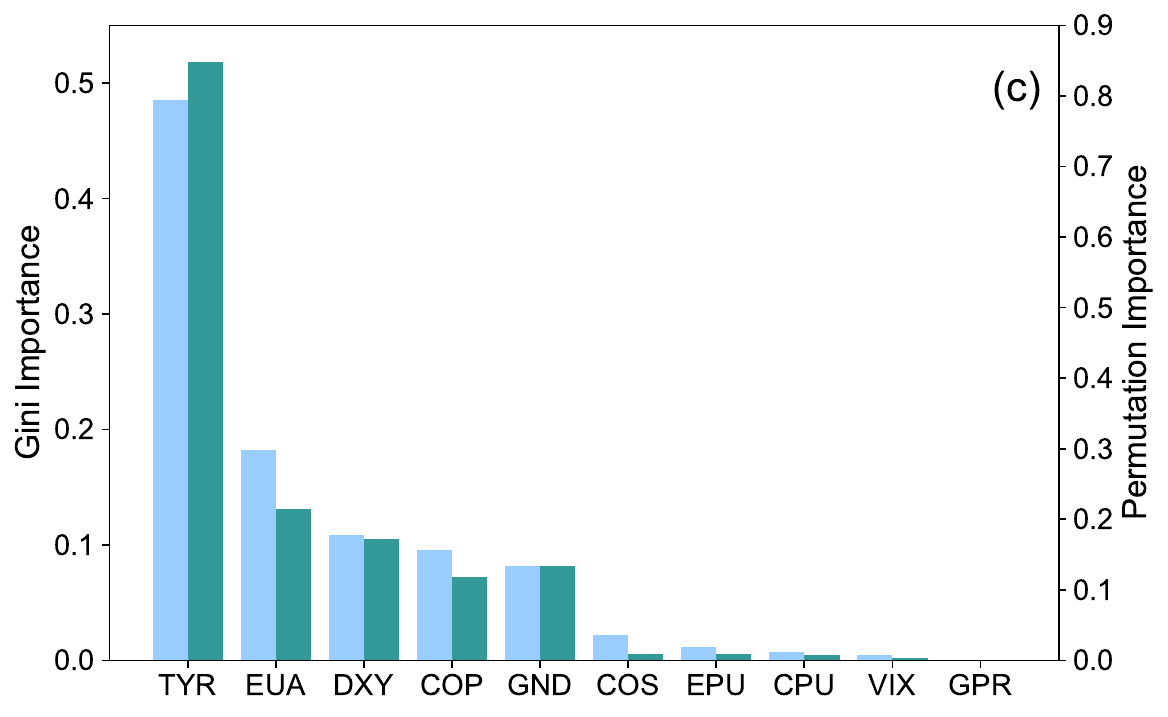}
  \includegraphics[width=0.45\linewidth]{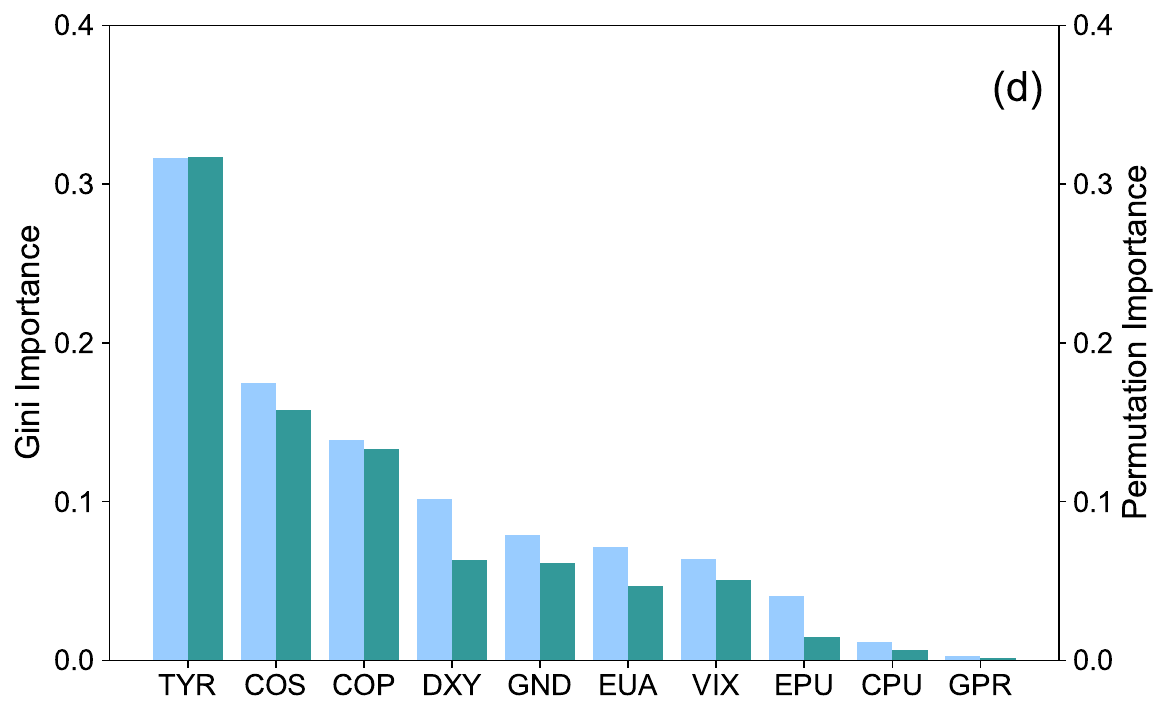}
  \caption{Relative (Gini) and absolute (Permutation) importance of different factors in explaining TCIs for return (a), volatility (b), skewness (c), and kurtosis (d) of staple food and energy markets.}
\label{Fig:RF_FactorsImportance_Time}
\end{figure}

\begin{figure}[!ht]
  \centering
  \includegraphics[width=0.321\linewidth]{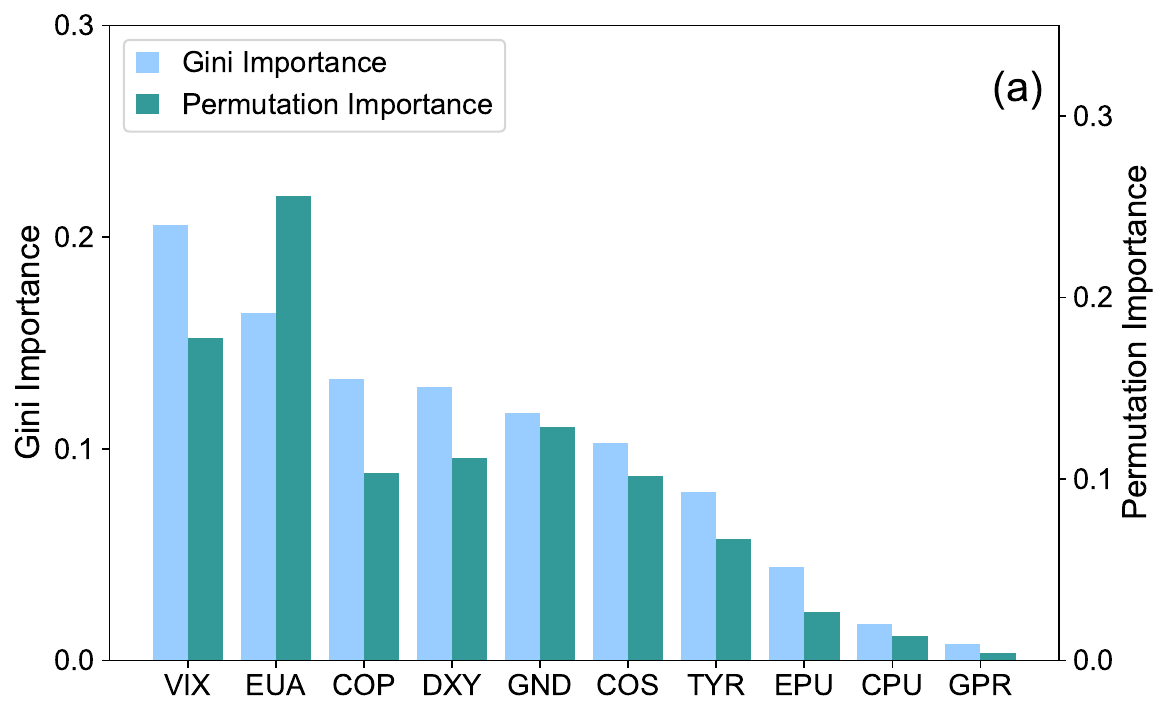}
  \includegraphics[width=0.321\linewidth]{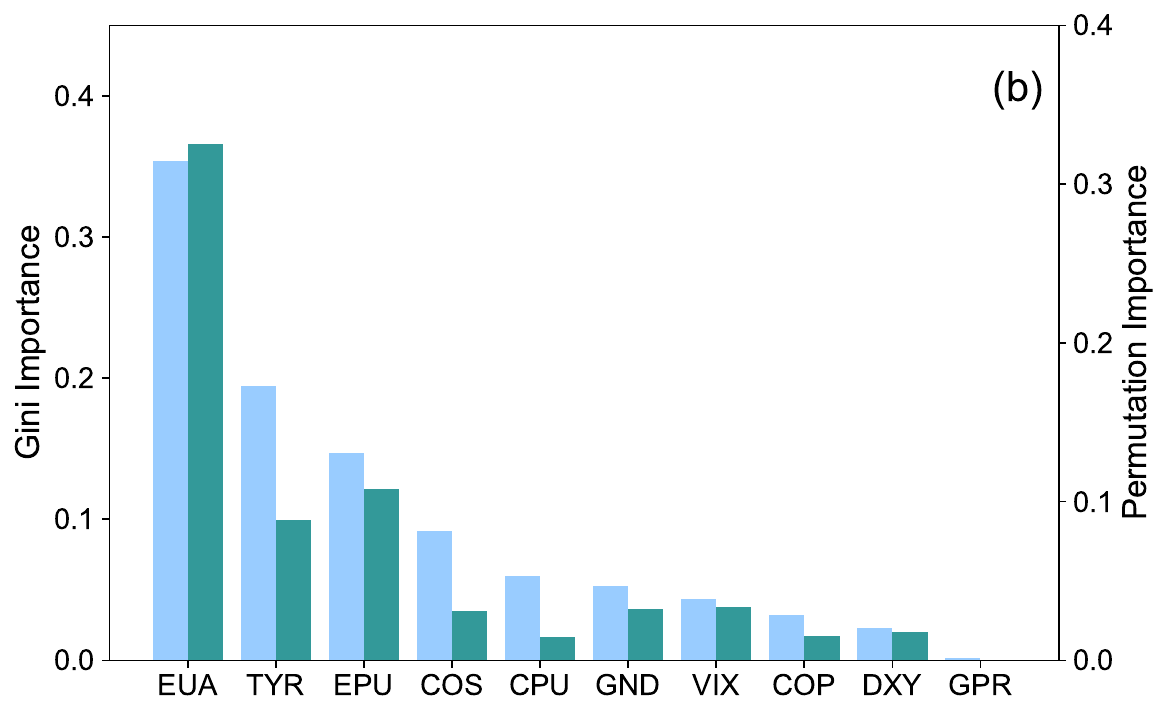}
  \includegraphics[width=0.321\linewidth]{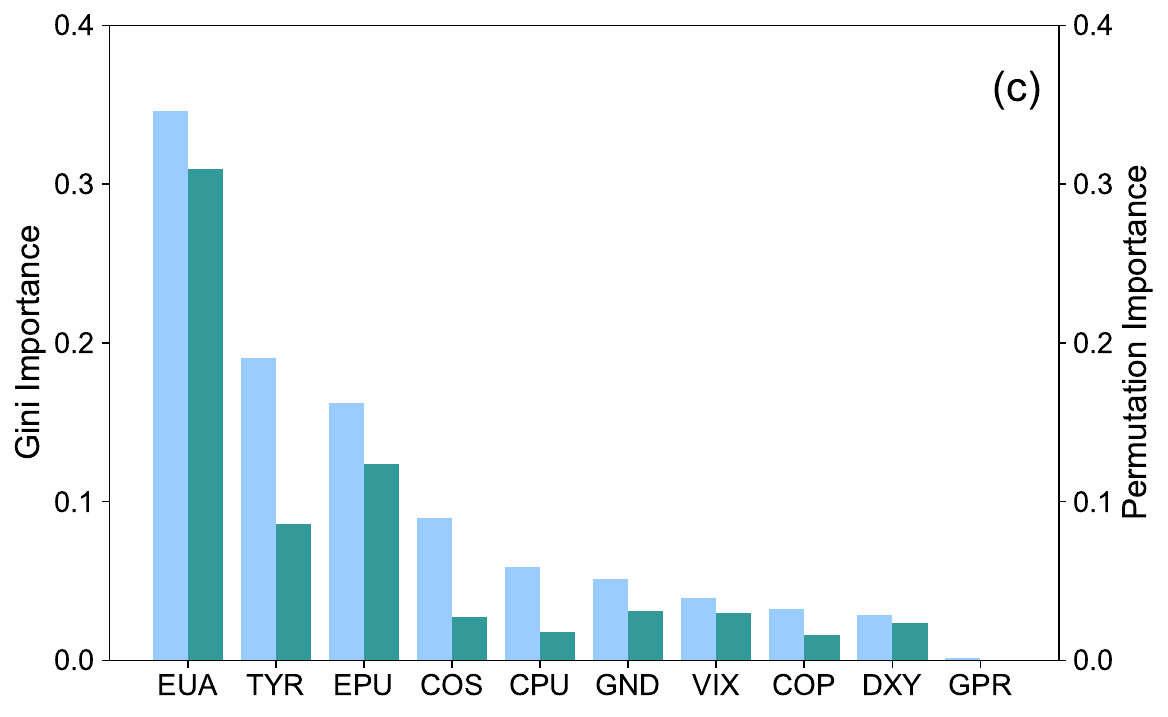}
  \includegraphics[width=0.321\linewidth]{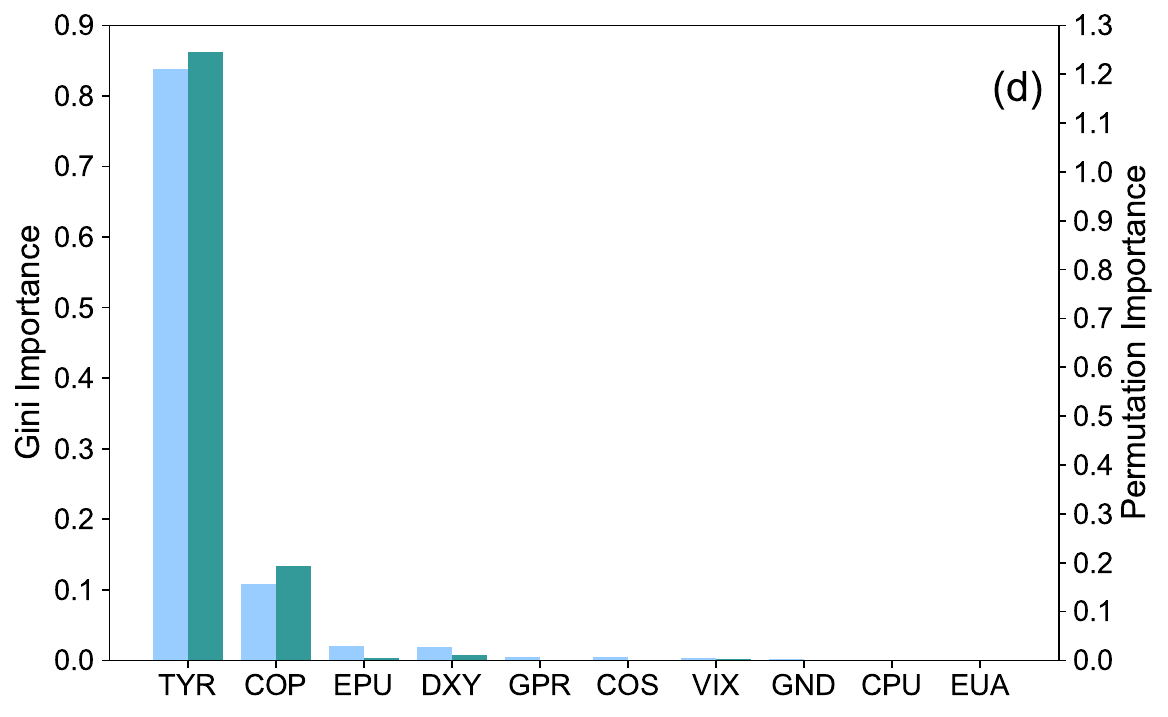}
  \includegraphics[width=0.321\linewidth]{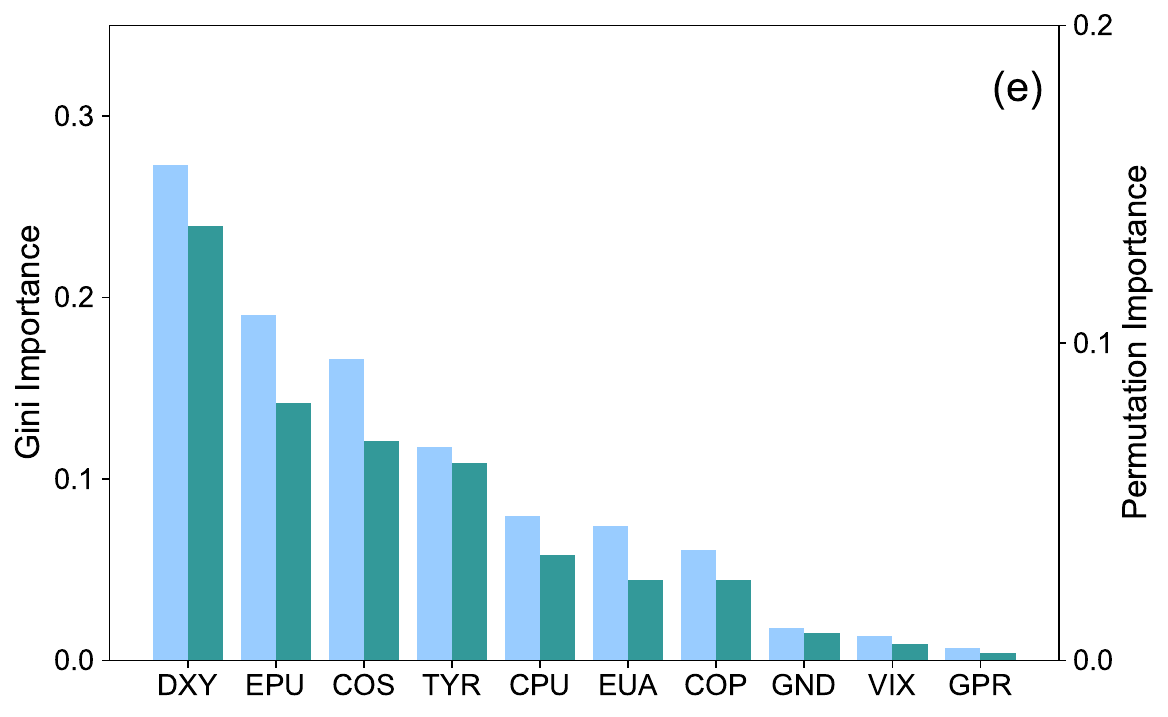}
  \includegraphics[width=0.321\linewidth]{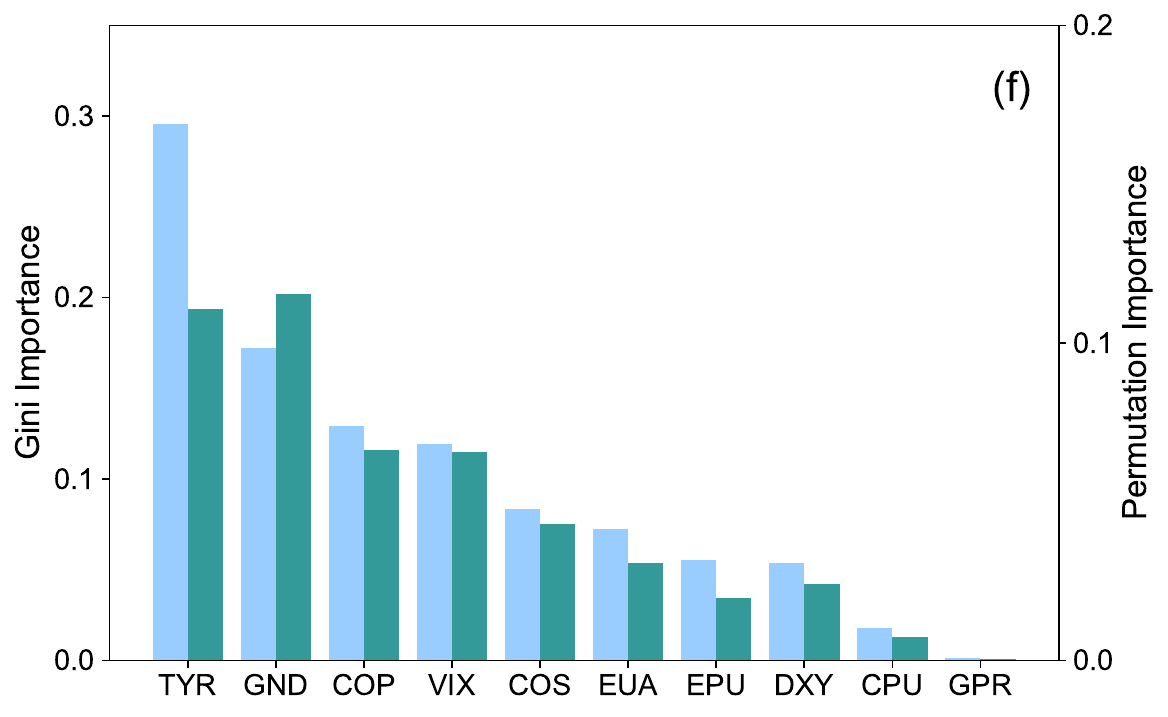}
  \includegraphics[width=0.321\linewidth]{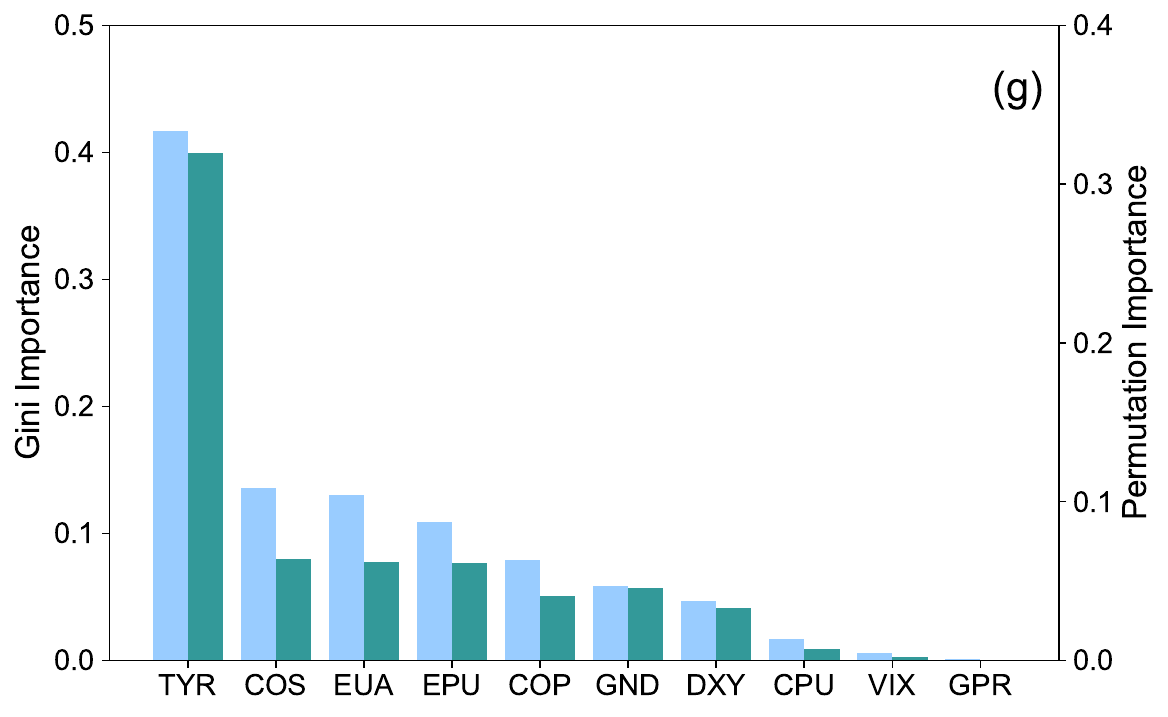}
  \includegraphics[width=0.321\linewidth]{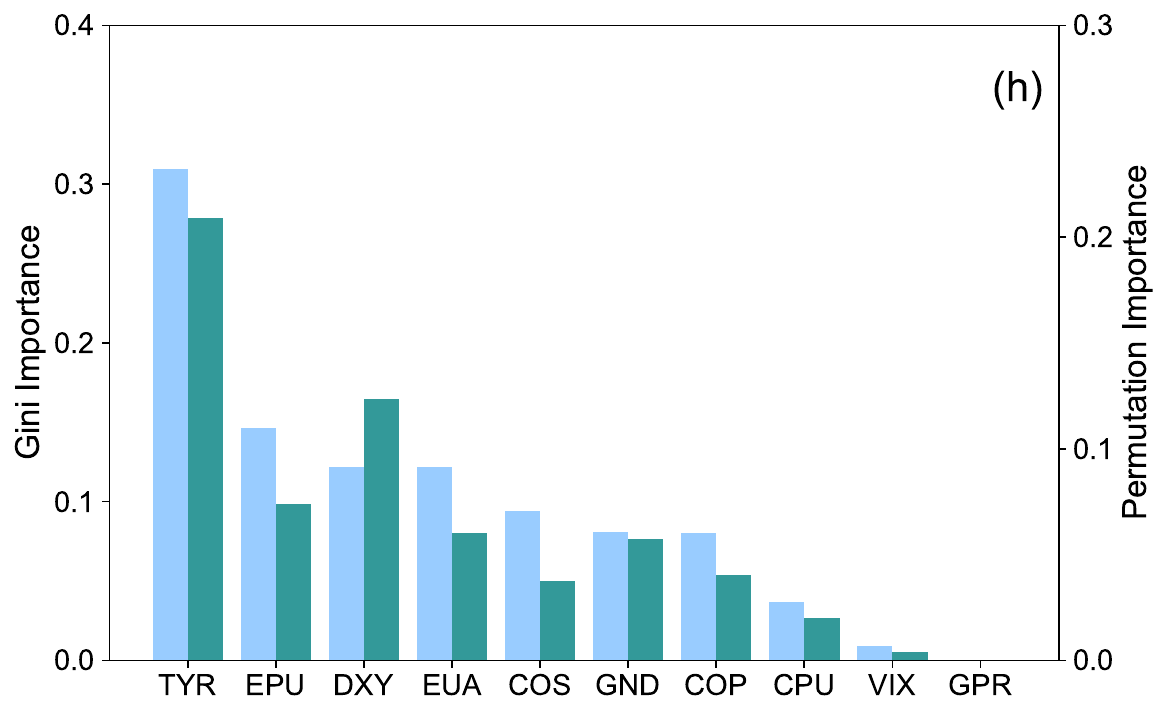}
  \includegraphics[width=0.321\linewidth]{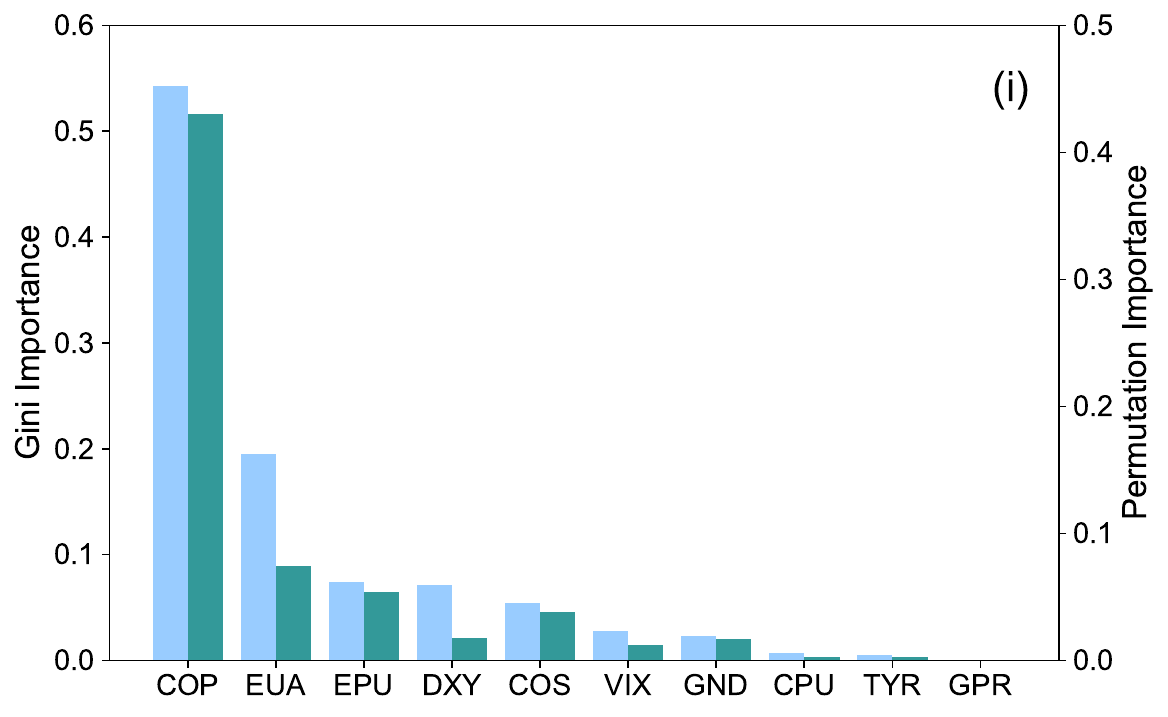}
  \includegraphics[width=0.321\linewidth]{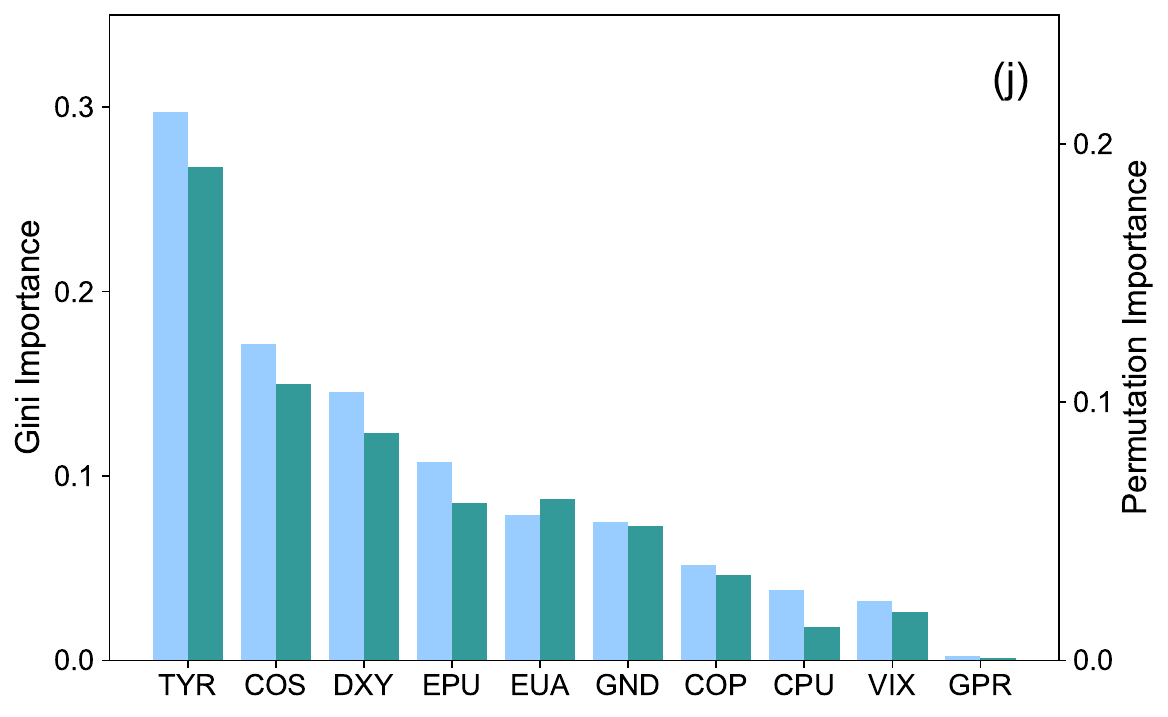}
  \includegraphics[width=0.321\linewidth]{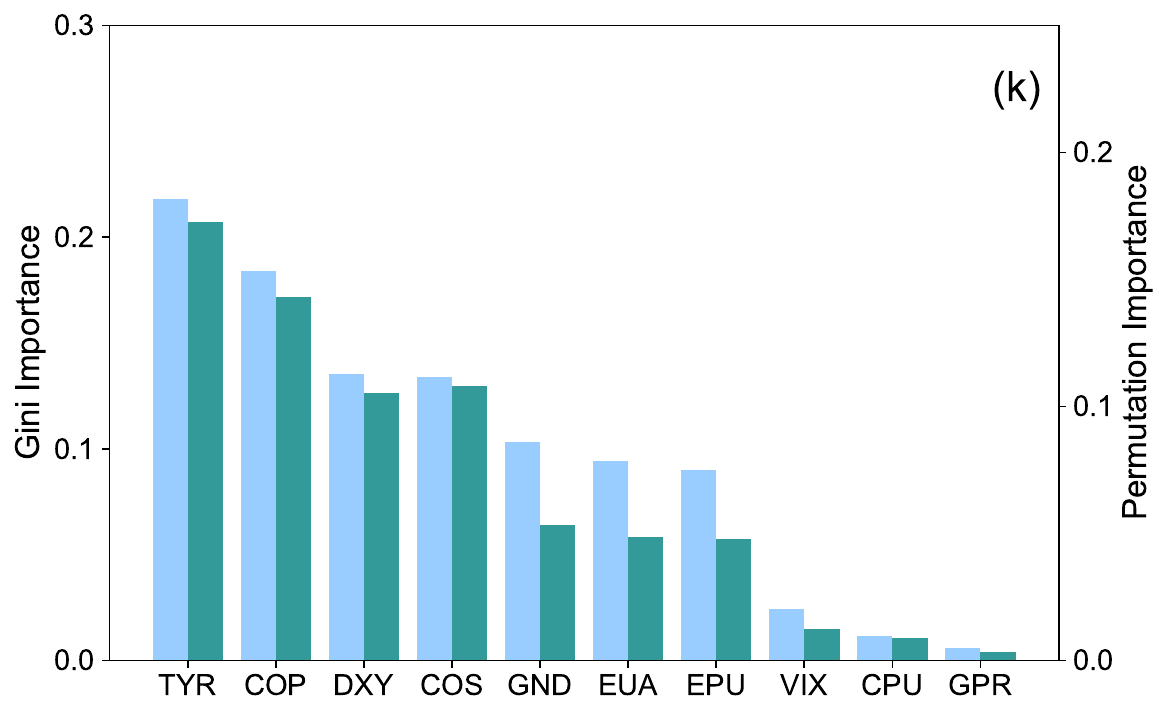}
  \includegraphics[width=0.321\linewidth]{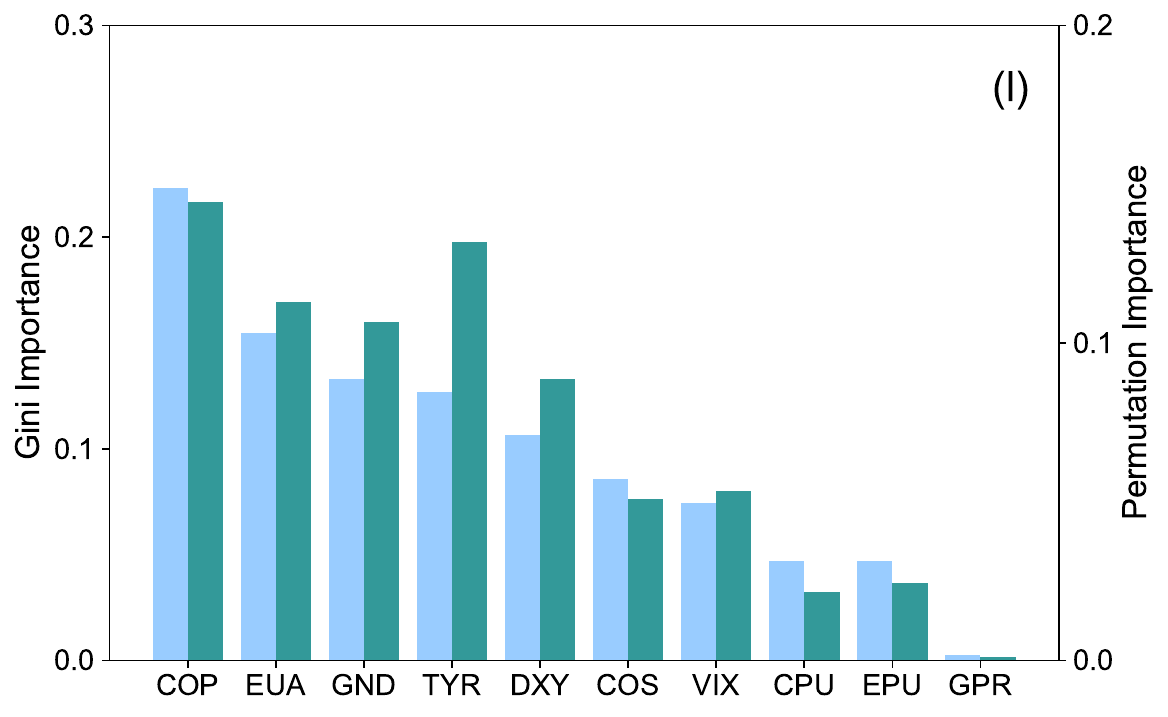}
  \caption{Relative (Gini) and absolute (Permutation) importance of different factors in explaining short-, medium-, and long-term TCIs corresponding to return (a–c), volatility (d–f), skewness (g–i), and kurtosis (j–l) of staple food and energy markets.}
\label{Fig:RF_FactorsImportance_Frequency}
\end{figure}

Both Gini and permutation importance are calculated for the ten selected factors to identify the key drivers of various risk spillovers. This dual approach provides a comprehensive and reliable assessment of the relative and absolute importance of each factor in explaining risk spillovers between energy and food markets. Figures~\ref{Fig:RF_FactorsImportance_Time} and~\ref{Fig:RF_FactorsImportance_Frequency} display the factor importance for the time- and frequency-domain TCIs of return, volatility, skewness, and kurtosis, respectively, with variables ranking in descending order of Gini importance. Upon comparison, we find that, despite minor numerical differences, the Gini and permutation importance measures consistently highlight the same key determinants in most cases, thereby reinforcing the credibility and robustness of our factor importance analysis. Moreover, the importance of these factors varies significantly across different moments and timescales of the connectedness.

As shown in the time-domain factor importance in Figure~\ref{Fig:RF_FactorsImportance_Time}, TYR emerges as the most critical factor of the volatility, skewness, and kurtosis connectedness. Especially, both Gini and permutation importance values for skewness and kurtosis TCIs are markedly higher than those of the other variables. As a core macroeconomic indicator reflecting the business cycle and monetary policy stance, the interest rate serves as a crucial conduit for risk transmission between energy and food markets. Changes in interest rates affect spillovers of volatility, asymmetry, and tail risks not only through financing costs and asset allocations, but also through shifts in market expectations. In addition, EUA exhibits high importance in explaining return and skewness TCIs, implying that under the backdrop of global carbon neutrality and green transition, carbon market policies have become a key variable affecting energy production costs and agricultural input structures. Carbon prices can influence co-movements of energy and food prices through cost-push channels and expectation-driven mechanisms, thereby driving the return and skewness connectedness.

VIX and EPU demonstrate high importance in explaining return and volatility TCIs, respectively, suggesting differentiated roles of market sentiment and policy uncertainty in driving return and volatility spillovers. Particularly during periods of elevated market stress, such as the financial crisis or the COVID-19 pandemic, sharp increases in the VIX are often accompanied by stronger connectedness between energy and food markets. Additionally, GND has strong explanatory power for both return and volatility TCIs, highlighting the amplifying effect of natural disasters on supply-demand expectations and market price fluctuations. COS and COP mainly influence the kurtosis and return TCIs, reflecting the impact of structural shifts in crude oil supply and demand, such as inventory shocks or production cut agreements, on return and kurtosis spillovers. This again underscores the central role of crude oil in the energy-food nexus. In contrast, GPR consistently shows relatively low importance across all models, which may be related to the weak sensitivity of moment connectedness to geopolitical risks over the full sample period or to the regional heterogeneity of geopolitical risk impacts.

Turning to the frequency-domain factor importance in Figure~\ref{Fig:RF_FactorsImportance_Time}, we observe notable differences in the core drivers of moment connectedness across different timescales. For return connectedness, the short-term TCI is primarily driven by market sentiment and carbon prices, while medium- and long-term TCIs are more influenced by macroeconomic variables. Specifically, in the short term, VIX and EUA play important roles, indicating that sentiment shocks and carbon market dynamics can rapidly affect return spillovers between energy and food markets. In addition, COP, DXY, and GND also show strong explanatory power, suggesting that crude oil supply, U.S. dollar strength, and natural disasters also contribute significantly to short-term return connectedness. In the medium and long term, the importance of macroeconomic indicators such as TYR and EPU increases substantially, implying that market responses to the interest rate changes and policy uncertainty become more pronounced over time, thereby enhancing the influence of these factors on the medium- and long-term return connectedness.

Unlike the short-term return TCI, the short-term volatility TCI exhibits clear single-factor dominance, being almost entirely explained by TYR. This suggests that in the short run, interest rate shocks may strongly influence volatility spillover effects through financing costs and risk pricing. In the medium term, economic and financial indicators such as DXY and EPU, along with supply-demand variables like COS, emerge as major drivers, implying that the sources of medium-term volatility transmission gradually extend to fundamental factors. For the long-term volatility TCI, the explanatory power of factors becomes more balanced, with TYR, GND, COP, and VIX all playing significant roles, reflecting a multi-factor-driven structure.

Regarding skewness connectedness, TYR ranks as the most important factor in both short and medium terms. COS, EUA, and EPU also show high importance, suggesting that interest rate shocks, inventory dynamics, carbon pricing, and economic policy uncertainty may trigger asymmetric market reactions on energy and food prices, thus contributing to skewness spillovers. In the long term, COP emerges as the most critical factor, indicating that structural changes in the crude oil market, such as OPEC+ production cuts and energy transitions, can cause a substantial impact on the long-term spillovers of asymmetry risks between energy and food markets.

Similar to skewness connectedness, TYR remains the most important driver for the short- and medium-term kurtosis TCIs, and COP is dominant for the long-term kurtosis TCI. However, kurtosis spillovers across different timescales are also influenced by several other variables, such as carbon prices, U.S. dollar strength, natural disaster shocks, and crude oil supply and inventories. Specifically, in the short and medium term, COS, COP, and DXY also demonstrate high importance, reflecting the significant impact of oil market fundamentals and exchange rate dynamics on short- and medium-term skewness spillovers. In the long term, in addition to the important drivers COP, TYR, and DXY, both EUA and GND also rank highly in importance, which means that tail risk spillovers between energy and food markets are closely tied to energy structure transitions, carbon market dynamics, and natural disaster risks.

Overall, the factor importance analysis from both time- and frequency-domain perspectives reveals that the moment connectedness within the energy-food system is driven by multiple factors, including macroeconomic and financial conditions, carbon pricing, market sentiment, natural disasters, and supply-demand fundamentals. Different types of risk spillovers exhibit distinct sensitivities to different variables, underscoring the importance of distinguishing between risk types and adopting targeted policy interventions. Furthermore, the dominant drivers of moment connectedness vary across timescales, highlighting the necessity of incorporating time-scale considerations into regulatory frameworks, policy design, and cross-market risk management strategies.

\subsection{Robustness checks}

\begin{figure}[!ht]
  \centering
  \includegraphics[width=0.985\linewidth]{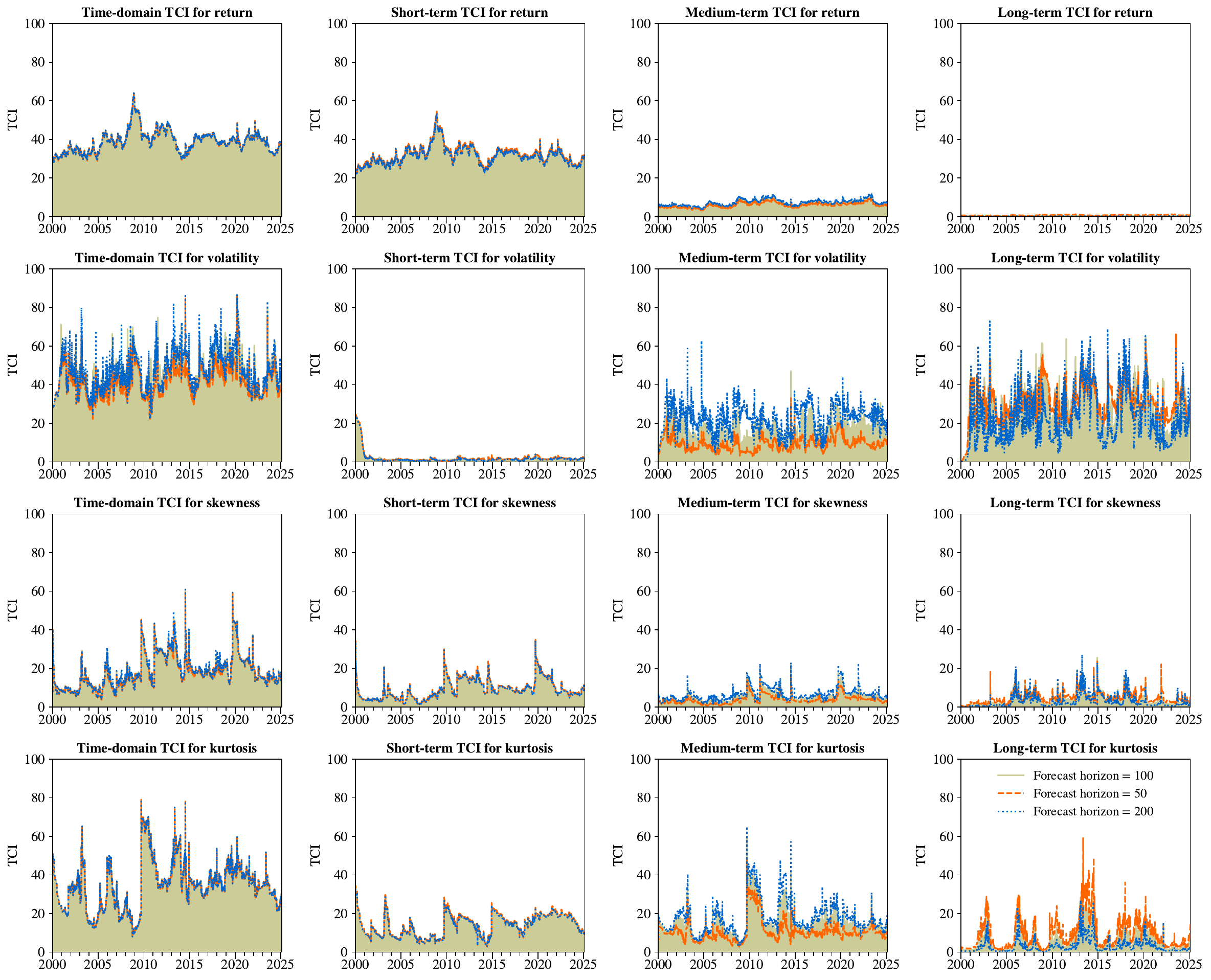}
  \caption{Robustness check of forecast horizon for dynamic total connectedness.}
\label{Fig:RobustnessCheck_WindowSize}
\end{figure}

In the time-frequency connectedness analysis, we set the forecast horizon $H$ to 100, following \cite{Chatziantoniou-Gabauer-Gupta-2023-ResourPolicy} and \cite{Zhou-Wu-Liu-Rognone-2023-NatCommun}. To ensure the reliability and robustness of our empirical results, we further re-estimate the TVP-VAR-DY and TVP-VAR-BK models by setting the forecast horizon to 50 and 200 trading days, respectively. We then compute both time- and frequency-domain TCIs for return, volatility, skewness, and kurtosis. As illustrated in Figure~\ref{Fig:RobustnessCheck_WindowSize}, the dynamic time-frequency TCIs obtained under different forecast horizons exhibit only minor numerical differences, with highly consistent overall patterns being maintained. This confirms that our main findings are robust to alternative forecast horizon specifications.

When identifying the key drivers of different risk spillovers using the random forest model, we calculate both the Gini importance and the permutation importance of each factor to comprehensively assess their relative and absolute importance. This dual verification of factor importance provides strong support for the reliability of our findings on the key determinants of multidimensional risk transmission in the energy-food system.

\section{Conclusions}
\label{S1:Conclude}

Amid rising macroeconomic uncertainty, accelerating climate change, and proliferating geopolitical tensions, risk spillovers within and across global energy and food markets have become increasingly dynamic, complex, and multidimensional. To address these mounting challenges, this paper systematically investigates the energy-food connectedness at four moments -- return, volatility, skewness, and kurtosis -- from both time- and frequency-domain perspectives, and identifies the key drivers of different moment connectedness to enhance our understanding of the underlying risk sources in the energy-food nexus.

We compute the conditional volatility, skewness, and kurtosis for four representative energy commodities and four staple foods using the GJRSK model to account for time-varying higher-order moments. The parameter estimates from the mean equations indicate potential mean-reversion behavior in the energy and soybean markets. The variance equations reveal the high volatility in the crude oil market, and significantly positive leverage coefficients suggest that volatility in both energy and food markets is more sensitive to bad news. Similarly, the skewness leverage coefficients for wheat, soybean, rice, WTI oil, and heating oil are significantly positive, implying more frequent and severe negative returns in these markets, whereas the opposite holds for corn, Brent oil, and natural gas. Furthermore, except for Brent oil and natural gas, most kurtosis equations exhibit significantly positive leverage coefficients, reflecting asymmetric kurtosis effects that adverse shocks lead to higher tail risks. These findings highlight the presence of strong leverage effects in the volatility, skewness, and kurtosis of energy and food markets, thus supporting the suitability of the GJRSK model for capturing higher-order moment dynamics.

The novel TVP-VAR-DY and TVP-VAR-BK methods are adopted to comprehensively examine the time-frequency dynamics of different moment connectedness among energy and food markets. Overall, we find strong evidence of multidimensional risk spillovers within the energy-food system, with intra-market spillovers generally exceeding cross-market spillovers. Importantly, connectedness at different moments and timescales is time-varying and highly responsive to major crises. Specifically, return connectedness remains high and rises sharply during events such as the global financial crisis, the COVID-19 pandemic, and the Russia-Ukraine conflict, with short-term return spillovers playing a dominant role. Volatility connectedness exhibits greater fluctuations and shows phased peaks during multiple crises, with long-term spillovers outweighing short-term ones. Skewness connectedness is typically low but surges during market turmoil, especially in the short term, suggesting that asymmetries in return distributions are mainly transmitted over short horizons during turbulent periods. Kurtosis connectedness is highly volatile and reacts strongly to extreme events, reflecting elevated tail risks, with spillovers more prominent in the short and medium term.

The net pairwise time-frequency connectedness networks provide further insights into the roles of different commodities in risk transmission at various moments and timescales, revealing the complex spillover relationships between energy and food markets. Brent and WTI oil consistently occupy central positions in the networks for return, volatility, skewness, and kurtosis connectedness, emphasizing the pivotal role of crude oil in the energy-food nexus. In the return networks, energy markets dominate the short-term structure, while medium-term linkages are weaker, and long-term connections are more dispersed. For volatility connectedness, no clear spillover paths are observed in the short term, whereas intra-energy linkages strengthen in the medium term, and both intra- and cross-market spillovers intensify in the long run, with crude oil persistently transmitting volatility to food markets. In the skewness networks, food markets exhibit tighter short-term linkages, which shift directionally in the medium term before weakening in the long term. In the kurtosis networks, energy markets show stronger short-term connectedness, cross-market interactions rise in the medium term, and long-term spillovers gradually decline. These results indicate the significant heterogeneity in connectedness structures across diverse moments and timescales, shedding light on the multidimensional risk transmission between energy and food markets.

The random forest model, along with Gini and permutation importance measures, is utilized to determine the key drivers of various types of moment connectedness. The selected factors encompass macroeconomic and financial conditions, external uncertainties, crude oil fundamentals, and climate and geopolitical shocks, all of which are closely related to risk spillovers between energy and food markets. The evaluation metrics indicate that, despite variations in model performance, all models outperform the benchmark, with particularly strong explanatory power for return and skewness connectedness. Notably, short-term connectedness is generally more interpretable than medium- and long-term connectedness, likely because short-term spillovers are better captured by the selected variables, while longer-term spillovers involve more complex dynamics. The factor importance analysis reveals clear multifactorial influences on the energy-food nexus, with variables such as TYR, DXY, EUA, VIX, GND, COP, COS, and EPU demonstrating high importance. Moreover, the core determinants vary significantly across moments and timescales, further uncovering the multidimensionality and heterogeneity of spillover effects.

Our findings provide valuable insights into risk transmission dynamics and underlying risk sources of energy and food markets, with important implications for policymakers, regulators, and market participants. First, policymakers should enhance the dynamic monitoring of risk spillovers and establish multi-level early-warning systems, especially during major crises, to improve the timeliness and effectiveness of policy responses. Given the central role of crude oil in multidimensional risk transmission, regulators need to strengthen macroprudential oversight of crude oil markets to mitigate the adverse effects of energy shocks. Second, interdepartmental coordination is essential, as the key drivers of moment connectedness span various aspects. Synergies among monetary, fiscal, energy, and agricultural policies should be promoted to prevent systemic risks and safeguard energy and food security. Moreover, given the evident heterogeneity of risk spillovers across moments and timescales, policy tools ought to be tailored accordingly, with differentiated and layered interventions based on specific risk types and timescales. Finally, market participants are expected to incorporate higher-moment risks, such as asymmetry and tail risks, into their risk management frameworks to optimize cross-asset and cross-market hedging strategies and enhance portfolio resilience.

Due to data availability and methodological constraints, this study focuses on a representative set of energy and food futures. Future research could expand the sample to include a broader range of commodities and regional markets, thereby improving the extensiveness and applicability of the findings. Regarding the analysis of driving factors, while this study emphasizes economic and financial conditions, market supply-demand dynamics, and climate and geopolitical shocks, future studies may consider incorporating additional variables such as microstructural market indicators to further deepen the understanding of spillover mechanisms. In addition, although the random forest model effectively identifies factor importance, it falls short in capturing complex interdependencies among variables. Future work could adopt more advanced modeling approaches to uncover risk transmission channels and multi-factor interactions, offering richer insights for policy formulation and investment decisions.

\section*{Acknowledgment}

This work was supported by the National Natural Science Foundation of China (Grant Numbers: 72201099, 72171083), the China Scholarship Council, the Fundamental Research Funds for the Central Universities, and the Central Universities' Program for Building World-Class Universities (Disciplines) and Special Development Guidance--Cultural Heritage and Innovation.

\section*{Data availability}

The price data for energy and staple foods in this paper are sourced from the \href{https://www.wind.com.cn}{Wind Database}, and the data on influencing factors are collected from the \href{https://www.wind.com.cn}{Wind Database}, the \href{https://fred.stlouisfed.org}{Federal Reserve Economic Data}, the \href{http://www.policyuncertainty.com}{Policy Uncertainty Website}, the \href{https://www.eia.gov}{U.S. Energy Information Administration}, and the \href{https://www.emdat.be}{EM-DAT Database}.

%

\end{document}